\documentclass[11pt]{elsarticle}
\usepackage{fullpage}
\usepackage{graphicx}
\usepackage{amsmath}
\usepackage{bbm}
\usepackage{amsfonts}
\usepackage{amssymb}
\usepackage{amsthm}
\usepackage{url}
\biboptions{sort&compress}
\usepackage{subfigure}

\begin{document}
\title{NLSEmagic:  Nonlinear {Schr\"odinger} Equation Multidimensional Matlab-based GPU-accelerated Integrators using Compact High-order Schemes}
\author{R. M. Caplan\footnote{Corresponding author.  Present address:  Predictive Science Inc.  9990 Mesa Rim Rd, Suite 170, San Diego, CA 92121. email: caplanr@predsci.com, phone: 858-225-2314}\\[1.0ex]
Nonlinear Dynamical System Group\footnote{\texttt{URL}: http://nlds.sdsu.edu},
Computational Science Research Center, and\\
Department of Mathematics and Statistics,
San Diego State University,\\
San Diego, California 92182-7720, USA\\
}
\date{\today}

\begin{abstract}
We present a simple to use, yet powerful code package called NLSEmagic to numerically integrate the nonlinear Schr{\"o}dinger equation in one, two, and three dimensions.  NLSEmagic is a high-order finite-difference code package which utilizes graphic processing unit (GPU) parallel architectures. The codes running on the GPU are many times faster than their serial counterparts, and are much cheaper to run than on standard parallel clusters. The codes are developed with usability and portability in mind, and therefore are written to interface with MATLAB utilizing custom GPU-enabled C codes with the MEX-compiler interface. The packages are freely distributed, including user manuals and set-up files.
\end{abstract}
\maketitle

{\bf PROGRAM SUMMARY}
\begin{small}
\noindent
{\em Manuscript Title:} NLSEmagic:  Nonlinear {Schr\"odinger} Equation Multidimensional Matlab-based GPU-accelerated Integrators using Compact High-order Schemes                                       \\
{\em Authors:} R. M. Caplan                                                \\
{\em Program Title:} NLSEmagic                                          \\
{\em Journal Reference:}                                      \\
{\em Catalogue identifier:}                                   \\
{\em Licensing provisions: none}                                   \\
{\em Programming language:} C, CUDA, MATLAB                                   \\
{\em Computer:} PC, MAC                                               \\
{\em Operating system:} Windows, MacOS, Linux                                       \\
{\em RAM:} Highly dependent on dimensionality and grid size.  For typical medium-large problem size in three dimensions, 4GB is sufficient.                                            \\
{\em Number of processors used:} Single CPU, number of GPU processors dependent on chosen GPU card (max is currently 3072 cores on GeForce GTX 690)                             \\
{\em Supplementary material:} Setup guide, Installation guide                                
{\em Classification:} 4.3 differential equations, 7.7 Other condensed matter                                         \\
{\em Subprograms used:} Ezyfit, nsoli, vol3d                                      \\
{\em Nature of problem:}\\
  Integrate solutions of the time-dependent one-, two-, and three-dimensional cubic nonlinear Schr{\"o}dinger equation.
   \\
{\em Solution method:}\\
  The integrators utilize a fully-explicit fourth-order Runge-Kutta scheme in time and both second- and fourth-order differencing in space.  The integrators are written to run on NVIDIA GPUs and are interfaced with MATLAB including built-in visualization and analysis tools.
   \\
{\em Restrictions:}\\
  The main restriction for the GPU integrators is the amount of RAM on the GPU as the code is currently only designed for running on a single GPU.
   \\
{\em Unusual features:}\\
  Ability to visualize real-time simulations through the interaction of MATLAB and the compiled GPU integrators.
   \\
{\em Additional comments:}\\
  Program has a dedicated web site at \url{www.nlsemagic.com}.
   \\
{\em Running time:}\\
  A three-dimensional run with a grid dimension of 87x87x203 for 3360 time steps (100 non-dimensional time units) takes about one and a half minutes on a GeForce GTX 580 GPU card.
   \\
\end{small}

\section{Introduction}
\label{s:intro}
The nonlinear Schr{\"o}dinger equation (NLSE) is a universal model describing the evolution and propagation of complex field envelopes in nonlinear dispersive media.  As such, it is used to describe many physical systems including Bose-Einstein condensates \cite{BEC_RCBOOK}, nonlinear optics \cite{ME_MI}, the evolution of water waves, thermodynamic pulses, nonlinear waves in fluid dynamics, and waves in semiconductors \cite{NLSE_nlpdebook}.  The general form of the NLSE can be written as
\begin{equation}
i\,\frac{\partial \Psi}{\partial t} + a\,\nabla^2\Psi - V({\bf r})\,\Psi + s\,|\Psi|^2\,\Psi = 0,
\label{NLSE}
\end{equation}
where $\Psi({\bf r},t) \in \mathbbm{C} $ is the value of the wavefunction, $\nabla^2 = \frac{\partial^2}{\partial x^2} + \frac{\partial^2}{\partial y^2} + \frac{\partial^2}{\partial z^2}$ is the Laplacian operator, and where $a>0$ and $s$ are parameters defined by the system being modeled.  $V(\bf{r})$ is an external potential term, which when included, makes Eq.~(\ref{NLSE}) known as the Gross-Pitaevskii equation \cite{BEC_RCBOOK}.   


Graphical processing units were first developed in order to allow graphics cards to parallelize their massive computations to speed up video games and image processing.  After realizing that such hardware could be adapted to run scientific codes as well, companies such as NVIDIA have developed APIs (such as the compute unified device architecture (CUDA) and OpenCL) to allow GPUs to be used for general computing.

For many problems, GPU computing is seen as a large improvement over previous parallel techniques since access to a large cluster can be expensive or not available.  A typical GPU (as of this writing) can have up to $3072$ processing cores and up to 12GB of RAM, with a throughput of over a Tera-FLOP on a single card.  The price of the GPUs is another major factor, as one (as of this writing) can purchase an off-the-shelve GPU with $1536$ cores, and $4$GB of RAM for under \$450.  This allows for super-computing capabilities to be realized on a single desktop PC for medium-sized problems.

GPUs have been used for various finite-difference PDE integrator codes with good results \cite{CUDA_FD_2004,CUDA_FD_maxwell,CUDA_FD_2008,CUDA_FD_2009,CUDA_FD_3Dwave,CUDA_FD_GL}.  In Ref.~\cite{CUDA_FD_2004}, the authors wrote code to simulate the two-dimensional Maxwell equations yielding a speedup of up to ten times when compared to the CPUs of the day (2004).  They did this before CUDA was developed using assembly code directly.  Ref.~\cite{CUDA_FD_maxwell} also did not utilize CUDA for their simulations of the three-dimensional Maxwell equations, yielding speedups of over $100$ times.  The two-dimensional Maxwell equations were simulated in Ref.~\cite{CUDA_FD_2008} using CUDA code where speedups of $50$ were reported.  In  Ref.~\cite{CUDA_FD_2009}, the authors simulate the three-dimensional wave equation with an eighth-order finite-difference scheme using CUDA on multiple GPUs.  A similar multi-GPU code was shown in Ref.~\cite{CUDA_FD_3Dwave} for the three-dimensional wave equation using a fourth-order finite-difference scheme where speedups of $60$ were reported.  In Ref.~\cite{CUDA_FD_GL}, the authors use CUDA to simulate the multi-dimensional complex Ginzburg-Landau equation (a generalization of the NLSE), but speedups versus CPU codes were not reported. All these studies indicate that a GPU treatment of the NLSE would be beneficial.

In this paper we describe our implementation of a code package called NLSEmagic (Nonlinear Schr{\"o}dinger Equation Multidimensional Matlab-based GPU-accelerated Integrators using Compact High-order Schemes) which integrates the NLSE in one, two, and three dimensions.  Previous codes have been published which integrate the NLSE in multi-dimensions \cite{CODE_GP_FORTRAN} including a recent C-based code which implements OpenMP parallelism for multi-core CPU systems \cite{CODE_GP_C}.  The code presented here differs from the previously published codes in two major ways. One, the NLSEmagic codes utilize NVIDIA GPUs for integrating the NLSE which can yield large speedups compared to a serial-CPU code, as well as decent speedups compared to typical multi-core CPUs (see Sec.~\ref{s:speedup}). Second, the codes presented here are MATLAB-based allowing for user-friendly setup of problems, ease of use, built-in MATLAB analysis and optimization functions, and on-the-fly visualizations drastically reducing the need for post-processing results.

In order to allow the MATLAB-based codes to run at speeds equal to typical compiled codes, we write the main NLSE integrators as custom C codes which connect to MATLAB through the MEX interface compiler.  The compiled integrators are called from a MATLAB script code and are up to $10$ times faster than writing the integrators as native MATLAB script codes (as shown in Sec.~\ref{s:mexspeedup}).  The C codes written for the MEX interface have an additional advantage as they can be written to only have custom MATLAB-based code for the input and output of data from MATLAB allowing them to be portable to non-MATLAB codes if desired.

Since the native language for GPU-compatible CUDA codes is C, adding GPU functionality to MATLAB is possible using the MEX compiler interface.  The newest versions of MATLAB have GPU-compatibility built-in \cite{CUDA_matlabgpu}, and ways to compile CUDA C code segments into callable MATLAB functions.  However, because many researchers do not have the newest versions of MATLAB, and in the interest of portability, it is preferable not to use these built-in functions, but rather write portable C integrator codes which contain CUDA code directly, and compile them with the CUDA-capable MEX compiler called NVMEX \cite{CUDA_matlabgpuOLD}.

Both serial-CPU and GPU-accelerated MEX routines are included in NLSEmagic and are called from MATLAB equivalently.  They integrate the NLSE over a specified number of time-steps (the chunk-size) before returning the current solution $\Psi$ to MATLAB.  They utilize fully explicit finite-difference schemes, meaning that the solution at the next time-step only relies on values of the solution at the previous time-step.  This eliminates the need for solving linear systems and nonlinear iterations at each time-step (which most implicit schemes do).  The drawback is that a conditional stability criteria exists that limits the time-step size based on the spatial grid-spacing and the scheme used to approximate the spatial derivatives.  This relationship is discussed in Sec.~\ref{s:stb}, and is implemented into the driver scripts of NLSEmagic.  The spatial schemes used are a standard second-order central difference and a two-step high-order compact scheme (2SHOC) \cite{ME_2SHOC} which is fourth order accurate.

This paper is organized as follows:  In Sec.~\ref{s:gpgpu}, we describe general purpose GPUs, specifically those produced by NVIDIA.  We discuss the physical structure of the GPUs as well as the CUDA API logical structure.  Compatibility and portability issues are also discussed. In Sec.~\ref{s:numalg}, we discuss the numerical algorithms used in NLSEmagic including boundary conditions and stability.  The example solutions to the NLSE that are used throughout the simulations in this paper are described in Sec.~\ref{s:examples}.  The serial code implementation of the integrators is described in Sec.~\ref{s:serial}, including the script integrators and a detailed description of the MEX code integrators.  Speedup results between the MEX codes and the equivalent C codes are shown.  Sec.~\ref{s:cudamex} describes in detail our implementation of the NLSEmagic integrators in CUDA MEX codes.  The speedup results of the CUDA MEX codes compared to the equivalent serial MEX codes are shown in Sec.~\ref{s:speedup}.  An overview of the NLSEmagic code package is described in Sec.~\ref{s:pack} including distribution details.

\section{General purpose GPU computing with NVIDIA GPUs}
\label{s:gpgpu}
Although there are other companies that produce GPU graphics cards (such as ATI), NVIDIA is by far the most established when it comes to general-purpose GPUs (GPGPU).  Their compute-only Tesla GPGPU cards as well as their native compute unified device architecture (CUDA) API and enormous support and development infrastructure make NVIDIA the most logical choice for developing and running GPU codes (compatibility and portability concerns will be discussed in Sec.~\ref{s:port}).  In this section we focus on the physical structure of the GPUs as well as the logical structure defined by the CUDA API.

\subsection{NVIDIA GPU physical structure}
\label{s:gpuphys}
\begin{figure}[hbt]
\centering
\includegraphics[width=5in]{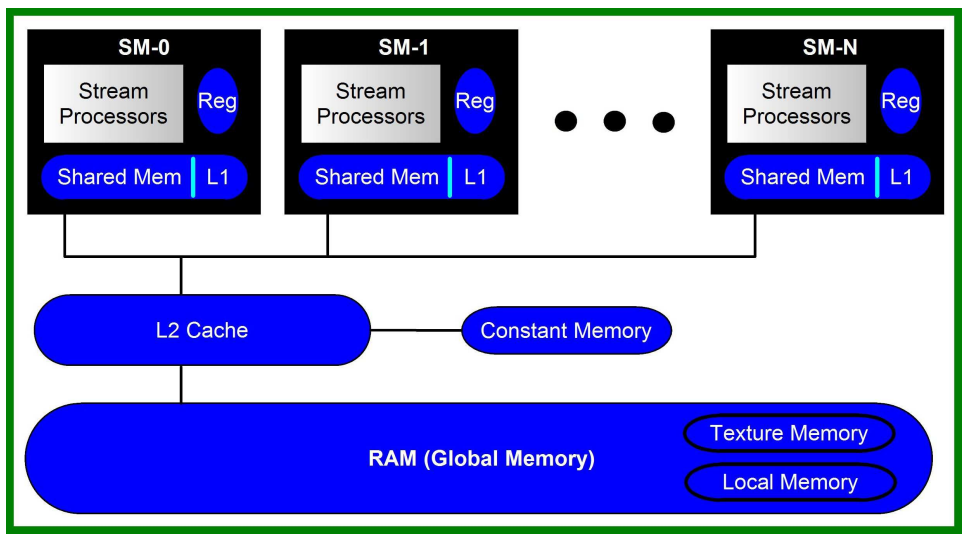}
\caption[Schematic of an NVIDIA GPU.]{(Color online) Simplified schematic of an NVIDIA GPU with Fermi architecture (A Tesla architecture looks the same but without caches).\label{f:cudastruct}}
\end{figure}

Since the programming model for NVIDIA GPUs is very closely related to their physical design, it is important to have an overview of their physical structure.  A simplified schematic of a Fermi architecture NVIDIA GPU is shown in Fig.~\ref{f:cudastruct}.
The GPU contains a number of stream multiprocessors (SM) which each contain a number of stream processors (SP) which are the compute cores of the GPU.  Each SM performs computations independently and simultaneously from the other SMs and has no connectivity or communication with other SMs.  Each SM has a small, fast memory which is configured to allocate some space for a level 1 (L1) cache and the rest of the space for shared memory.  The shared memory is shared between all SPs, so each core can access any part of shared memory.  Each SM also has fast registers for the SPs to use for computations and storage of local variables.

The main memory of the GPU is a large amount of DRAM called the \emph{global memory space} which, depending on the GPU, can have a capacity from 512MB up to 12GB.  Accesses to the global memory from the SPs are much slower than accesses to shared memory \cite{cudadoc_pg}.  Parts of the global memory space are used for local variables (when they cannot fit into registers or cache) and texture memory.  Between the DRAM and the SMs is a level 2 (L2) cache which improves memory performance when SMs access global memory.  Additional hardware details are beyond the required scope of this paper.  

A typical GPU code transfers data from the host computer's RAM to the GPU RAM, performs computations using the data on the SMs of the GPU, and then when completed, transfers the resulting data back to the host computer's RAM.  The memory transfer has large latency associated with it, and so it is ideal to compute as much as possible on the GPU before transferring the data back.   However, since the host computer cannot `see' the data until it has been transfered back from the GPU, a trade-off between maximum performance and usability is often encountered for time-stepping problems (see Sec.~\ref{s:chunk-size}).  


\subsection{CUDA API and logical structure}
\label{s:gpulogic}
NVIDIA allows programmers to utilize its GPUs through an API called the compute unified device architecture (CUDA) API. CUDA is a code extension to C/C++ which gives low-level access to the GPU's memory and processing abilities.  The CUDA codes are compiled by a free compiler provided by NVIDIA called {\tt nvcc} (a FORTRAN CUDA compiler is also available through PGI).

There are two sections of a CUDA C code (usually written in a file with a {\tt .cu} extension).  One section of code is the host code which is executed on the CPU.  This section contains setup commands for the GPU, data transfer commands to send data from the CPU to the GPU (and vice-versa), and code to launch computations on the GPU.  The second part of a CUDA code contains what are known as CUDA \emph{kernel} routines. These are routines which get compiled into binaries which are able to be executed on the GPU by calls invoked by the host code.

\begin{figure}[hbt]
\centering
\includegraphics[width=5in]{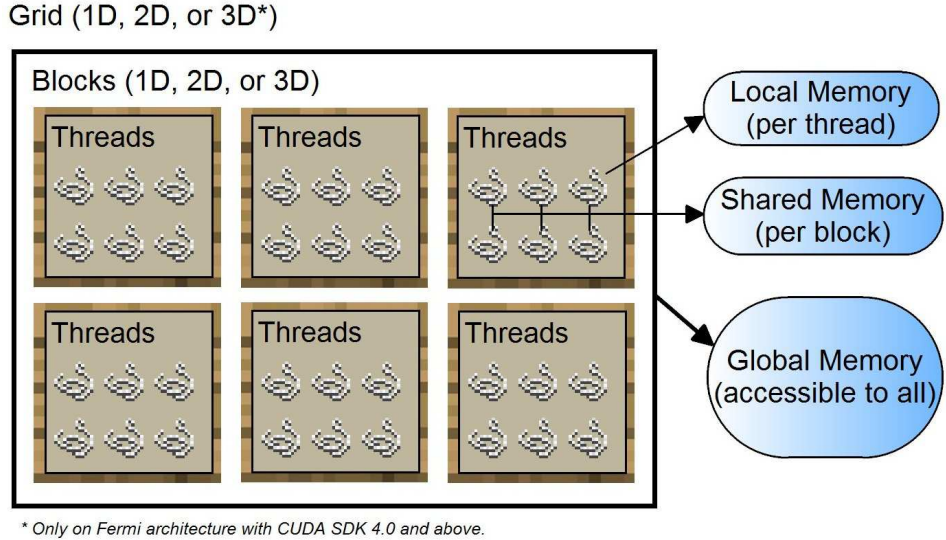}
\caption[Schematic of the CUDA API logic structure.]{(Color online) Schematic of the CUDA API logic structure.\label{f:cudalogstruct}}
\end{figure}

The CUDA programming model is based on a logical hierarchical structure of \emph{grids}, \emph{blocks}, and \emph{threads}.  This hierarchy is shown in Fig.~\ref{f:cudalogstruct}.
When a kernel is launched, it is launched on a user-defined compute-grid which can be one-, two-, or three-dimensional.  This grid contains a number of blocks where the computations to be performed are distributed.  Each block must be able to be computed independently as no synchronization is possible between blocks.  Each block contains a number of threads, which are laid out in a configuration that is one-, two-, or three-dimensional.  The threads perform the computations and multiple threads are computed simultaneously on the SM.  Synchronization of threads within a block is possible, but for best performance, should be kept to a minimum.  Each thread has its own local memory and all threads in a block have access to a block-wide shared memory.  All threads in all blocks on the grid have access to the global memory.  

Each thread has access to intrinsic variables (stored in registers) which are used to identify its position in the compute-grid and block.  The grid dimensions are stored in variables called {\tt gridDim.x}, {\tt gridDim.y}, and {\tt gridDim.z} and the block dimensions are stored in {\tt blockDim.x}, {\tt blockDim.y}, and {\tt blockDim.z}.  The block that a thread is contained in is given in coordinates by the variables {\tt blockIdx.x}, {\tt blockIdx.y}, and {\tt blockIdx.z}.   The thread's position in the block is given by the variables {\tt threadIdx.x}, {\tt threadIdx.y}, and {\tt threadIdx.z}.  All these variables are used to determine what part of the problem the specific thread should compute.

The logical structure relates directly to how the hardware of the GPU executes a kernel code.  Each block is executed by an SM, where each thread is executed by an SP.  This is why the threads within a block have access to a shared memory, but blocks cannot access each others shared memory.  The local memory of each thread is stored in the registers of the SM, unless the register space is used up in which case the local variables get stored in cache, and, if the cache's are full, to global memory (reducing performance).

\subsection{Compatibility}
\label{s:compat}
Since new and updated GPU cards are being developed and released continuously (such as the recently released Kepler-based GPUs), compatibility can become a major concern.  To help with this issue, NVIDIA has maintained a compatibility scheme known as the major/minor compute capability of a GPU.  This is a number in the form of $y.x$ where $y$ and $x$ are the major and minor compute capabilities respectively.  Whether code written and compiled for one GPU will work on a different model GPU depends on the compute capabilities of the two cards and on how the code was compiled (see Ref.~\cite{cudadoc_pg} for details).

CUDA codes can be compiled into a binary executable for a specific GPU known as `cubin' objects, and/or be compiled into parallel thread execution (PTX) assembly code which is compiled into binary at runtime for the GPU being used (which can have a slight performance impact).

The compiler options for the NLSEmagic code package described in this paper have been set to compile PTX code for 1.0, 1.3, and 2.0, as well as cubin objects for compute capability 2.0 (The codes have also been tested to compiled on the new compute capability 3.x architectures, which requires modification to the compile scripts).

\subsection{Portability}
\label{s:port}
A major reason many scientific programmers are hesitant to write their programs for GPUs and using CUDA to do so, is the issue of portability.  As a solution to these concerns, a new API called OpenCL \cite{CUDA_OpenCL} has been developed by the Khronos Group \cite{CUDA_OpenCL} which can function across multiple GPU vendors, as well as other CPU architectures.  Although code written in OpenCL is guaranteed to work on all multi-core architectures, specific optimization is necessary for the codes to work efficiently.  A number of studies have been done comparing the performance of OpenCL versus CUDA \cite{CUDA_OpenCLvsCUDA1,CUDA_OpenCLvsCUDA2,CUDA_OpenCLvsCUDAold}.  In general, the CUDA codes outperform the OpenCL equivalent codes (sometimes by a significant percentage).  However, with further OpenCL-specific performance tunings, it is possible to get the OpenCL codes to run nearly as fast as the CUDA codes.  OpenCL and CUDA are very similar in terms of logical structure and code design.  Therefore, codes written in CUDA (such as NLSEmagic) can easily be changed into OpenCL codes (in fact, an \emph{automatic} CUDA-to-OpenCL converter called {\tt cu2cl} has recently been developed \cite{CUDA_CU2CL}).

Many other portability options exist for CUDA codes and GPU codes in general including the Portland Group's x86 CUDA compiler \cite{CUDA_x86}, the MCUDA compiler created by the University of Illinois \cite{CUDA_MCUDA}, and the Ocelot project from Georgia Tech \cite{CUDA_PTXonAMDandIntel} (see references for details).   

\section{Finite-difference schemes}
\label{s:numalg}
There are many methods to numerically integrate the NLSE.  The scheme we use in NLSEmagic is the classic fourth-order Runge-Kutta (RK4) in time and a choice of either a standard second-order central-difference (CD) in space for the Laplacian operator, or a fourth order two-step high order compact scheme (2SHOC) \cite{ME_2SHOC}.  These schemes are some of the simplest explicit finite-difference schemes for the NLSE that are conditionally stable.  Once the principles of the CUDA implementation are understood, more advanced schemes could be added to the code package (such as Mimetic operators \cite{NUM_Mimetic}).

The basic methods used in the code presented here are different than those used in Refs.~\cite{CODE_GP_FORTRAN,CODE_GP_C}.  There, the authors use a semi-implicit, second-order in space and time split-step Crank-Nicholson (SSCN) method.  The major difference between the methods are that the RK4 used here is conditionally stable (see Sec.~\ref{s:stb}), while the SSCN scheme is unconditionally stable, which in some cases can be more efficient depending on the accuracy and resolution requirements of the problem.  However, since the schemes used in NLSEmagic are up to fourth-order accurate in both space and time, the equivalent accuracy of a simulation can often be realized on a coarser grid, in which case the problem of bounds on the time-step can be greatly reduced.

In three dimensions, the computational grid consists of $N\times M\times L$ grid points ($N\times M$ in two dimensions, $N$ in one dimension).  The wave function of the NLSE is discretized as
\begin{equation}\label{disc}
\Psi(x,y,z,t) \equiv \Psi_{i,j,k}^n,
\end{equation}
where $n$ is the current time step and $i,j,k$ are the spatial positions on the grid. The time-step is denoted as ${\bf k}$ and the spatial step-size is the same in each direction and defined by $h$.  Therefore, $t=n{\bf k}$, $x=x_0+ih$, $y=y_0+jh$, and $z=z_0+k\,h$.  We define the end-time of the simulation as $t_{\mbox{\scriptsize end}} = {\bf k}\,t_{\mbox{\scriptsize res}}$ where $t_{\mbox{\scriptsize res}}$ is the number of time-steps taken.

\subsection{Fourth-order Runge-Kutta}
In order to avoid (especially in three-dimensional problems) the need to solve a linear system and run nonlinear iteration routines at each time step, we use a fully explicit time-difference scheme.  The classic four-stage, fourth-order Runge-Kutta (RK4) scheme \cite{RK4} is the simplest single-stage explicit time-stepping that can be used for the NLSE directly (i.e, not using a real-imaginary staggered step as in Ref.~\cite{FD_STBNLSEDISS}) and is (conditionally) stable \cite{ME_RK4STB}.  The RK4 scheme can be written in algorithmic form as
\begin{alignat}{5}
\label{RK4}
& 1)\; K_{\mbox{\scriptsize tot}} = F(\Psi^n) 
&\qquad & 6)\;  K_{\mbox{\scriptsize tmp}} = F(\Psi_{\mbox{\scriptsize tmp}})
\\
& 2)\;  \Psi_{\mbox{\scriptsize tmp}} = \Psi^n + \frac{{\bf k}}{2}\, K_{\mbox{\scriptsize tot}}   
&\qquad & 7)\;  K_{\mbox{\scriptsize tot}} = K_{\mbox{\scriptsize tot}} + 2\, K_{\mbox{\scriptsize tmp}} \notag
\\
&3)\;  K_{\mbox{\scriptsize tmp}} = F(\Psi_{\mbox{\scriptsize tmp}})  
&\qquad &  8)\;  \Psi_{\mbox{\scriptsize tmp}} = \Psi^n + {\bf k}\,K_{\mbox{\scriptsize tmp}} \notag
\\
&4)\;  K_{\mbox{\scriptsize tot}} = K_{\mbox{\scriptsize tot}} + 2\, K_{\mbox{\scriptsize tmp}} 
&\qquad &9)\;  K_{\mbox{\scriptsize tmp}} = F(\Psi_{\mbox{\scriptsize tmp}}) \notag
\\
& 5)\;  \Psi_{\mbox{\scriptsize tmp}} = \Psi^n + \frac{{\bf k}}{2}\, K_{\mbox{\scriptsize tmp}} 
&\qquad &10)\;  \Psi^{n+1} = \Psi^n + \frac{{\bf k}}{6}\, (K_{\mbox{\scriptsize tot}} + K_{\mbox{\scriptsize tmp}}), \notag 
\end{alignat}   
where
\[
F(\Psi) = \frac{\partial \Psi}{\partial t} = i\left[a\nabla^2\Psi + \left(s\,|\Psi|^2 - V(\bf{r})\right)\Psi\right].
\]
In NLSEmagic, the Laplacian of $\Psi$ in $F(\Psi)$ is evaluated using either the CD or 2SHOC scheme. We denote the resulting combined scheme as RK4+CD and RK4+2SHOC respectively. 

The RK4 has an error which is fourth-order in the time step ($O({\bf k}^4)$).  In NLSEmagic, this error will always be much less than the error in computing the Laplacian of $\Psi$ because, as we will see in Sec.~\ref{s:stb}, in order for the RK4 scheme to be stable we must have ${\bf k} \propto h^2$, the proportionality constant depending on the dimensionality of the problem and spatial scheme. Thus, an error of $O({\bf k}^4)$ in our case is proportional to an error of $O(h^8)$, and since our spatial schemes are accurate only to $O(h^2)$ or $O(h^4)$, the time-step error resulting from the RK4 is negligible.

\subsection{Difference schemes for the Laplacian operator}
\label{s:numlap}
NLSEmagic has two options for computing the Laplacian.  The first is the standard second-order central difference scheme (CD), and the other is a two-step high-order compact scheme (2SHOC) which is fourth-order accurate \cite{ME_2SHOC}.  

The 2SHOC scheme allows for the use of courser grids with equivalent accuracy to the CD scheme, or for use with simulations that need high accuracy. Usually, high-order schemes have a wide stencil, which are not ideal for parallel applications due to extra communication and/or memory latency. Also, the grid points near the boundary are difficult to deal with.  The 2SHOC schemes avoid these problems because in each step, the computation is compact (relying only on adjacent grid points) making it easier to code.  The drawbacks of the scheme are that they require the storage of an extra temporary variable and an increase in the number of floating point operations per grid point when compared to the standard fourth-order wide stencils.  However, for our GPU implementation, we feel the advantages of using the 2SHOC scheme outweigh the disadvantages.

In one-dimension, the two steps of the 2SHOC scheme are defined as
\begin{alignat}{3}
&1) \qquad &D_i &= \frac{1}{h^2}\left(\Psi_{i+1} - 2\Psi_i + \Psi_{i-1}\right), \label{2shoc1d}\\
&2) \qquad &\nabla^2\Psi_i &\approx \frac{7}{6}D_i - \frac{1}{12}\left(D_{i+1} + D_{i-1}\right).\label{2shoc1d2}
\end{alignat}
In two dimensions, the 2SHOC scheme is given by
\begin{alignat}{3}
&\begin{tabular}{ll} 
$1)$ & $D_{i,j} = \dfrac{1}{h^2}$ 
\begin{tabular}{|c|c|c|} \hline
  &  1 &   \\ \hline
1 & -4 & 1 \\ \hline
  &  1 &  \\ \hline
\end{tabular}
$\Psi_{i,j}$
\end{tabular} \label{2d2shocs1}
\\
\;&\; \notag
\\
&\begin{tabular}{ll} 
$2)$ & $\nabla^2\Psi_{i,j} \approx -\dfrac{1}{12}$
\begin{tabular}{|c|c|c|} \hline
  &   1 &   \\ \hline
1 & -12 & 1 \\ \hline
  &   1 &   \\ \hline
\end{tabular}
$D_{i,j} + \dfrac{1}{6\,h^2}$  
\begin{tabular}{|c|c|c|} \hline
1 &    & 1 \\ \hline
  & -4 &   \\ \hline
1 &    & 1 \\ \hline
\end{tabular}
$\Psi_{i,j},$
\end{tabular} \label{2d2shocs2}
\end{alignat}
and in three dimensions,
\begin{alignat}{5}
&\begin{tabular}{ll} 
$1)$ & $D_{i,j,k} = \dfrac{1}{h^2}\left(\;
\begin{tabular}{|c|c|c|} \hline
  &   &   \\ \hline
  & 1 &   \\ \hline
  &   &   \\ \hline
\end{tabular}
\;\Psi_{i,j+1,k} +
\begin{tabular}{|c|c|c|} \hline
  &  1 &   \\ \hline
1 & -6 & 1 \\ \hline
  &  1 &   \\ \hline
\end{tabular}
\;\Psi_{i,j,k} +
\begin{tabular}{|c|c|c|} \hline
  &   &   \\ \hline
  & 1 &   \\ \hline
  &   &   \\ \hline
\end{tabular}
\;\Psi_{i,j-1,k}\;\right),$
\end{tabular}
\label{3d2shocs}
\\
\;&\; \notag
\\
&\begin{tabular}{ll}
$2)$ & $\nabla^2\Psi_{i,j,k} \approx -\dfrac{1}{12}\left(\;
\begin{tabular}{|c|c|c|} \hline
  &   &   \\ \hline
  & 1 &  \\ \hline
  &   &   \\ \hline
\end{tabular}
\;D_{i,j+1,k} +
\begin{tabular}{|c|c|c|} \hline
  &   1 &   \\ \hline
1 & -10 & 1 \\ \hline
  &   1 &   \\ \hline
\end{tabular}
\;D_{i,j,k} +
\begin{tabular}{|c|c|c|} \hline
  &   &   \\ \hline
  & 1 &   \\ \hline
  &   &   \\ \hline
\end{tabular}
\;D_{i,j-1,k} \; \right)$
\end{tabular}
\label{3d2shocs2} \\
&\begin{tabular}{lll}
$\qquad$ & $\qquad$ & $+ \dfrac{1}{6\,h^2}\left(\;
\begin{tabular}{|c|c|c|} \hline
  & 1 &   \\ \hline
1 &   & 1 \\ \hline
  & 1 &   \\ \hline
\end{tabular} 
\;\Psi_{i,j+1,k} +
\begin{tabular}{|c|c|c|} \hline
1 &     & 1  \\ \hline
  & -12 &    \\ \hline
1 &     & 1  \\ \hline
\end{tabular}
\;\Psi_{i,j,k} +
\begin{tabular}{|c|c|c|} \hline
  & 1 &   \\ \hline
1 &   & 1 \\ \hline
  & 1 &   \\ \hline
\end{tabular}
\;\Psi_{i,j-1,k} \; \right).$
\end{tabular} \notag
\end{alignat}
The standard second-order central differencing in each dimension is simply given by step one of the 2SHOC schemes shown above.

\subsection{Boundary conditions}
\label{s:bc}
NLSEmagic includes three boundary condition options:  1) Dirichlet (D), 2) Modulus-Squared Dirichlet (MSD), and 3) Laplacian-zero (L0).  For one-dimensional simulations, NLSEmagic1D also contains code for a one-sided (1S) boundary condition.

For each boundary condition, it is necessary to define the time-derivative at a boundary point so that it can be implemented in each step of the RK4 scheme.  In order to use the boundary conditions with the 2SHOC scheme, they additionally need to be expressed in terms of the Laplacian in order to compute proper boundaries in the first step of the 2SHOC.

Dirichlet boundary conditions are defined as when the value of the function at the boundaries is fixed to be a constant value, i.e. 
\begin{equation}
\label{BCDdef}
\Psi_b = B, 
\end{equation}
where the subscript $b$ represents a boundary point, and $B$ is a real constant.  The time-derivative formulation of this condition is simply
\begin{equation}
\label{BCDdt}
\frac{\partial \Psi_b}{\partial t} = 0.
\end{equation}
For the NLSE, the Dirichlet condition in terms of the Laplacian of the wavefunction is
\begin{equation}
\label{BCDlap}
\nabla^2\Psi_b = -\frac{1}{a} (s|\Psi_b|^2 - V_b) \Psi_b.
\end{equation}

The Modulus-squared Dirichlet boundary condition is defined to be where the modulus-squared of the wavefunction at the boundary is set to a fixed constant, i.e.
\begin{equation}
\label{BCMSDdef}
\left|\Psi_b\right|^2=B,
\end{equation}
where $B$ is a real constant.  The MSD boundary condition is useful for many problems, especially those with a constant density background (such as the examples described in Sec.~\ref{s:examples}).  The time-derivative formulation of the MSD boundary condition can be approximated as \cite{ME_MSD}
\begin{equation}
\label{msd}
\frac{\partial \Psi_b}{\partial t} \approx i\,\mbox{Im}\left[  \frac{1}{\Psi_{b-1}}  \frac{\partial \Psi_{b-1}}{\partial t}  \right]\Psi_b,
\end{equation}
where $\frac{\partial \Psi_{b-1}}{\partial t}$ is precomputed using the internal finite-difference scheme.  This reliance on the previously computed value of the time derivative of the interior point requires special treatment in the CUDA codes of NLSEmagic (see Sec.~\ref{s:cudamex} for details).  The Laplacian form of the MSD for the NLSE is
\begin{equation}
\label{BCMSDlap}
\nabla^2\Psi_b \approx \left[ \mbox{Im}\left(i\, \frac{\nabla^2\Psi_{b-1}}{\Psi_{b-1}}\right) + \frac{1}{a}\,\left(N_{b-1} - N_b\right)\right]\Psi_b,
\end{equation}
where
\begin{equation}
\label{nbnb1}
N_b = s\,|\Psi_b|^2 - V_b, \qquad N_{b-1} = s\,|\Psi_{b-1}|^2 - V_{b-1}. \notag
\end{equation}

The Laplacian-zero boundary condition is defined as setting the Laplacian of the wavefunction at the boundary to zero.  The time-derivative formulation of the L0 boundary condition for the NLSE is given by
\begin{equation}
\label{BCL0dt}
\frac{\partial \Psi_b}{\partial t} = i\,(s|\Psi_b|^2 - V_b)\,\Psi_b,
\end{equation}
while the Laplacian form is, by definition,
\begin{equation}
\label{BCL0lap}
\nabla^2 \Psi_b = 0.
\end{equation}

The boundary conditions included in NLSEmagic are chosen due to their relatively easy implementation and broad applicability.  Additional boundary conditions could be added to the NLSEmagic codes if needed.

\subsection{Stability}
\label{s:stb}
The finite-difference schemes used in NLSEmagic are fully explicit and, like most explicit schemes, are conditionally stable.  This means that for a given spatial step size $h$, there exists a maximum value of the time-step ${\bf k}$ that can be used without the scheme becoming unstable (i.e, blow-up).  In Ref.~\cite{ME_RK4STB}, we performed a full linearized stability analysis for the schemes in NLSEmagic.  The results are coded into the scripts of NLSEmagic where, given the spatial step size $h$, the largest stable time-step ${\bf k}$ is automatically set.  However, from experience it is observed that the purely linear bounds found in Ref.~\cite{ME_RK4STB} are typically very close to the linearized bounds, and since the bounds must be artificially lowered (due to nonlinear effects), using the purely linear bounds is often all that is needed (an exception to this would occur in the presence of large external potential values).  Therefore, the basic driver scripts of NLSEmagic only compute the linear bounds.  

The linear stability bounds of the RK4+CD scheme for the NLSE with $V({\bf r})=0$ and $s=0$ using Dirichlet or periodic boundary conditions are 
\begin{equation}
\label{stblincd}
{\bf k} < \frac{h^2}{d\,\sqrt{2}\, a},
\end{equation}
where $d$ is the dimensionality of the problem (1, 2, or 3).  The equivalent linear bound for the RK4+2SHOC scheme are given by
\begin{equation}
\label{stblin2shoc}
{\bf k} < \left(\frac{3}{4}\right)\frac{h^2}{d\,\sqrt{2}\, a}.
\end{equation}
In the NLSEmagic driver scripts, these bounds are multiplied by $0.8$ to avoid instability due to nonlinearities, boundaries, and/or the external potential.

\section{Example test problems}
\label{s:examples}
Before describing the implementation of the NLSEmagic codes, we show here the example test problems that we will make use of in testing the speedup of the codes.

\begin{figure}[hbt]
\centering
\begin{center}
\includegraphics[width=2in]{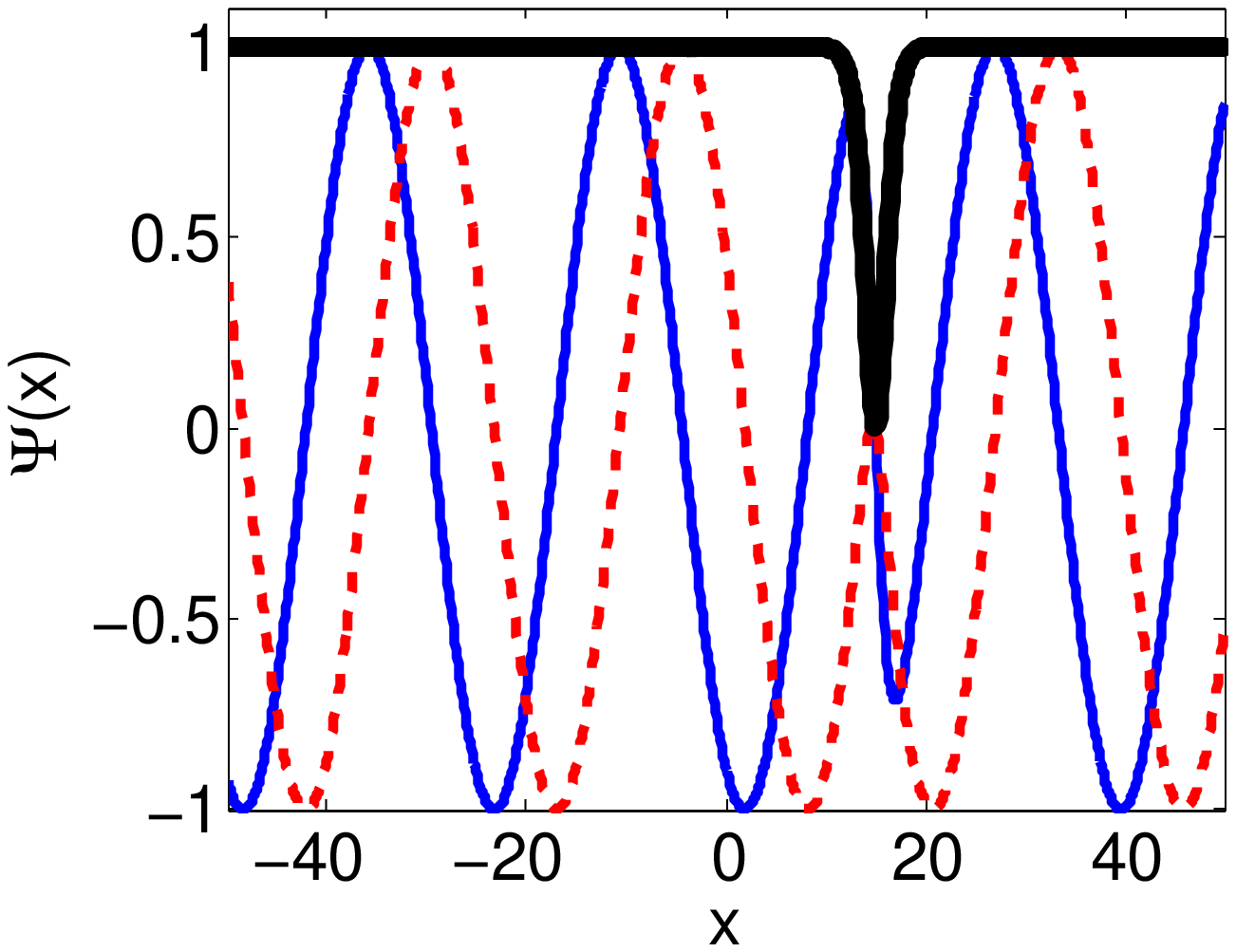}
\includegraphics[width=2.2in]{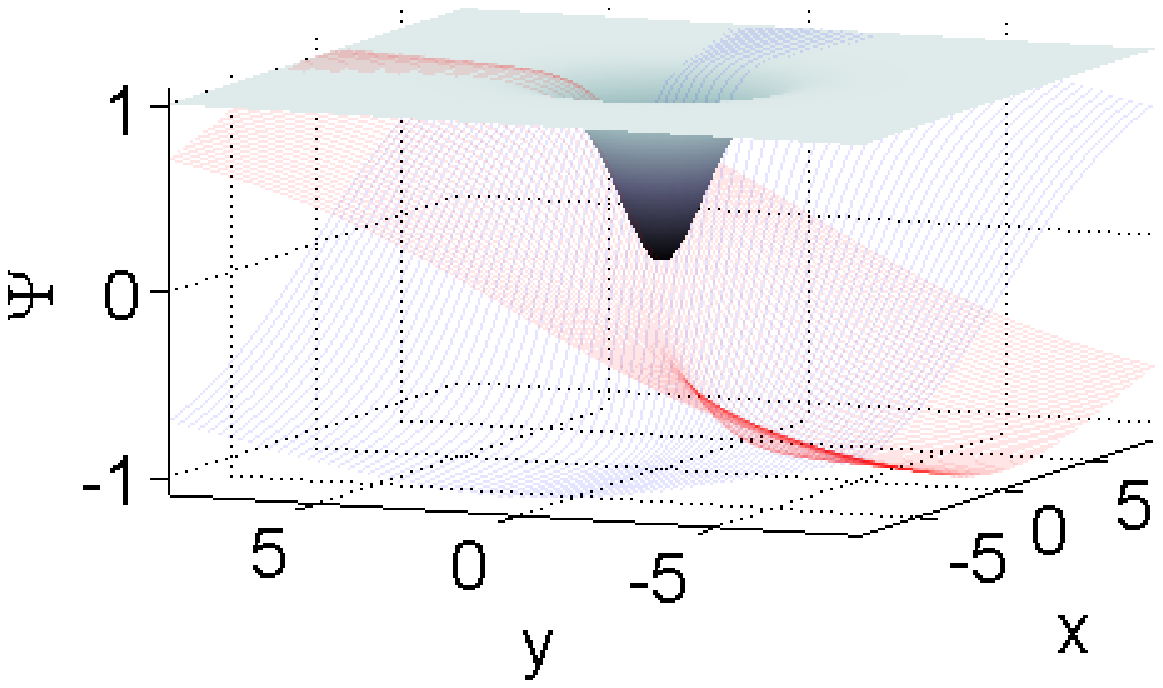} 
\includegraphics[width=2in]{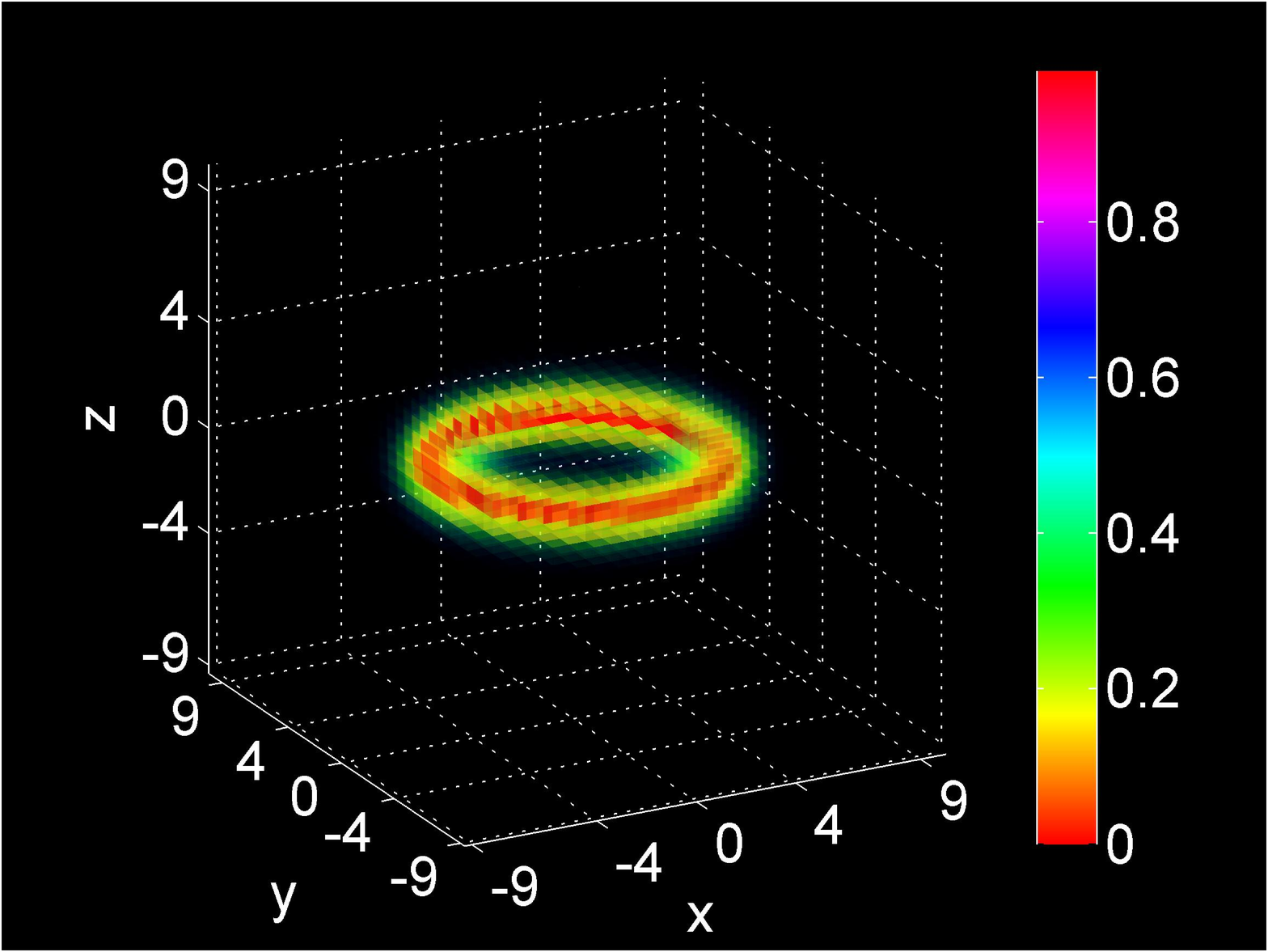} 
\caption{(Color online) Left:  Example of the one-dimensional dark soliton of Eq.~(\ref{soliton}) at time $t=30$.   The thin solid (blue) line and the thin dashed (red) line are the real and imaginary parts of $\Psi$ respectively. The thick solid (black) line is the modulus-squared of the wave-function $|\Psi|^2$.  The grid size here is $N=1000$.  Middle:  Example of the approximate two-dimensional dark vortex of Eq.~(\ref{exmp2Dvort}) at time $t=0$.   The (blue) and (red) mesh lines are the real and imaginary parts of $\Psi$ respectively. The solid (gray) surface is the modulus-squared of the wave-function $|\Psi|^2$.  The grid-size is $N=M=70$.  Right:  Example of the approximate three-dimensional dark vortex ring of Eq.~(\ref{3dvr1}) at time $t=0$.  The image shows an inverted-volumetric rendering of the modulus-squared of $\Psi$.  The grid size is $N=M=L=29$.\label{f:examples}}
\end{center}
\end{figure}

In one-dimension, we use the following exact co-moving dark soliton solution to the NLSE with $V(x)=0$ and $s<0$ \cite{SOL_Bright_Gray_Dark_Opt}:
\begin{equation}
\label{soliton}
\Psi(x,t) = \sqrt{\left|\frac{\Omega}{s}\right|}\,\mbox{tanh}\left[\sqrt{\frac{|\Omega|}{2a}}\,(x-c\,t)\right]\,\mbox{exp}\left(i\left[\frac{c}{2a}\,x + \left(\Omega - \frac{c^2}{4a}\right)t\right]\right),
\end{equation}
where $c$ is the velocity of the soliton and $\Omega$ is the frequency.  The soliton describes a localized rarefaction curve in the modulus squared of $\Psi$ which propagates without dispersion or dissipation amidst a constant density background.  For our simulations we use $s=-1$, $a=1$, $c=0.5$, and $\Omega = -1$and use a spatial grid-size of $h=0.01$.  The computed linear and linearized stability bounds for the RK4+CD scheme yield a maximum available time-step of ${\bf k}=0.0070711$ and ${\bf k}=0.0070534$ respectively.  For the RK4+2SHOC scheme, the stability bounds are ${\bf k}=0.0053033$ and ${\bf k}=0.0052934$ respectively.  Therefore, we use a time-step of ${\bf k}=0.005$ for all simulations to ensure stability.  A depiction of such a soliton within our simulations is shown in Fig.~\ref{f:examples}.  The solution has a constant modulus-square value at the boundaries (far from the soliton) and we therefore use the MSD boundary condition (\ref{msd}).

In two dimensions, we use an approximation to a steady-state dark vortex solution to the NLSE with $V(x)=0$ and $s<0$ given by \cite{BEC_DYN_NONLIN}
\begin{equation}
\label{exmp2Dvort}
\Psi(r,\theta,t) = f(r)\,\mbox{exp}[i\,(m\,\theta + \Omega\,t)],
\end{equation}
where $m$ is the topological charge of the vortex (in our case, we use $m=1$) and where $f(r)$ is a real-valued radial profile which can only be found exactly through numerical optimization routines.  Since we will be simulating the vortex solution of on large grids (up to nearly $2000 \times 2000$), inserting the interpolated exact numerical solution of $f(r)$ onto the two-dimensional grid can take excessive time to formulate.  Due to this, and since we are not interested in the exact solutions, we do not use the numerically exact profile but rather an approximation to it given by the initial condition of the one-dimensional dark soliton of Eq.~(\ref{soliton}) for $x>0$ with $c=0$.  

We use a spatial step-size of $h=0.25$.  Although this allows for a maximum time step of approximately ${\bf k}=0.016$, in order to allow for comparison between simulations of different dimensionality, we set the time-step to that of the one-dimensional examples, ${\bf k}=0.005$.  A depiction of the dark vortex within our simulations is shown in Fig.~\ref{f:examples}.

In three dimensions, we use an approximation to a dark vortex ring solution to the NLSE amidst a co-moving back-flow which causes the vortex ring to be a steady-state solution (see Ref.~\cite{ME_DISS} for details).  The form of the initial condition in cylindrical coordinates is 
\begin{equation}
\label{3dvr1}
\Psi(r,z,\theta,0) = g(r,z)\,\mbox{exp}\left[i\left(\frac{c}{2a}\,z\right)\right],
\end{equation}
where $g(r,z)$ is taken to be a numerically-exact two-dimensional $m=1$ dark vortex (which, since we only are using two-dimensional interpolations of resolutions up to $144\times 144$, is able to be formulated efficiently in this case) at position $r=d$ in the $r \mbox{--} z$ half-plane, where $d$ is the radius of the vortex ring, and $c$ is its intrinsic transverse velocity (given to asymptotic approximation in Ref.~\cite{VR_NLSE_VEL_71}).  In our example, we choose $d=5$ and use a spatial step of $h=3/2$.  Once again, for comparison purposes, we use a time-step of ${\bf k}=0.005$ even though the scheme would be stable with a larger time-step (as large as $0.045$).  A depiction of the vortex ring within our simulations is shown in Fig.~\ref{f:examples}.

\section{Serial code implementations}
\label{s:serial}
The serial codes in NLSEmagic are included for two reasons.  One, they allow calculations of the speedup of the CUDA codes and validate their output.  Second, the MEX serial codes are quite usable on their own, as they are very simple to install/compile and run, and, as shown in Sec.~\ref{s:mexspeedup}, are much faster than their equivalent MATLAB script codes.  Script integrators are also included as they are useful as a platform for testing new numerical schemes and boundary conditions.

All NLSE integrators in NLSEmagic are set-up to simulate a `chunk' of time-steps of the NLSE and return the resulting solution $\Psi$.  The `chunk-size' is set based on the desired number of `frames' (i.e. the number of times $\Psi$ is accessed for plotting and analysis).  For the serial integrators, this chunk-based approach is not necessary, but as we will see in Sec.~\ref{s:chunk-size}, the chunk-based approach is very important for the CUDA integrators.

\subsection{MATLAB script code integrators}
\label{s:mscripts}
The script implementations of the NLSE integrators are used to quickly develop and test new methods, and for comparison of the speedup of the MEX integrators.  They are split into two script files.  The main driver scripts ({\tt NLSE*D.m}) compute the chunk of time-steps in a {\tt for} loop, while the computations of $F(\Psi)$ with the desired boundary conditions in the RK4 scheme of Eq.~(\ref{RK4}) are done by calling a separate script file ({\tt NLSE*D\underline{ }F.m}).  All of the scripts utilize MATLAB vectorized operations wherever possible for efficiency [i.e. using {\tt A+B;} instead of {\tt for i=1:N; A(i)+B(i); end;}].  The script integrators are very straight-forward implementations of the numerical schemes and have been validated by numerous simulations of known solutions to the NLSE and LSE.

\subsection{MATLAB MEX code integrators}
\label{s:mexcode}
When using a MEX file, it is important to handle complex values correctly.  Since MEX files are written in C, one must either use a complex-number structure, or split the real and imaginary parts of the variables into separate vectors and write the numerical schemes accordingly.  As shown in Ref.~\cite{CUDA_FD_GL}, for CUDA codes, using complex structures is less efficient than splitting the real and imaginary parts directly.  To keep the serial and CUDA codes as similar to each other as possible, the split version of the numerical schemes are used.  Writing $F(\Psi)$ of Eq.~(\ref{RK4}) in this way yields
\begin{alignat}{2}
\label{fsplit}
F(\Psi)_{i,j,k}^R &= -a\,\nabla^2\Psi_{i,j,k}^I - s\left[\left(\Psi_{i,j,k}^R\right)^2 + \left(\Psi_{i,j,k}^I\right)^2\right]\,\Psi_{i,j,k}^I + V_{i,j,k}\,\Psi_{i,j,k}^I\\ 
F(\Psi)_{i,j,k}^I &= \;\;a\,\nabla^2\Psi_{i,j,k}^R + s\left[\left(\Psi_{i,j,k}^R\right)^2 + \left(\Psi_{i,j,k}^I\right)^2\right]\,\Psi_{i,j,k}^R - V_{i,j,k}\,\Psi_{i,j,k}^R,\notag
\end{alignat}
where $\Psi^R$ and $\Psi^I$ are the real and imaginary parts of the solution respectively.  Additional operations are straight-forward (for the split version of the MSD boundary condition, see Ref.~\cite{ME_MSD}).

As mentioned, a MEX code in C is very similar to a regular C code, but uses MATLAB-specific interfaces for input and output data, and special MATLAB versions of memory allocation functions.  Instead of having a {\tt main()} function as in standard C, the primary program is within a function called {\tt mexFunction(nlhs, plhs[],nrhs, prhs[])}.  The pointer {\tt prhs[]} (pointer-right-hand-side) gives the C code access to the input data being sent from MATLAB, where {\tt nrhs} is the number of input variables (including arrays and scalars).  The pointer {\tt plhs[]} (pointer-left-hand-side) is used in allocating memory for output arrays and scalars, allowing the MEX code to return data to MATLAB.

Special MEX functions ({\tt mxGetPr()}, {\tt mxGetPi()} and {\tt mxGetScalar()}) are used to extract the data from the {\tt prhs[]} array of MATLAB inputs. The function {\tt mxGetScalar()} is used to extract a scalar value while the functions {\tt mxGetPr()} and {\tt mxGetPi()} retrieve the pointer to the real and imaginary part of the selected input array respectively.   If the input array is fully real (which can be often the case, especially for initial conditions), then the {\tt mxGetPi()} is null-valued which causes a segmentation fault if accessed.  To get around this problem, code has been added to the NLSEmagic integrators that checks to see if there is an imaginary part of the input array of $\Psi$ (using the function {\tt mxIsComplex()}), and if not, allocates an all-zero imaginary array manually (which is then freed at the end of the MEX code).

The MEX functions {\tt mxCreateDoubleMatrix()} (for 1D and 2D arrays) and {\tt mxCreateNumericArray()} (for higher dimensionality arrays) are used to allocate memory space for output arrays which are then pointed to by {\tt plhs[i]}.  This memory is then seen as a one-dimensional array by the C code, but in MATLAB will be the desired dimensionality. 

\begin{figure}[t]
\centering
\includegraphics[width=4in]{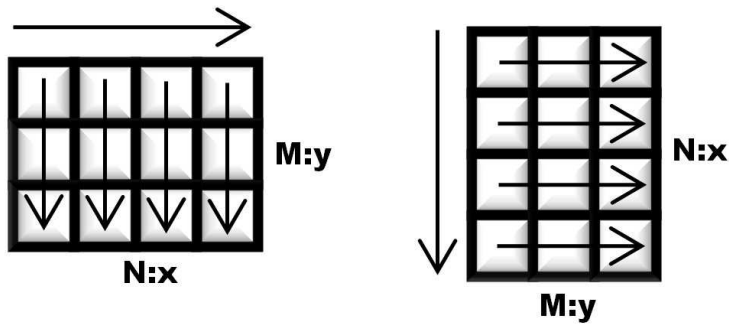}
\caption[Representation of a two-dimensional array in MATLAB and C.]{Representation of a two-dimensional array in MATLAB (left) and in C (MEX) (right). The arrows indicate the linear access pattern.  The $x$ and $y$ dimensions shown are the $x$ and $y$ dimensions of the solution $\Psi$.\label{f:2DMEXtranspose}}
\end{figure}

In order to use the input and output arrays in C, proper indexing is essential.  The MEX functions {\tt mxGetN()} and {\tt mxGetM()} are used to get the dimensions of one-dimensional and two-dimensional arrays (see below for three-dimensional array handling).  However, because MATLAB stores arrays column-order, while C stores them row-order, the $M$ value obtained from {\tt mxGetM()} in the C code is actually the number of columns, while the $N$ value obtained from {\tt mxGetN()} is the number of rows (see Fig.~\ref{f:2DMEXtranspose}).  
Therefore, the input array is seen by the MEX file as transposed from the way it is seen in MATLAB (of course no actual transpose operation is performed in the MATLAB to C passing of arrays, only the order in which the different languages access the elements is different).  Thus, while in MATLAB, the column length of the $\Psi$ solution array is the number of grid points in the $x$-direction of $\Psi$ and the row length is the number in the $y$-direction, inside the MEX file, this is reversed. (As will be seen in Sec.~\ref{s:cudamex}, this makes the `$x$' CUDA grid dimension to be along the $y$-direction of the solution and vice-versa).  The difference between the row-order access of C and the column-order access of MATLAB must be taken into account to correctly access the elements in an input array (for our discussion here, denoted $A$).  To access array element $(i,j)$ (where $i\in[0,N-1]$ and $j\in[0,M-1]$) of array $A$ in the MEX file, one uses {\tt A[M*i + j]}, where $M$ is the number of columns obtained from {\tt mxGetM()}. 

\begin{figure}[t]
\centering
\includegraphics[width=4in]{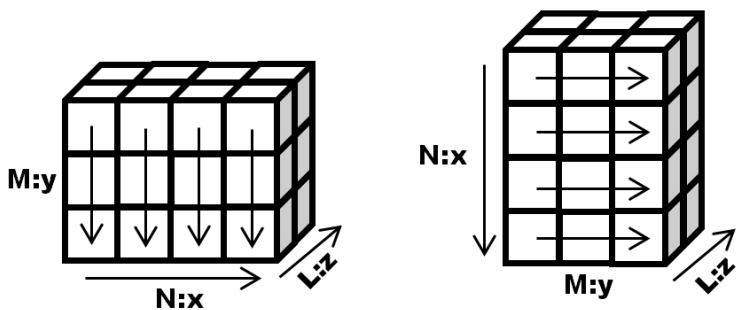}
\caption[Representation of a three-dimensional array in MATLAB and C.]{Representation of a three-dimensional array in MATLAB (left) and in C (MEX) (right).  The arrows indicate the linear access pattern.  The $x$, $y$ and $z$ dimensions shown are the $x$, $y$ and $z$  dimensions of the solution $\Psi$.  Here, the $z$ dimension maintains its orientation in the access pattern, while the $x$ and $y$ dimensions are transposed.\label{f:3DMEXtranspose}}
\end{figure}

In three dimensions, the MEX functions {\tt mxGetNumberOfDimensions()} and its counterpart {\tt mxGetDimensions()} are used to get the dimensions of the three-dimensional input arrays.  The {\tt mxGetDimensions()} returns an array of dimension lengths which, in our codes, are stored as $L$, $M$, and $N$.  As in the two-dimensional case, the access pattern differs in MATLAB and C.  Once again, the $M$ and $N$ dimensions of the MATLAB array are seen by the MEX file as swapped (see Fig.~\ref{f:3DMEXtranspose}).  
Thus, in Sec.~\ref{s:cudamex}, the CUDA grid dimensions for $x$ will be along the $y$-direction and vice-versa but the $z$ grid dimension will correspond to the $z$-direction of the solution.  To access array element $(i,j,k)$ (where $i\in[0,L-1]$, $j\in[0,N-1]$, and $k\in[0,M-1]$) of array $A$ in the MEX file, one uses {\tt A[M*N*i + M*j + k]}.  

Within a MEX file, memory allocations and deallocations are performed using MEX comparable functions to the C functions {\tt malloc()} and {\tt free()} called {\tt mxMalloc()} and {\tt mxFree()}.  The MEX function {\tt mxMalloc()} allocates the desired memory in a MATLAB memory-managed heap.  The advantage of using {\tt mxMalloc()} instead of {\tt malloc()} is that if a MEX file fails, or is interrupted during its execution, MATLAB can handle and free the memory to better recover from the failure \cite{matlabprimer8th}.

The compilation of a MEX file is performed within MATLAB using the {\tt mex} command as {\tt mex mymexcode.c}.  The MEX compiler uses whatever C compiler on the current machine is selected by MATLAB.  By default, MATLAB uses an included LCC compiler, but this can be changed using the {\tt mex -setup} command in MATLAB.  For the CUDA MEX codes on Windows, the compiler must be set to Microsoft Visual C++ while on Linux, the standard GCC compiler can be used (for our serial MEX codes, we also use these compiler options).  After a MEX file is compiled, the resulting binary file has a special extension which depends on the operating system being used ({\tt .mexw32} and {\tt .mexw64} for 32-bit and 64-bit Windows respectively, and {\tt .mexglx} and {\tt .mexa64} for 32-bit and 64-bit Linux respectively).

Both single and double precision versions of the NLSEmagic serial MEX integrator codes are included in order to properly compare the single-precision versions of the CUDA MEX codes to their serial counterparts (see Sec.~\ref{s:speedup}).  The single precision MEX codes convert the input arrays using the cast {\tt (float)}, and then convert the output arrays back to double precision (MATLAB's native format) after the computation of the desired number of time-steps.

\subsection{Speedup of serial MEX codes}
\label{s:mexspeedup}
\begin{figure}[hbtp]
\centering
\includegraphics[width=2.8in]{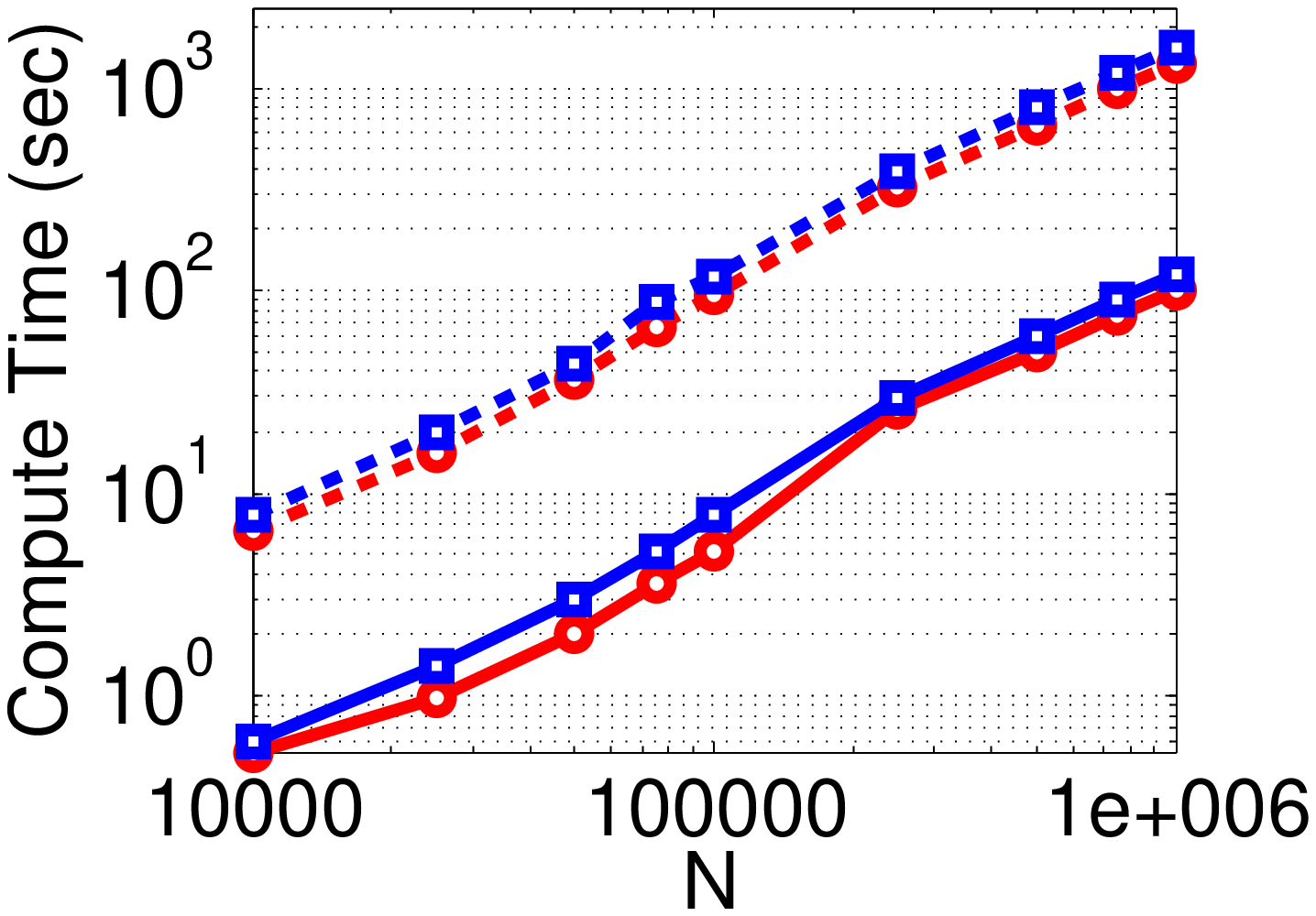}
\includegraphics[width=2.8in]{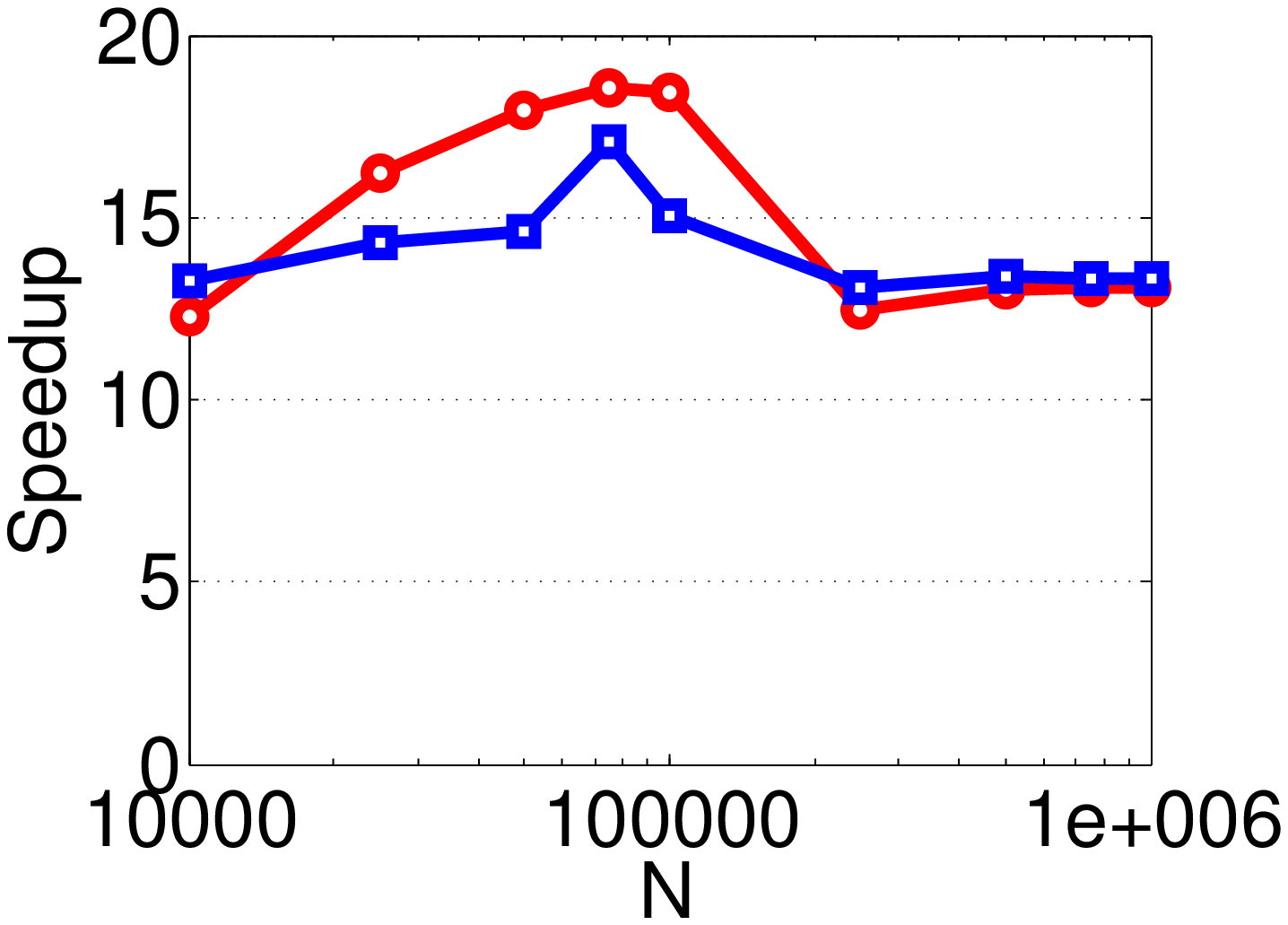} \\
\includegraphics[width=2.8in]{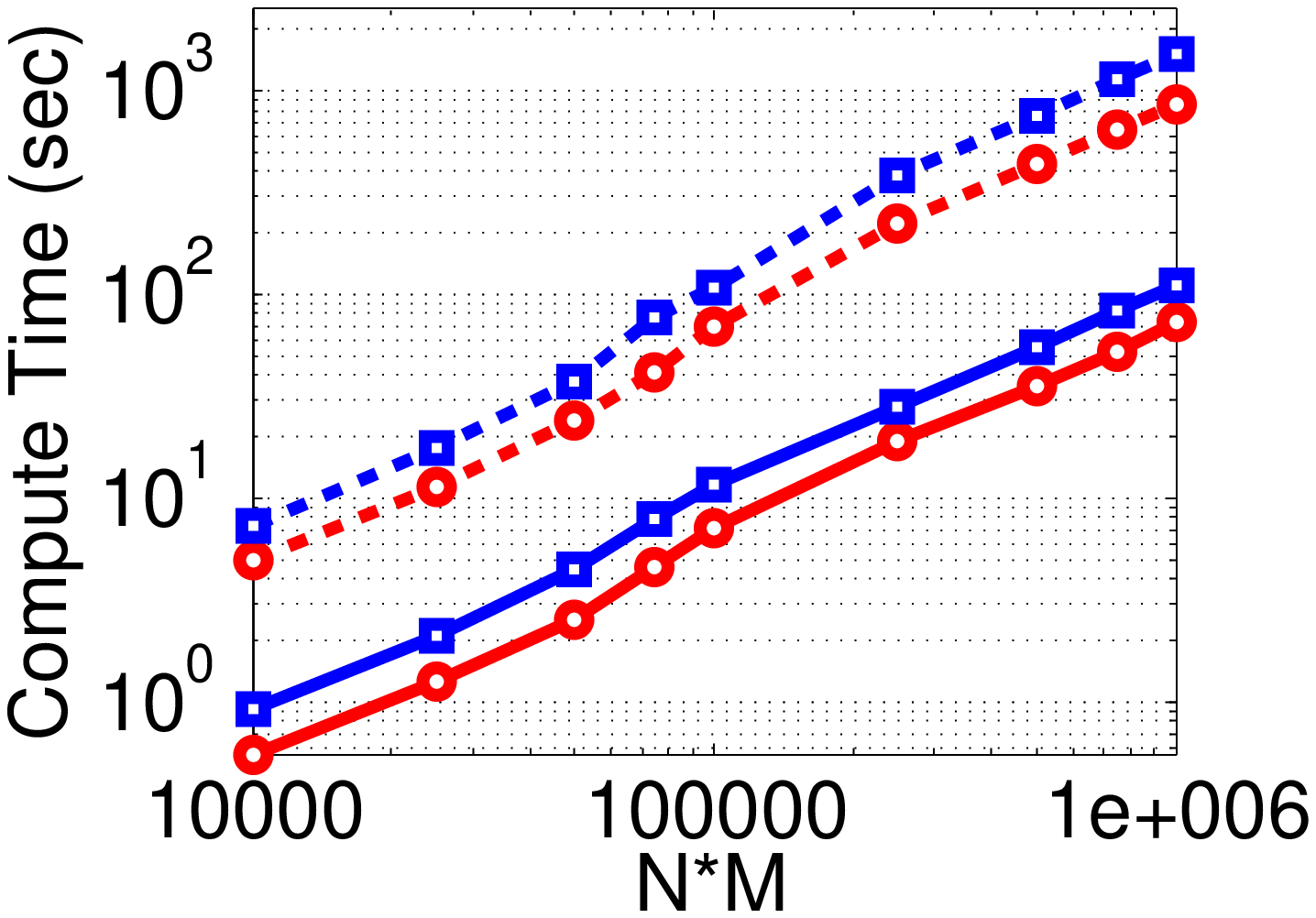}
\includegraphics[width=2.8in]{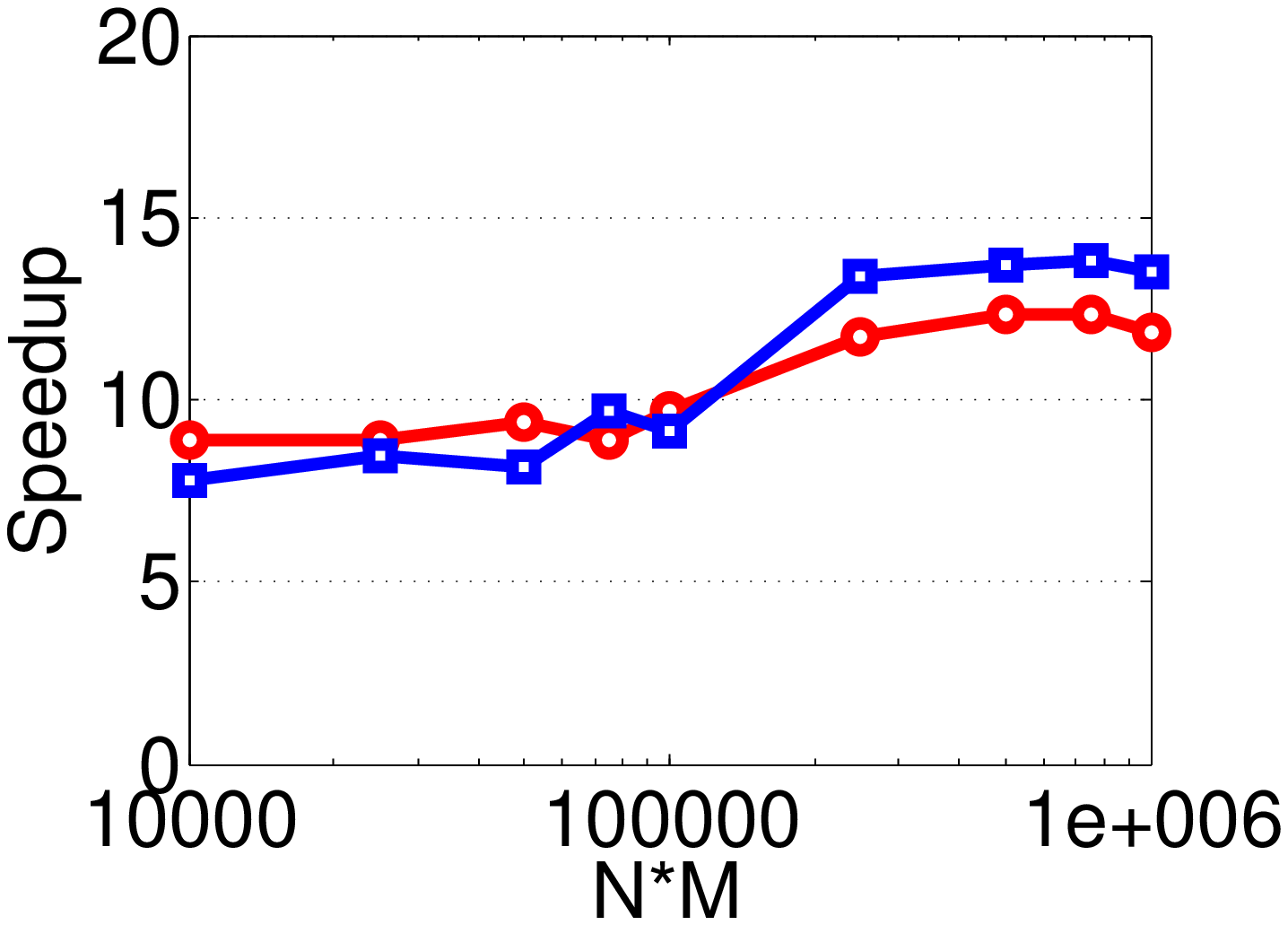} \\
\includegraphics[width=2.8in]{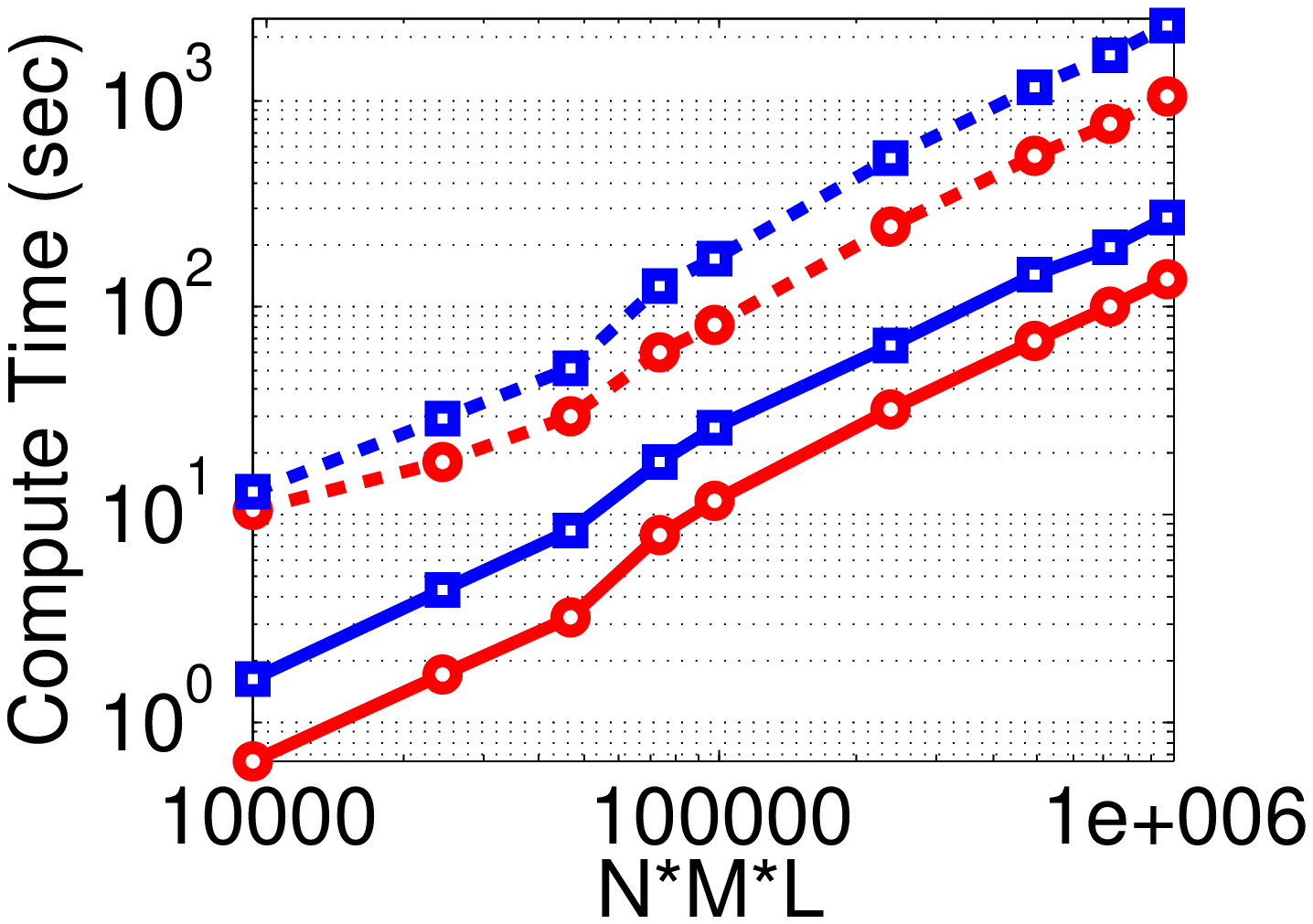}
\includegraphics[width=2.8in]{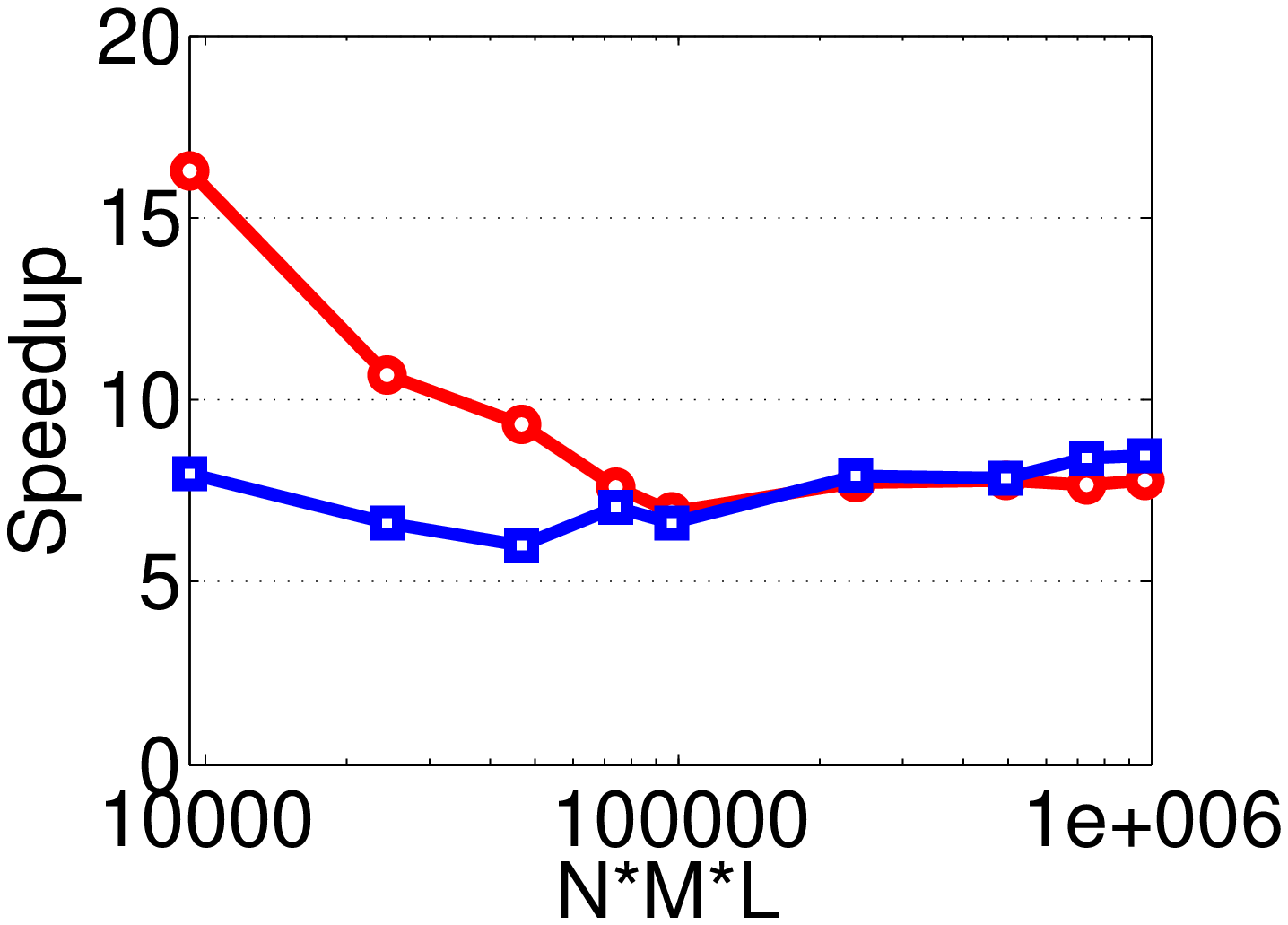}
\caption[Timing results of MATLAB scripts versus MEX integrators.]{(Color online) Timing results comparing script (dashed lines) and C MEX (solid lines) NLSE double-precision integrators for both the RK4+CD (red circles) and RK4+2SHOC (blue squares) schemes using the example problems of Sec.~\ref{s:examples} with an end-time of $t=5$ and a number of frames of $10$.  The simulations are performed for a total number of grid points of ${\bf N} = 10000$ to ${\bf N} = 1000000$.  The results for the one-, two-, and three-dimensional codes are shown top to bottom.  The left column displays the simulation times and the right column shows the corresponding speedup of the MEX codes.\label{f:mexspeedup}}
\end{figure}

To illustrate the advantage of using MEX codes over script codes in MATLAB to integrate the NLSE, we show timing results comparing the performance of the script integrators of NLSEmagic to the equivalent MEX codes using the example problems described in Sec.~\ref{s:examples}.  As mentioned in Sec.~\ref{s:mscripts}, our script codes are sufficiently optimized by utilizing MATLAB's built-in vector operations wherever possible.  We set an end-time of $t=5$, the number of frames to $10$, and use double precision for all simulations as it is MATLAB's native format for the script codes.  The simulations are performed for a total number of grid points of ${\bf N} = 10000$ to ${\bf N} = 1000000$.  The results are shown in Fig.~\ref{f:mexspeedup}.
We see that using MEX codes are, on average, eight to fifteen times faster than the equivalent script codes.  Although there is variation in the speedup (more for lower dimensions and variable over problem size), these variations are small and therefore the speedup of the MEX integrators can be considered approximately constant for all dimensions and problem sizes.  Therefore, the serial MEX codes included in NLSEmagic are valuable in their own right (especially for researchers who are currently using script codes) since no special setup is needed to compile them and their use is straight-forward.

\section{GPU-accelerated CUDA MEX code integrators}
\label{s:cudamex}
In this section we describe in detail the design of the CUDA MEX NLSE integrators in NLSEmagic.  Both single and double precision versions of all integrators are produced in order for the codes to be compatible with older GPU cards, as well as to increase performance when double precision accuracy is not required (as will be shown in Sec.~\ref{s:speedup}, the single precision GPU codes can be much faster than the double precision codes).

As mentioned in Sec.~\ref{s:gpulogic}, there are two main sections of a CUDA code, the host code and the kernel codes.  The host code is run on the CPU and is used to set up the problem, handle memory allocations and transfers to the GPU, and run the GPU kernel code.  The kernels are the routines that are run on the GPU. 

The basic outline of the integrators is that the input solution $\Psi$ and potential $V({\bf r})$ arrays obtained through the MEX interface are transfered to the GPU's memory.  Then, the desired number of time-steps are computed on the GPU using kernel functions, and the resulting solution is transfered back to the CPU memory, where it is outputted to MATLAB through the MEX interface.  Since the memory transfers from CPU to GPU and vice-versa are quite slow, the larger the number of time-steps the GPU computes before sending the solution back to the CPU  (i.e. the larger the chunk-size), the better the expected speedup.  In Sec.~\ref{s:chunk-size} we investigate this in detail for our simulations.  

\subsection{Specific code design for all dimensions}
\label{s:cudamexspec}
We now discuss the details of the CUDA implementation which are the same for the one-, two-, and three-dimensional integrator codes (in the sections that follow, we will discuss dimension-specific code design).  We focus on the CUDA design aspects of the codes, and therefore do not mention any MEX interfaces (as they are all equivalent to the serial MEX codes previously described).  Thus, we assume that the solution $\Psi$ and potential $V({\bf r})$ have already been obtained through the MEX interface along with all scalar parameters.

The first step in the code is to allocate arrays on the GPU's global memory to store the solution, potential, and RK4 temporary arrays.  This is accomplished by using the {\tt cudaMalloc()} or the {\tt cudaMallocPitch()} (see Sec.~\ref{s:cudacode2d}) function.  The next step in the code is to transfer the solution $\Psi$ and potential $V({\bf r})$ arrays from the CPU's memory to the memory locations allocated for them on the GPU's global memory.  This is done through the {\tt cudaMemcpy()} or the {\tt cudaMemcpy2D()} (see Sec.~\ref{s:cudacode2d}) function.  

Before calling the compute kernel routines to integrate the NLSE, it is first necessary to define the CUDA compute-grid and block sizes for use in the kernel calls.  The choice of block size is not trivial, and is dependent on the size of the problem and the GPU card being used.  In order for a GPU to be most efficient, it requires that the problem have good \emph{occupancy} on the specific card.  This means that as many SMs are running as many SPs during the program simultaneously.  In NLSEmagic, the block sizes are chosen to provide efficiency for typical sized problems in each dimension (see the following sections for details).  The block size in each dimension is set using the {\tt dimBlock()} function.  

The compute-grid size is found by taking the {\tt ceil} of the length of the dimension of the specified problem array divided by the block size of that dimension.  For example, the $x$-direction grid size in two dimensions would be found by {\tt ceil(M/dimBlock.x)} where {\tt dimBlock.x} is the $x$-direction block-size (it is important to note that, as per the discussion in Sec.~\ref{s:mexcode}, the $x$- and $y$-direction of the CUDA blocks are actually along the $y$- and $x$-direction respectively of the solution $\Psi$ within the C code).  Once the values for the grid dimensions are found, the grid is set up using the function {\tt dimGrid()}.

Once the CUDA compute-grid and block sizes are set up, the code computes a chunk-size number of time-steps of the RK4 scheme using kernels which run on the GPU.  The more computations a single kernel call performs, the better the performance of the GPU code.  Therefore, we try to combine as many steps in the RK4 algorithm of Eq.~(\ref{RK4}) as possible.  Since each compute block must be able to run independently of all the other blocks within a single kernel call, synchronization and race conditions will limit how many steps of the RK4 can be combined.  For example, each computation of $F(\Psi)$ in Eq.~(\ref{RK4}) (after the first one), for each grid point, requires the previously completed computation of $\Psi_{\mbox{\scriptsize tmp}}$ of the neighboring points from the previous step in order to compute the Laplacian of $\Psi$.  Therefore, even with block-wide synchronizations, grid points along the boundary of compute-blocks will have no guarantee that the $\Psi_{\mbox{\scriptsize tmp}}$ values it needs will have been computed yet or not (or possibly, even overwritten).  Therefore, multiple separate kernels are required to run the steps in Eq.~(\ref{RK4}).  When using the RK4+2SHOC scheme, an additional kernel (called four times) is required due to the same synchronization issues involving neighboring points needed from one step to the other (in this case the values of the Laplacian $D$ in the first step of the 2SHOC scheme).

Although the computations of $F(\Psi)$ and $D$ require separate kernels, the other steps in the RK4 algorithm \emph{can} be combined with the kernels for $F(\Psi)$.  Thus, the RK4 can be computed using only four kernel calls (8 when the additional kernel needed in the 2SHOC scheme is included) which contain steps 1--2, 3--5, 6--8, and 9--10 of Eq.~(\ref{RK4}).  The only problem is that steps 3--5 and 6--8 can cause a synchronization issue.  This is because step 5 (and 8) overwrites the $\Psi_{\mbox{\scriptsize tmp}}$ values, which are the inputs to step 3 (and 6).  Thus, when one block of threads finishes the step 3--5 computation, the adjacent block could still require the old values of $\Psi_{\mbox{\scriptsize tmp}}$ along its neighboring edge to compute the Laplacian.  To solve this, we introduce a storage array called $\Psi_{\mbox{\scriptsize out}}$ which allows step 5 to output its values to $\Psi_{\mbox{\scriptsize out}}$ so that the other blocks still have access to the original values of $\Psi_{\mbox{\scriptsize tmp}}$ .  Then in step $6$, the code evaluates $F(\Psi_{\mbox{\scriptsize out}})$, while step $8$ outputs to $\Psi_{\mbox{\scriptsize tmp}}$.  Adding this vector does not increase the overall global memory storage requirements of the RK4 scheme because since the steps are now combined into four kernels, the $K_{\mbox{\scriptsize tmp}}$ array no longer needs to be stored in global memory and can instead be stored in the per-block shared memory.  The RK4 algorithm can therefore be described on the GPU as
\begin{alignat}{6}
\label{RK4_GPU}
&1)\; k_{\mbox{\scriptsize tmp}} = F(\Psi) 
& \qquad 
&7)\; k_{\mbox{\scriptsize tmp}} = F(\Psi_{\mbox{\scriptsize {\bf out}}})
\\
&2)\; K_{\mbox{\scriptsize tot}} = k_{\mbox{\scriptsize tmp}}   
&\qquad 
&8)\; K_{\mbox{\scriptsize tot}} = K_{\mbox{\scriptsize tot}} + 2\, k_{\mbox{\scriptsize tmp}} \notag
\\
&3)\; \Psi_{\mbox{\scriptsize tmp}} = \Psi + \frac{{\bf k}}{2}\, k_{\mbox{\scriptsize tmp}}
&\qquad 
&9)\; \Psi_{\mbox{\scriptsize tmp}} = \Psi + {\bf k}\,k_{\mbox{\scriptsize tmp}} \notag
\\
&4)\;  k_{\mbox{\scriptsize tmp}} = F(\Psi_{\mbox{\scriptsize tmp}})  
&\qquad 
&10)\;  k_{\mbox{\scriptsize tmp}} = F(\Psi_{\mbox{\scriptsize tmp}}) \notag
\\
&5)\;  K_{\mbox{\scriptsize tot}} = K_{\mbox{\scriptsize tot}} + 2\, k_{\mbox{\scriptsize tmp}}
&\qquad 
&11)\;  \Psi = \Psi + \frac{{\bf k}}{6}\, (K_{\mbox{\scriptsize tot}} + k_{\mbox{\scriptsize tmp}}), \notag 
\\
&6)\;  \Psi_{\mbox{\scriptsize {\bf out}}} = \Psi + \frac{{\bf k}}{2}\, k_{\mbox{\scriptsize tmp}},
&\qquad \notag 
\end{alignat}   
where $k_{\mbox{\scriptsize tmp}}$ is only stored in shared memory, while $\Psi$, $K_{\mbox{\scriptsize tot}}$, $\Psi_{\mbox{\scriptsize tmp}}$, and $\Psi_{\mbox{\scriptsize out}}$ are stored in global memory (although during computation, they are transfered to shared memory as discussed below).

All four kernel calls in the computation of a RK4 step are computed using a single kernel function named {\tt compute\underline{ }F()}.  One of its input parameters tells the kernel which step it is computing, and the corresponding computation is selected in a switch statement.  For the 2SHOC scheme, an additional kernel is written named {\tt compute\underline{ }D()} which computes the $D$ array before each call to {\tt compute\underline{ }F()}.

As discussed in Sec.~\ref{s:gpuphys}, thread accesses to the per-block shared memory are much faster than accesses to the global memory of the GPU.  Therefore if any array needed in the computations is accessed more than once by a thread in the block, it is worthwhile to copy the block's required values of the array from global memory into shared memory.  Then, after computing using the shared memory, the results can be copied back into global memory.  Since shared memory is block-based, each thread in the block needs to copy its own value from the global array into the shared memory space (some threads may need to copy more than one value as we will show in Secs.~\ref{s:cudacode1d}, \ref{s:cudacode2d}, and \ref{s:cudacode3d}).

In {\tt compute\underline{ }F()}, five shared memory arrays are required (seven for the 2SHOC scheme).  These consist of the real and imaginary parts of the $\Psi$ input and the $F(\Psi)$ result [called $k_{\mbox{\scriptsize tmp}}$ in the RK4 algorithm of Eq.~(\ref{RK4_GPU})], as well as the potential array $V$.  In the 2SHOC scheme, the real and imaginary parts of $D$ are also stored in shared memory arrays.  In the {\tt compute\underline{ }D()} kernel for the 2SHOC scheme, only two shared memory arrays are needed (for $\Psi$).

Each thread in the block copies the global memory values into the shared memory arrays.  Since the number of stream processors in the stream multi-processor that the block is being computed on almost always has less processors than the number of threads in the block, threads will typically not have access to all of the shared memory of the block after copying their own values.  This is a problem because each thread has to access neighbor elements in the shared memory block for computation of the Laplacian of $\Psi$.  Therefore, a {\tt \underline{ }\underline{ }syncthreads()} function is called after the copy which synchronizes all the threads in the block, ensuring that the entire shared-memory array is filled before using it.

After the synchronization, the threads on the boundary of the block have to copy additional values into shared memory since they require accessing points which are beyond the block boundary due to the finite-difference stencil (see Secs.~\ref{s:cudacode1d}, \ref{s:cudacode2d}, and \ref{s:cudacode3d} for details in each dimension).  These transfers are not done before the block-synchronization because in CUDA, when a group of threads all need to perform the same memory transfer, they are able to do it in a single memory copy instead of one-by-one.  Adding the boundary transfers before the synchronization may break up this pattern and cause the memory-copies not to be aligned for single-instruction copying.

After the transfers to shared memory are completed, the $F(\Psi)$ values are computed for all grid points within the boundaries of the solution $\Psi$.  Threads which happen to be on the boundary of $\Psi$ compute the boundary conditions of $F(\Psi)$.  When using the MSD boundary condition of Ref.~\cite{ME_MSD}, a block-synchronization is required to be sure that the interior value of $F(\Psi)$ has been pre-computed (since it is necessary for the computation of the MSD condition).  A problem exists if in any direction, the block is only one cell wide (since the interior point of $F(\Psi)$ will not be computed by that block).  To avoid this problem in the most efficient way (i.e, without adding extra error-checking code), we leave the detection and solution of this issue to the NLSEmagic driver scripts which automatically adjust the solution grid size to ensure that this condition does not occur. 

Once the boundary conditions of the $F(\Psi)$ array is completed, the kernel uses the result to compute the remaining sub-steps of the RK4 algorithm.  After the desired number of RK4 time-steps (the chunk-size) have been completed, a call to {\tt cudaDeviceSynchronize()} is made which ensures that all kernels are fully completed before continuing in the host code.  This is required because although the CUDA GPU-to-CPU memory copies are designed to be implicitly synchronizing, in practice for large problems, we found that this was not completely reliable.  

The current $\Psi$ arrays are then transfered to the CPU using the {\tt cudaMemcpy()} (or {\tt cudaMemcpy2D()}) function.  The next step is to free all the CUDA global memory spaces used (with {\tt cudaFree()}) and rest the device with {\tt cudaDeviceReset()}).  Typically one would not reset the device in a CUDA code, but due to some problems with MATLAB memory management on Linux platforms, this step is necessary and does not have any noticeable impact on performance.  Finally, the MEX file then returns the new value of $\Psi$ back to MATLAB.

\subsection{One-dimensional specific code design}
\label{s:cudacode1d}
In the one-dimensional NLSEmagic integrators, the CUDA compute-grid is set-up to be one-dimensional with a block size of 512 threads.  Although for small problems, a block size of 512 is not efficient in terms of occupancy (see Sec.~\ref{s:cudamex}), we feel that since one-dimensional problems typically run fast enough without any GPU-acceleration, the only time GPU-acceleration will be needed in one-dimensional problems is when the problem size is very large.  In such a case, a block size of 512 will allow for good occupancy in most situations. 

One of the key concerns in parallelizing finite-difference codes is the need for the cells on the edge of a compute-block to require values from their neighbor in another block.  Since in the CUDA codes all threads in all blocks have equal access to global memory, there seems to be no problem with threads on the block boundaries.  However, this is only true if the CUDA code only uses global memory (see Ref.~\cite{CUDA_FD_3Dwave} and \cite{CUDA_FD_GL}, where the authors take this approach).  

In order to greatly speedup the codes, it is vital to use the per-block shared memory in the computation whenever any global memory space is needed to be accessed by any thread more than once.  The typical way to use shared memory is to have each thread in the block copy one value of $\Psi$ (or $D$) and $V$ from global memory into shared memory arrays which are the same size as the block size.  Then the computation is done using the shared memory array, and at the end each thread stores the output back into the global memory array.  However, in this scenario, the block boundary cells do not have all their required $\Psi$ (or $D$) values available in shared memory.  There are a few different ways to deal with this.  A simple solution is to have the block boundary threads directly access the required neighbor point(s) of $\Psi$ (or $D$) from global memory when needed (as was done in Ref.~\cite{CUDA_FD_2008}).  Another solution is to set up the grid to overlap the boundary cells so that all threads copy values into shared memory as before, but only the interior threads of the block perform the computations, which can be performed using only shared memory.  Through testing, we have found that this solution is not efficient because a large number of threads (the boundary threads) are idle after the memory copy.   

\begin{figure}[hbtp]
\centering
\includegraphics[width=5in]{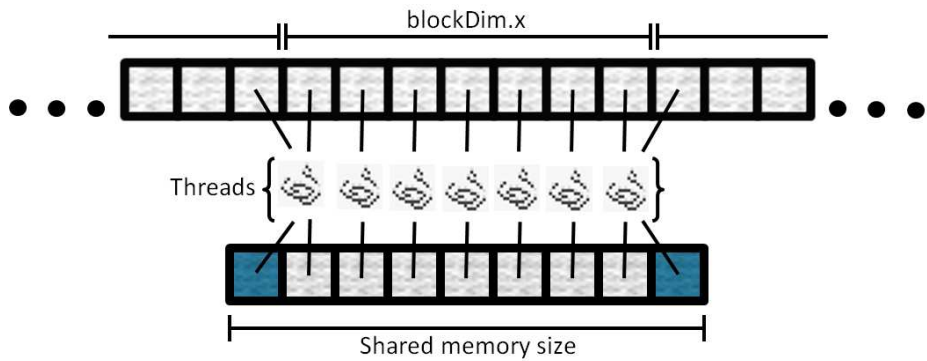}
\caption[One-dimensional shared memory structure used in the NLSEmagic CUDA integrators.]{(Color online) One-dimensional shared memory structure and transfer of global memory to shared memory in NLSEmagic.  Note that the threads on the boundary of the block must perform two global memory retrievals into shared memory. \label{f:CUDA1Dblockstruct}}
\end{figure}

Another solution is that instead of having the shared memory size be equal to the block size, the shared-memory size is allocated to be two cells greater than the block size.  In the stage of the kernel where the threads transfer $\Psi$ and $V$ from global memory into shared memory, the boundary threads also transfer the neighboring points into the extra cell space at the boundary of shared memory.  Therefore there are two global memory transfers performed by the block boundary cells instead of one.  This process is depicted in Fig.~\ref{f:CUDA1Dblockstruct}.
Later, in the computation part of the kernel, all threads can then compute using shared memory.  Although this seems to be equivalent to the first method in terms of the number of global memory accesses (since in the first solution, the boundary cells access global memory twice as well --- once to transfer their value to shared memory and once to access their required neighboring point), in practice, due to intricacies of the card hardware and software execution (for example, limiting the number of divergent branches), this method can be more efficient overall and therefore the one adopted in our codes.  

\subsection{Two-dimensional specific code design}
\label{s:cudacode2d}
In two-dimensional CUDA codes, the CUDA API provides special global memory allocation and transfer routines called {\tt cudaMallocPitch()} and {\tt cudaMemcpy2D()}.  The {\tt cudaMallocPitch()} allocates a global memory space which is expected to be accessed in a two-dimensional pattern.  It aligns the data with the card hardware, often padding the rows of the matrix.  Therefore, the memory allocation function returns a pitch value which is used to access the element ($i,j$) as {\tt pitch*i+j} instead of the usual {\tt M*i+j}.  The pitch value is also used in the {\tt cudaMemcpy2D()} to transfer data from the CPU to the GPU and vice-versa.  The special two-dimensional routines are somewhat optional as one could also just use a standard linear memory transfer of the data and access the elements normally, but NVIDIA strongly recommends against it \cite{cudadoc_pg}.  Therefore, the two-dimensional integrators of NLSEmagic use the specialized functions. 

The CUDA compute-grid is set to be two-dimensional with two-dimensional blocks with a block size of $16\times 16$.  Based on performance tests, this is the best overall block size to use even though newer cards would allow a larger block size.

\begin{figure}[hbtp]
\centering
\includegraphics[width=3in]{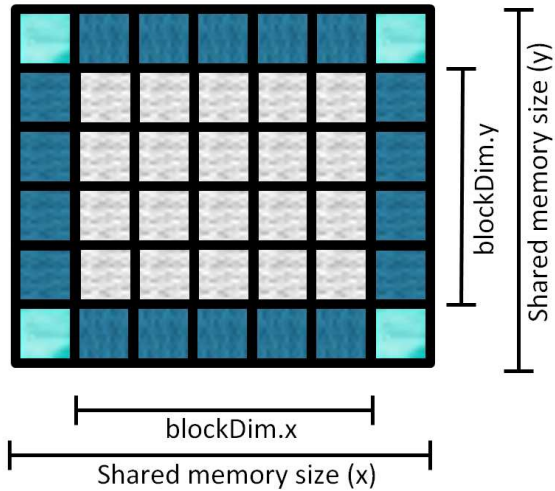}
\caption[Two-dimensional shared memory structure used in the NLSEmagic CUDA integrators.]{(Color online) Two-dimensional shared memory structure.  The threads on the boundary of the block must perform additional global memory retrievals into shared memory. Those on the corners perform a total of three transfers for the CD scheme, and four transfers for the 2SHOC (due to the required corner cells in the 2SHOC scheme).  All other boundary threads perform two total transfers.\label{f:CUDA2Dblockstruct}}
\end{figure}

As in the one-dimensional code, the shared memory space is allocated to be two cells larger in width and height of the block size as shown in Fig.~\ref{f:CUDA2Dblockstruct}.  
The boundary threads once again grab the neighboring points in addition to their designated points from global memory during the shared-memory transfer process.  For the CD scheme, the corner neighbor points are not needed, but for the 2SHOC scheme they are.  This adds an additional global memory access on the corner threads (in addition the extra accesses from the other two neighboring points), making the corner threads copy up to four global values per array into shared memory.    

When dividing the problem into CUDA blocks, the CUDA thread access ordering is an important consideration.  The CUDA grid and block indexing is column-major ordered, with the `$x$' direction along the columns and the `$y$' direction is along the rows, and the threads are scheduled in that order.  Thus, when copying arrays from global memory to shared memory, it is important that the access pattern of the memory being copied matches the access pattern of the thread scheduler \cite{CUDA_MemAccess}.  Since C arrays are row-major, a shared memory array {\tt  A} should be allocated (counter-intuitively) as {\tt A[BLOCK\underline{ }SIZEY][BLOCK\underline{ }SIZEX]}.  The memory copy from elements in a global array {\tt A\underline{ }global} is then given as 
\begin{center}
{\tt A[threadIdx.y][threadIdx.x] = A\underline{ }global[pitch*i+j]},
\end{center}
where the position of the current thread within the solution array is given by
\begin{center}
{\tt j = blockIdx.x*blockDim.x + threadIdx.x,\\
     i = blockIdx.y*blockDim.y + threadIdx.y.}
\end{center}
Since in MATLAB (when using {\tt meshgrid()} to form the $x$ and $y$ arrays of $\Psi$) the $x$ direction of $\Psi$ is along the rows of the $\Psi$ array, and the $y$ direction is along the columns, and when in the MEX file, this is transposed, then, as mentioned in Sec.~\ref{s:mexcode}, the `$x$' direction of the CUDA grid and blocks actually represent the $y$ direction in $\Psi$ and vice-versa.  Therefore, in defining the grid size, we use {\tt M/dimBlock.x} for the $x$ dimension of the grid and {\tt N/dimBlock.y} for the $y$ even though {\tt N} is the $x$-direction of $\Psi$ and {\tt M} is the $y$-direction.  This is very important to remember when checking the MSD boundary condition grid requirement mentioned in Sec.~\ref{s:cudamexspec}.  The thread access ordering described is not a trivial matter.  In our tests, using the incorrect ordering can slow down the codes by almost a half!

\subsection{Three-dimensional specific code design}
\label{s:cudacode3d}
In three dimensions, the CUDA API provides a hardware-optimized memory allocation and transfer functions similar to the two-dimensional ones mentioned in Sec.~\ref{s:cudacode2d}.  However, the implementation of the three-dimensional functions (called  {\tt cudaMalloc3D()} and {\tt cudaMemcpy3D()}) are not straight-forward and requires making structures with information about the arrays, and passing the structures to the functions.  Through testing of the two-dimensional NLSEmagic codes versus a naive two-dimensional implementation (without the specialized allocation and memory-copy), we did not observe any significant performance boost in using the hardware-optimized routines.  Therefore, in order to keep things as simple as possible, the three-dimensional integrators of NLSEmagic use the standard {\tt cudaMalloc()} and {\tt cudaMemcpy()} used in the one-dimensional codes in which case a point ($i,j,k$) in a global memory array {\tt A\underline{ }global} is simply given by {\tt A\underline{ }global[N*M*i + M*j + k]}.

In Refs.~\cite{CUDA_FD_2009} and \cite{CUDA_FD_3Dwave}, the three-dimensional finite-difference CUDA codes were set-up to do multiple two-dimensional sweeps and combine the results to form the three-dimensional derivatives.  Thus, only two-dimensional grid and block structures were used.  This was done because the limit of 16KB of shared memory of the older GPUs was felt to be too limiting for using three-dimensional blocks.  However, in our NLSE integrators, we use three-dimensional blocks which we note is efficient even when using older Tesla GPUs.  On the Fermi cards, we have tested our code for various block sizes and have used the current most efficient block sizes for a typical problem size depending on the numerical method and precision being used. 

An inherent difficulty with three-dimensional CUDA codes is that originally, even though blocks could be three-dimensional, the CUDA compute-grid could only be two-dimensional.  Thus, one could not implement three-dimensional codes in a straight-forward extension from two-dimensional grid structures.  This is an additional reason why previous studies (such as Refs.~\cite{CUDA_FD_2009} and \cite{CUDA_FD_3Dwave}) used two-dimensional slice algorithms.  

NVIDIA has solved this problem as of the release of the CUDA SDK 4.0.  With the new SDK, all Fermi based GPUs and higher can now have a full three-dimensional grid structure.  Our NLSE integrators in NLSEmagic were developed before this release, and therefore did not take advantage of this new feature.  Instead, a custom logical three-dimensional structure on a two-dimensional CUDA compute-grid was used.  In order to allow NLSEmagic to be compatible with older Tesla-based GPUs, the integrators are not currently updated to use the new three-dimensional grids.

\begin{figure}[hbtp]
\centering
\includegraphics[width=6in]{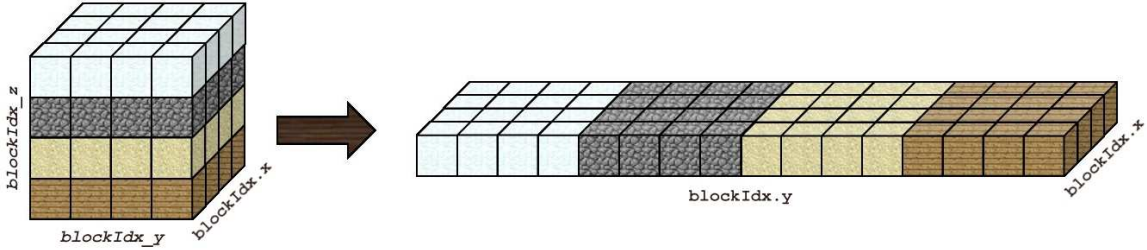}
\caption[Three-dimensional CUDA block reordering used in the NLSEmagic CUDA integrators.]{(Color online) Logical block reordering for a 2D CUDA grid from desired 3D grid structure. \label{f:CUDA3D}}
\end{figure}

The three-dimensional problem grid is viewed as a series of three-dimensional slabs, each being one block high.  These slabs are then seen as being placed side-by-side to create a two-dimensional grid of three-dimensional blocks as shown in Fig.~\ref{f:CUDA3D}.
We thus have a CUDA compute-grid which is two-dimensional with three-dimensional blocks, and a `virtual' compute-grid which is a three-dimensional grid with three-dimensional blocks.  In order to make the indexing as easy as possible within the actual CUDA compute-grid, we create new indexing variables per thread in order to use the full three-dimensional virtual compute-grid.  

We note here that, as in the two-dimensional case, the CUDA block/thread directions $x$, $y$, and $z$ do not all correspond to those of $\Psi$.  The CUDA block $x$-direction is actually along the $y$-direction of $\Psi$ ($M$) and the CUDA block $y$-direction is actually along the $x$-direction of $\Psi$ ($N$).  The $z$-direction of both are equivalent.  This is important to keep in mind for the computation of grid sizes and in checking the block requirements of the MSD boundary-condition.

The first step in setting up the three-dimensional grid is that a variable 
\begin{center}
{\tt gridDim\underline{ }y = ceil(N/dimBlock.y)}
\end{center}
is set to represent the virtual $y$-direction grid size.  Therefore, the actual CUDA grid is defined to be 
\begin{center}
{\tt dimGrid(ceil(M/dimBlock.x),gridDim\underline{ }y*ceil(L/dimBlock.z))}.  
\end{center}
The value of {\tt gridDim\underline{ }y} is passed into the kernels in order for the threads to have access to it.  Inside the kernel, a new variable representing the thread's $z$-direction block position within the virtual three-dimensional grid called {\tt blockIdx\underline{ }z} is created defined as 
\begin{center}
{\tt blockIdx\underline{ }z  = blockIdx.y/gridDim\underline{ }y},
\end{center}
which uses integer division to act as a {\tt floor} function.  The $y$-direction block position in the virtual grid is then found and stored in a new variable as 
\begin{center}
{\tt blockIdx\underline{ }y  = blockIdx.y - blockIdx\underline{ }z*gridDim\underline{ }y}.
\end{center}
The {\tt blockIdx\underline{ }y} variable is therefore replacing the intrinsic {\tt blockIdx.y} variable (which would give the $y$-direction block within the true CUDA grid).  These new block position variables are depicted in Fig.~\ref{f:CUDA3D}.

Once the new indexing variables are created, the computation of the thread's designated position within the solution array is straight-forward as if there were a true CUDA three-dimensional grid.  The value of the thread's indexes are
\begin{center}
{\tt k = blockIdx.x*blockDim.x + threadIdx.x,\\
     j = {\bf blockIdx\underline{ }y}*blockDim.y + threadIdx.y,\\
     i = {\bf blockIdx\underline{ }z}*blockDim.z + threadIdx.z,}
\end{center}
where the thread is associated with the global memory position {\tt A\underline{ }global[N*M*i + M*j + k]}.

Once the older Tesla GPU cards are no longer in general use, direct three-dimensional CUDA grids can be implemented into the code, which, for our implementation, would be very straight-forward (just setting up the grid based on $M$, $N$, and $L$ directly and replacing the new thread indexes with the intrinsic ones).

As was the case in the two-dimensional codes, the dimensions of the shared memory space must be done in accordance with the thread access patterns in order to be efficient.  (Note that this discussion is completely independent of the implementation of the three-dimensional compute-grid since, either way, three-dimensional blocks are being used and this issue is a block-wide problem).  Since the threads are accessed in column order, first along $x$, then $y$, then $z$, it is important that the shared memory space be allocated as {\tt A[BLOCK\underline{ }SIZEZ][BLOCK\underline{ }SIZEY][BLOCK\underline{ }SIZEX]} so that the memory copy is done as
\begin{center}
{\tt A[threadIdx.z][threadIdx.y][threadIdx.x] = A\underline{ }global[N*M*i + M*j + k]}.
\end{center}
As in the two-dimensional case, this correct ordering has a huge effect on performance (upwards of $50\%$) and is therefore vital.

As in the lower-dimensional integrator codes, the three-dimensional integrators allocate the shared memory space to be two cells larger in each dimension than the block size.  Thus, the threads on the block boundaries copy additional values from global memory into shared memory.  For each required shared memory array. the interior threads copy one value, the threads along the faces of the block boundaries copy two values, the threads along the edges of the block boundary copy three values (four for $\Psi$ in the 2SHOC scheme), and the threads on the corners of the block boundary have to copy four values (seven for $\Psi$ in the 2SHOC scheme).  Since the 2SHOC scheme never relies on the diagonal corner cells of $\Psi$ or $D$, as it does not use a full 27-points stencil as seen in Eq.~(\ref{3d2shocs2}), no additional copies are necessary.  Due to the large number of extra serialized memory copies for the boundary cell threads, the performance of the three-dimensional codes is expected to be less than the two-dimensional codes (this is indeed the case as shown in Sec.~\ref{s:speedup3d}).

\section{Speedup results}
\label{s:speedup}
Here we show the results of comparing the compute-time of the NLSEmagic CUDA integrators versus the serial MEX integrators.  The serial integrators are run on a single core of an Intel Corei3 CPU, and the GPU codes are run on a GeForce GTX 580 card (full specifications of both the CPU and GPU hardware are given in the Appendix).  Even though the double precision performance is artificially reduced in the GeForce cards, we show that this only effects our one-dimensional code, whereas our two- and three-dimensional codes have memory bottlenecks which make the reduced double precision performance negligible in practice.  (After possible future optimization of the code, this may not be the case and it would be beneficial to run the codes on a Tesla compute-only GPU).  Because the GeForce GPUs are many times cheaper than the Tesla (or Quadro) cards and the performance is not expected to be effected in a major way, we feel justified in running our codes on a GeForce card.

It is important to point out that our goal in the speedup timings is to compare the GPU integrators to a typical serial implementation, and demonstrate their efficiency.  As mentioned in the introduction, there are other options available to parallelize a serial code including OpenMP and MPI on desktop multi-core machines, as well as on large high-performance clusters.  Our focus is not on how the GPU integrators compare to other parallel implementations (such as the OpenMP code in Ref.~\cite{CODE_GP_C}), but how they perform compared to a serial implementation.  Although not the main focus, to further demonstrate the advantages of the GPU implementation, we give estimates on how the GPU integrators would perform versus the theoretical maximum parallel performance of an OpenMP code run on a quad-core CPU with the same specifications as the single CPU core used in the serial runs.

For all speedup tests, we use the example simulations of Sec.~\ref{s:examples} with an end-time of $t=50$ and vary the total grid-size from about $~1000$ to $~3,000,000$.  In each case we compute the computation time of the integrators, ignoring the small extra time required for generating the initial condition and outputting results.

Before computing the speedup results, it is first necessary to examine the effect that CPU-GPU memory transfers has on the performance of the integrators.

\subsection{Chunk-size Limitations}
\label{s:chunk-size}
As mentioned in Sec.~\ref{s:gpuphys}, memory transfers between the CPU and GPU are very slow, and therefore it is best to compute as much as possible on the GPU before transferring the data back.  For simulations where analysis results are cumulative, the entire simulation can be performed on the GPU with only two memory transfers (one at the beginning and one at the end of the computation).  For most studies of time-dependent problems such as the NLSE, it is desired (or essential) to have access to the computation data at regular intervals in order to save the data, display it for observation and animations, and to run intermediate analysis.  However, the more times the data is needed by the CPU, the slower the code will perform.  Therefore for NLSEmagic, the more frames the simulation is plotted and analyzed, the slower the overall code will be.  

In order to use the codes in the most efficient manner, it is necessary to see how much the simulation is slowed down as a function of the size of the chunk of time-steps performed by the CUDA integrators (which we have designated as the \emph{chunk-size}).  Once a chunk-size is found which exhibits acceptable lack of slow-down, the number of frames that the solution is viewable will be determined by the number of time-steps.

\begin{figure}[hbtp]
\centering
\includegraphics[width=2.8in]{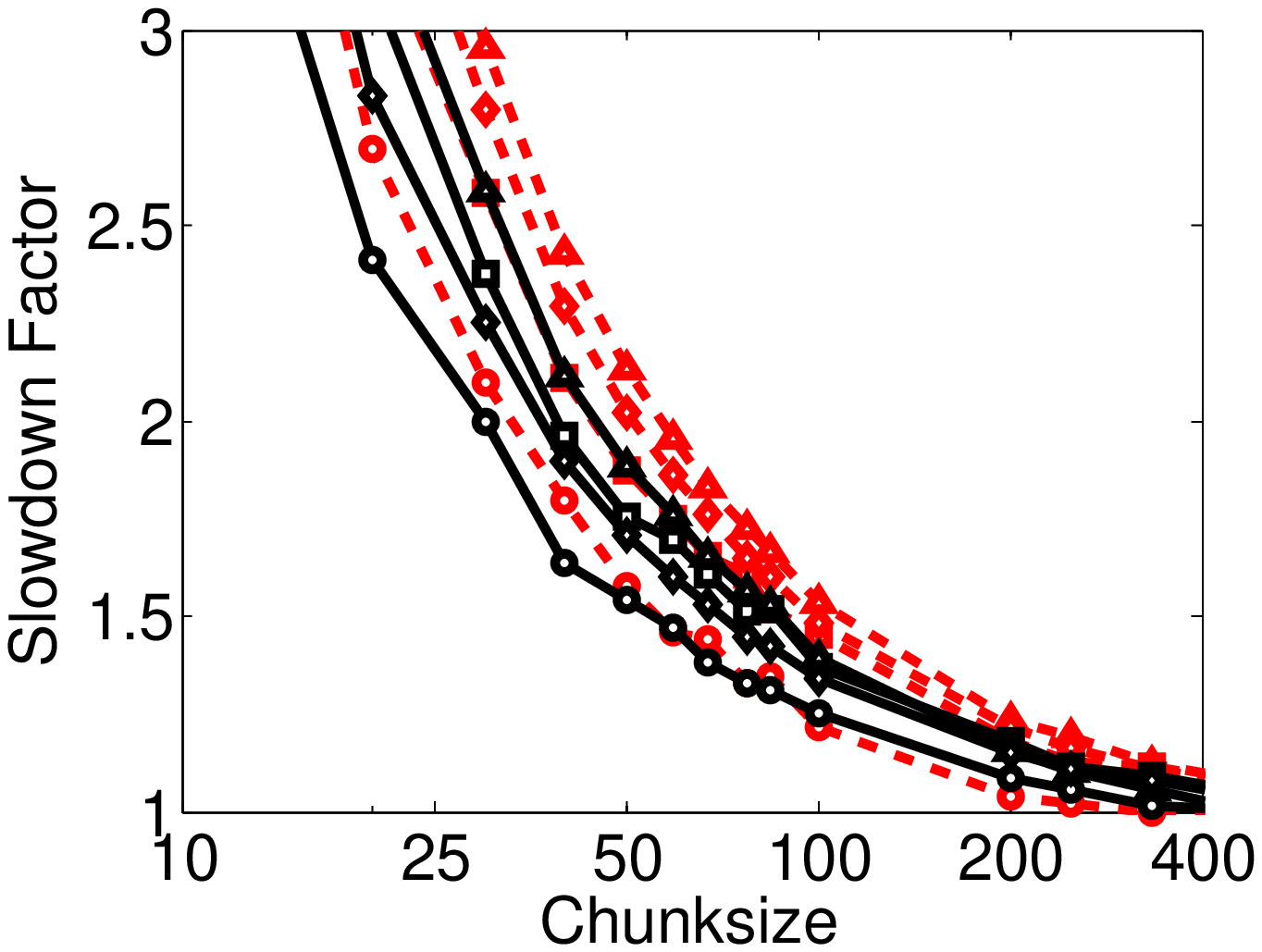}
\includegraphics[width=2.8in]{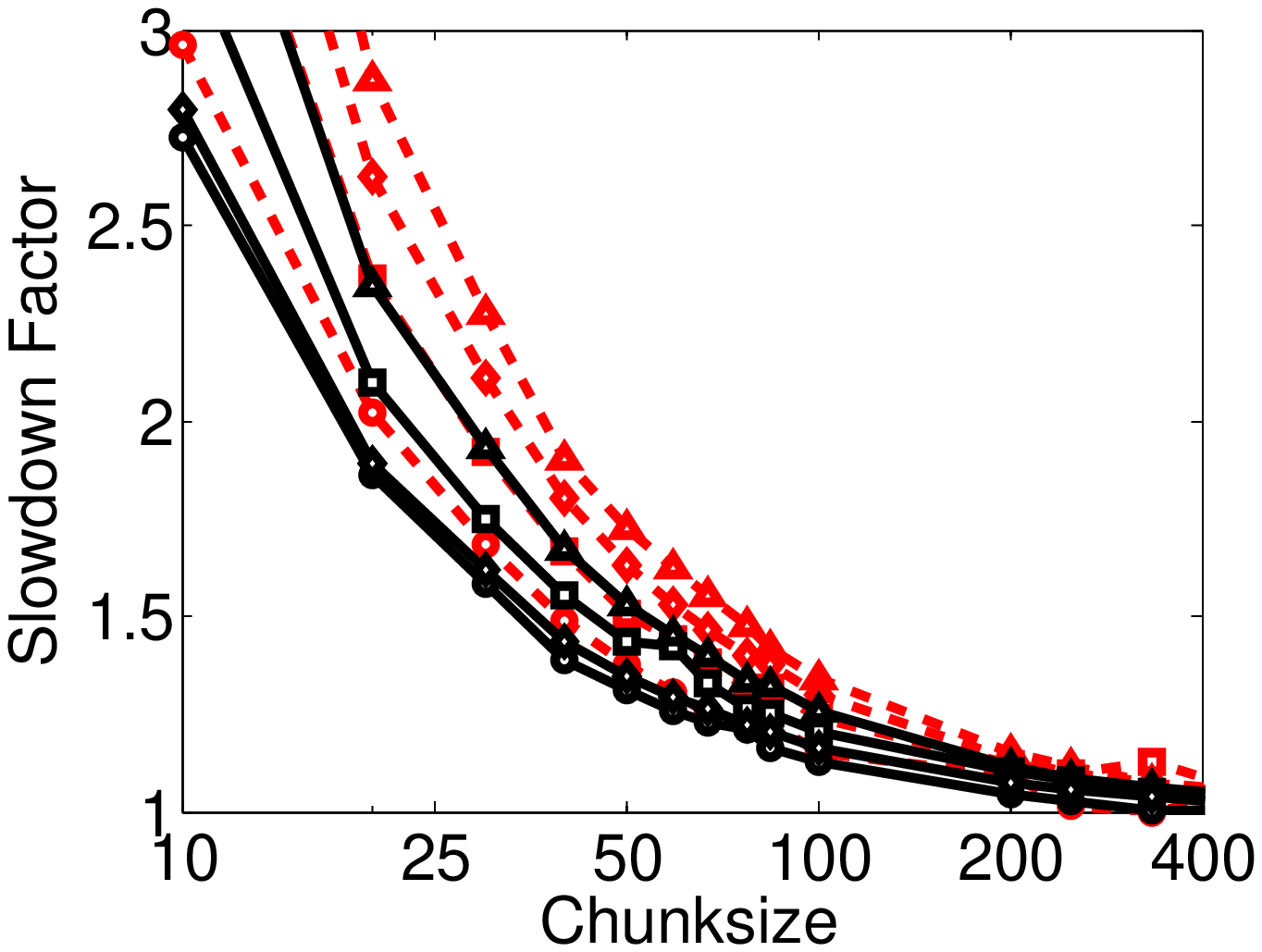} \\
\includegraphics[width=2.8in]{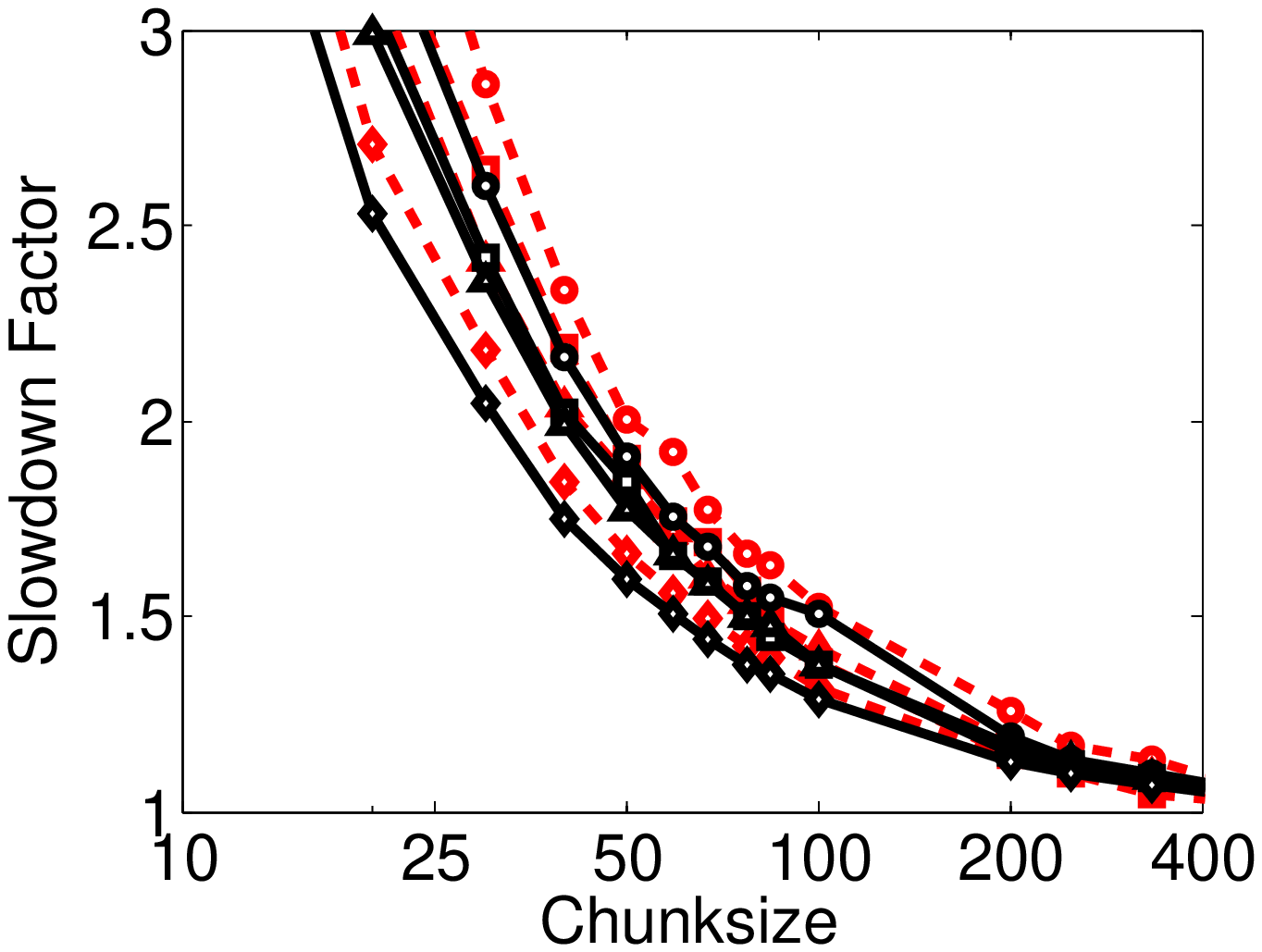}
\includegraphics[width=2.8in]{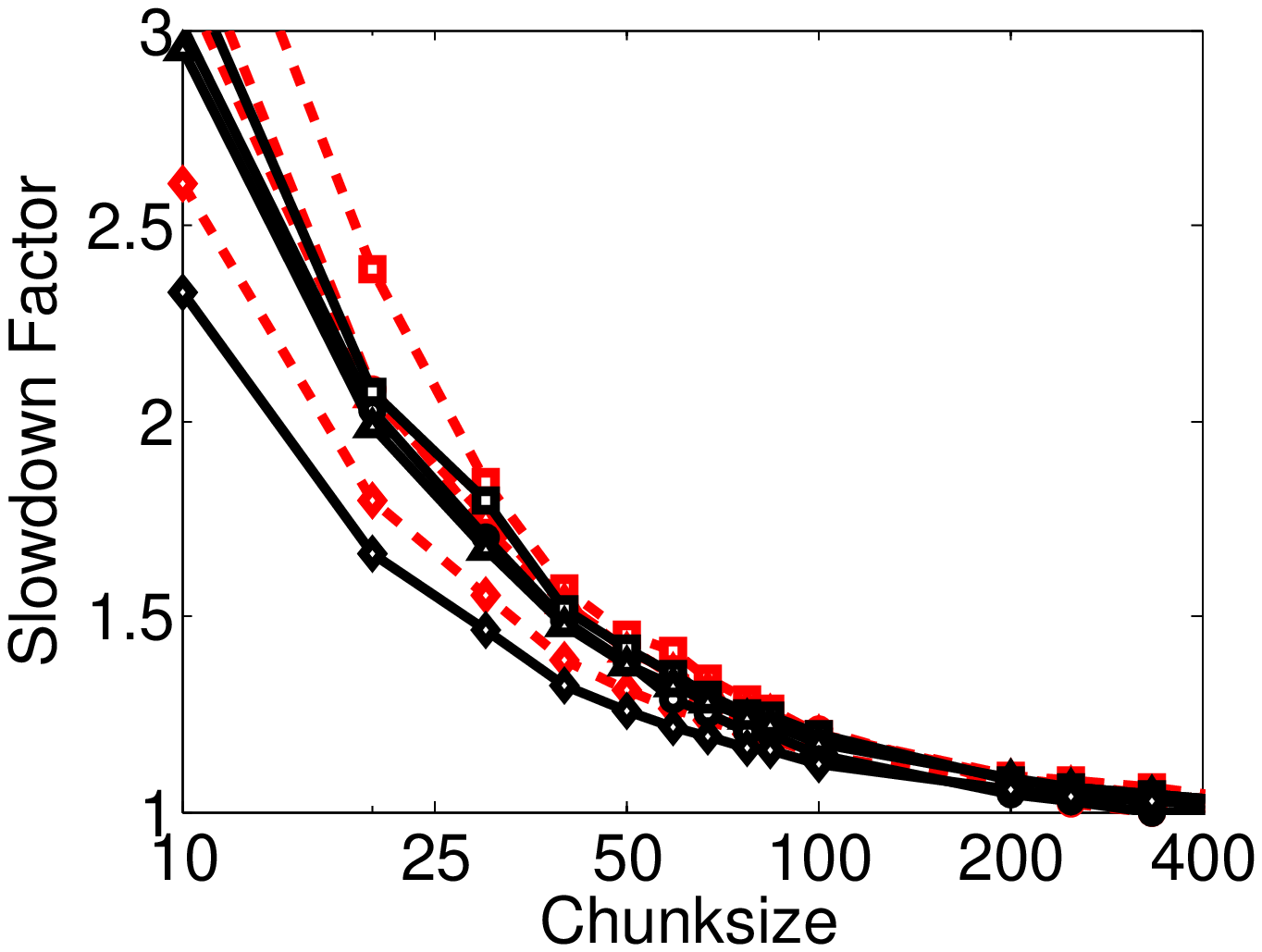} \\
\includegraphics[width=2.8in]{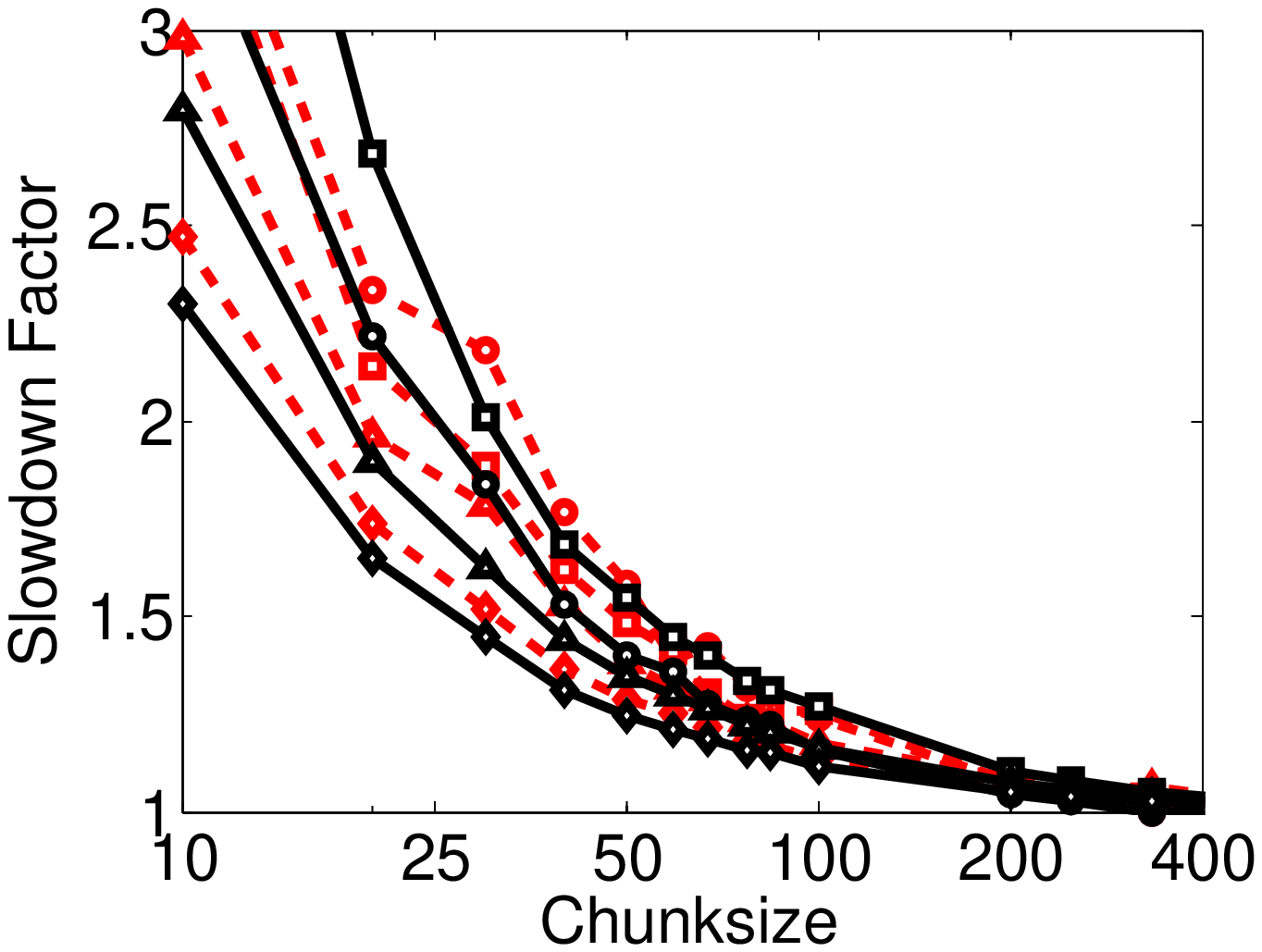}
\includegraphics[width=2.8in]{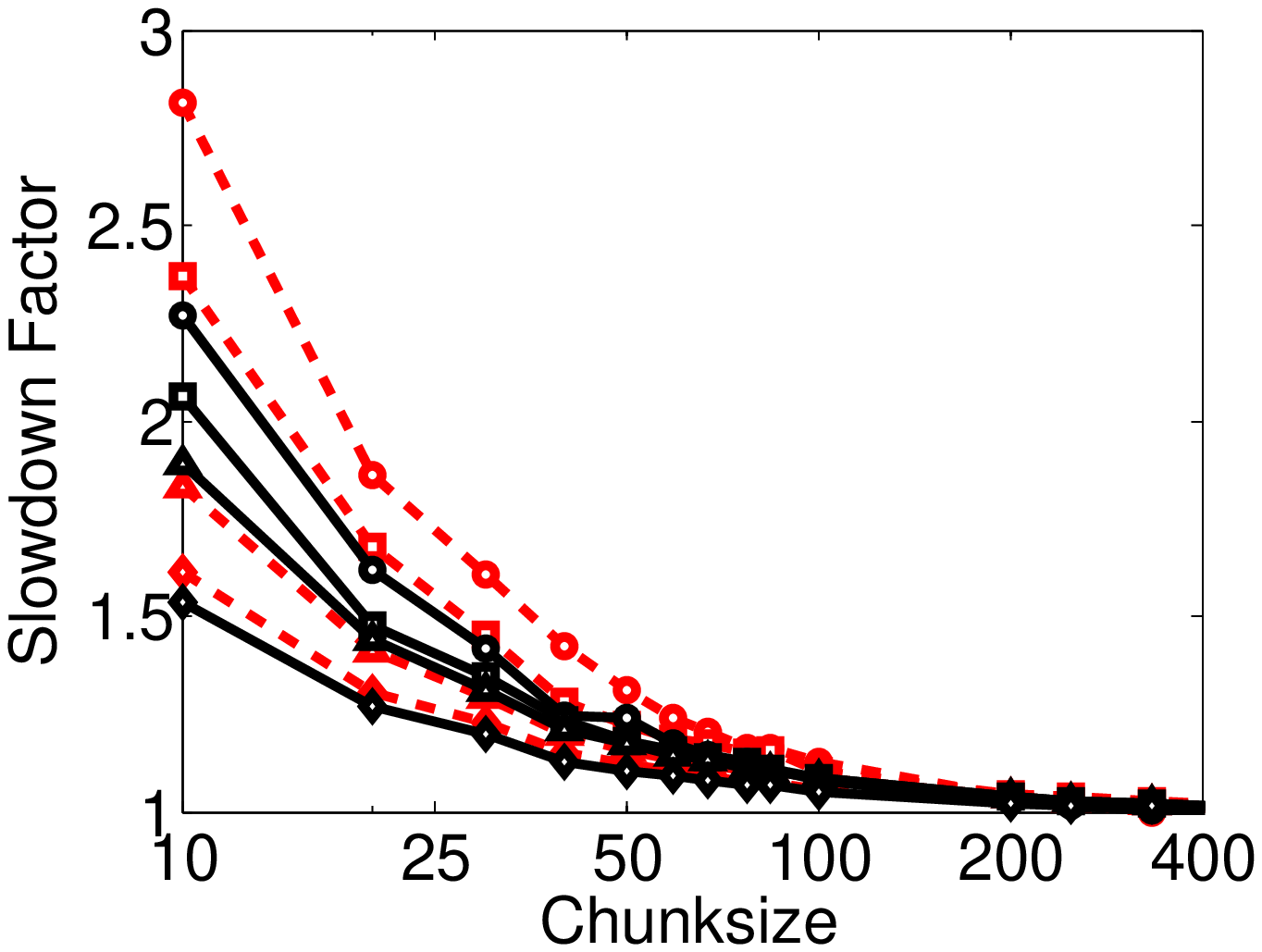}
\caption[Slowdown factor as a function of chunk-size of the NLSEmagic CUDA  integrators.]{(Color online) Slowdown factor as a function of chunk-size of the CUDA NLSEmagic integrators computed using the examples of Sec.~\ref{s:examples} with an end-time of $t=5$ with $1000$ time-steps.  The slowdown is defined as how much slower the simulation is compared to the fastest possible simulation (where the chunk-size equals the number of time-steps -- not shown).  The results are shown for the RK4+CD and RK4+2SHOC (left and right respectively), for one-, two-, and three-dimensional simulations (top to bottom respectively), and for single and double precision (dashed and solid lines respectively).  The circle, square, triangle, and diamond lines represent total grid-sizes ${\bf N}$ of $1000$, $10,000$, $100,000$, and $1,000,000$ respectively.  \label{f:chunk-size}}
\end{figure}

To see how the chunk-size affects performance of the NLSEmagic CUDA integrators, we run the examples described in Sec.~\ref{s:examples} for various chunk-sizes and compare the compute-times.  Since the performance of the codes due to chunk-size is not affected by the total number of time steps (as long as the number of time steps is larger than several chunk-sizes), we fix the end-time of the simulations to be $t=5$ and the time-step to be ${\bf k}=0.005$ yielding $1,000$ time-steps.  We run the examples in each dimension with single and double precision for both the RK4+CD and RK4+2SHOC schemes for total grid sizes (${\bf N}$) of around $1,000$, $10,000$, $100,000$, and $1,000,000$ (the two- and three-dimensional codes have slightly different resolutions due to taking the floor of the square- or cubed-root of ${\bf N}$ as the length in each dimension respectively).   To eliminate dependency of the results on the specific compute times of the given problem (which would make comparison of different grid sizes difficult), for each chunk-size, we compare the compute times of the simulation to the fastest time of all the simulations of that problem.  As expected, this fastest time nearly always occurs when the chunk-size is equal to the number of time-steps.  We thus compute a `slowdown' factor as the time of the fastest run divided by the time of the other runs.  This makes the results applicable to almost all simulations run with NLSEmagic.   The results are shown in Fig.~\ref{f:chunk-size}.
We see that in each case, the chunk-size has to be somewhat large for the codes to perform with top efficiency.  Simulations with lower chunk-sizes can be done with acceptable slowdown based on the figures.  In the higher dimensional cases, the larger the total grid size, the less the slowdown factor for a given chunk-size (in the one-dimensional case, the results are varied, with the lowest resolution tested outperforming the highest).  Also, in general, it is seen that the higher-dimensional codes have less slowdown for a given chunk-size than lower-dimensional codes.  Likewise, using the 2SHOC versus the CD scheme and double versus single precision, lower the slowdown for a given chunk-size as well.  This is understandable since the more work the CUDA kernels are doing, the higher percentage of the total time is spent on computations compared to  memory transfers to the CPU and back.  In either case, when running NLSEmagic, it is best to first look up on Fig.~\ref{f:chunk-size} where the problem being simulated falls, and then choose an appropriate chunk-size to get the best performance out of the codes (within the simulation requirements).

\subsection{One-Dimensional Speedup Results}
\label{s:speedup1d}
\begin{figure}[p!]
\centering
\includegraphics[width=2.8in]{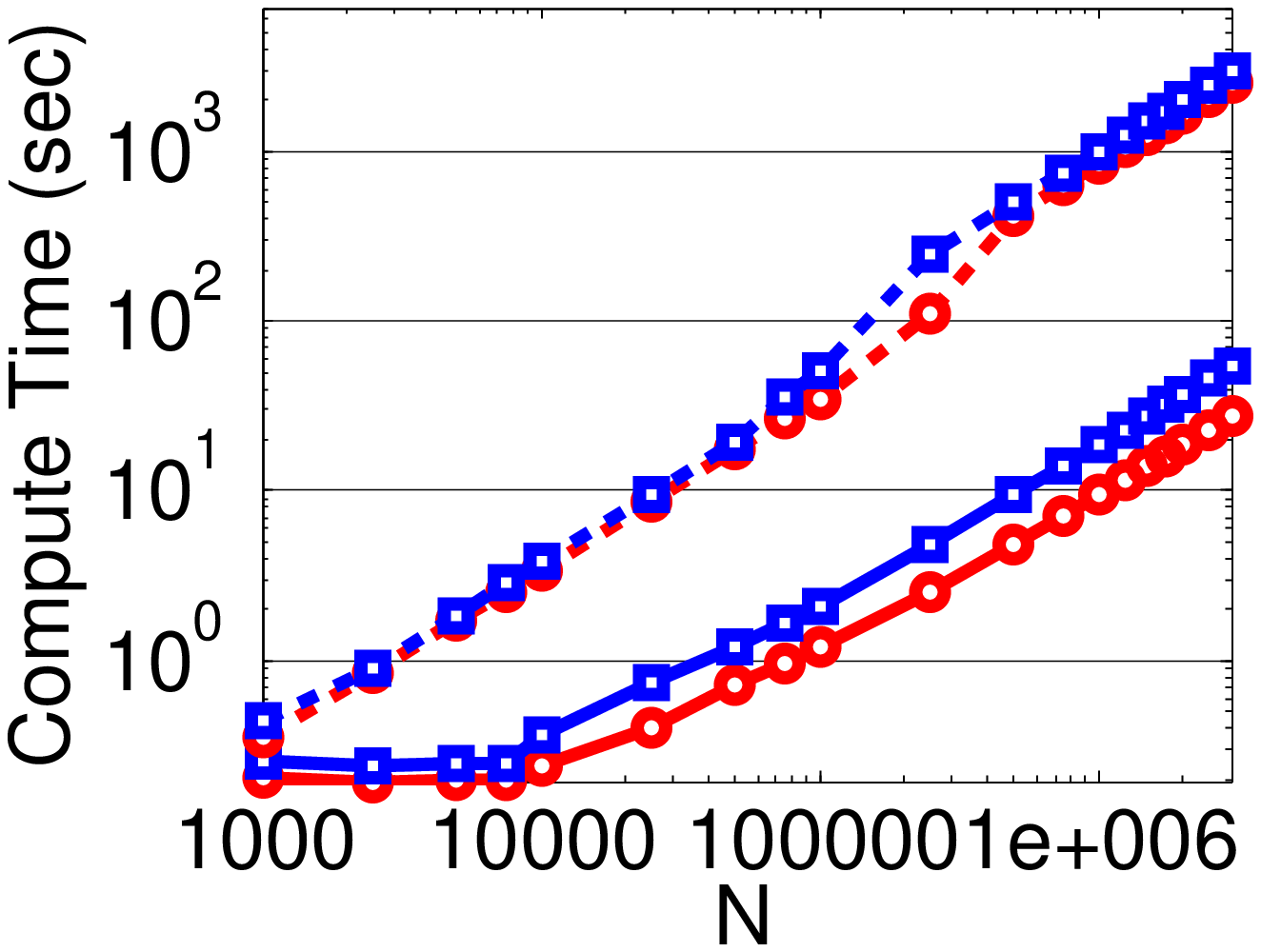}
\includegraphics[width=2.8in]{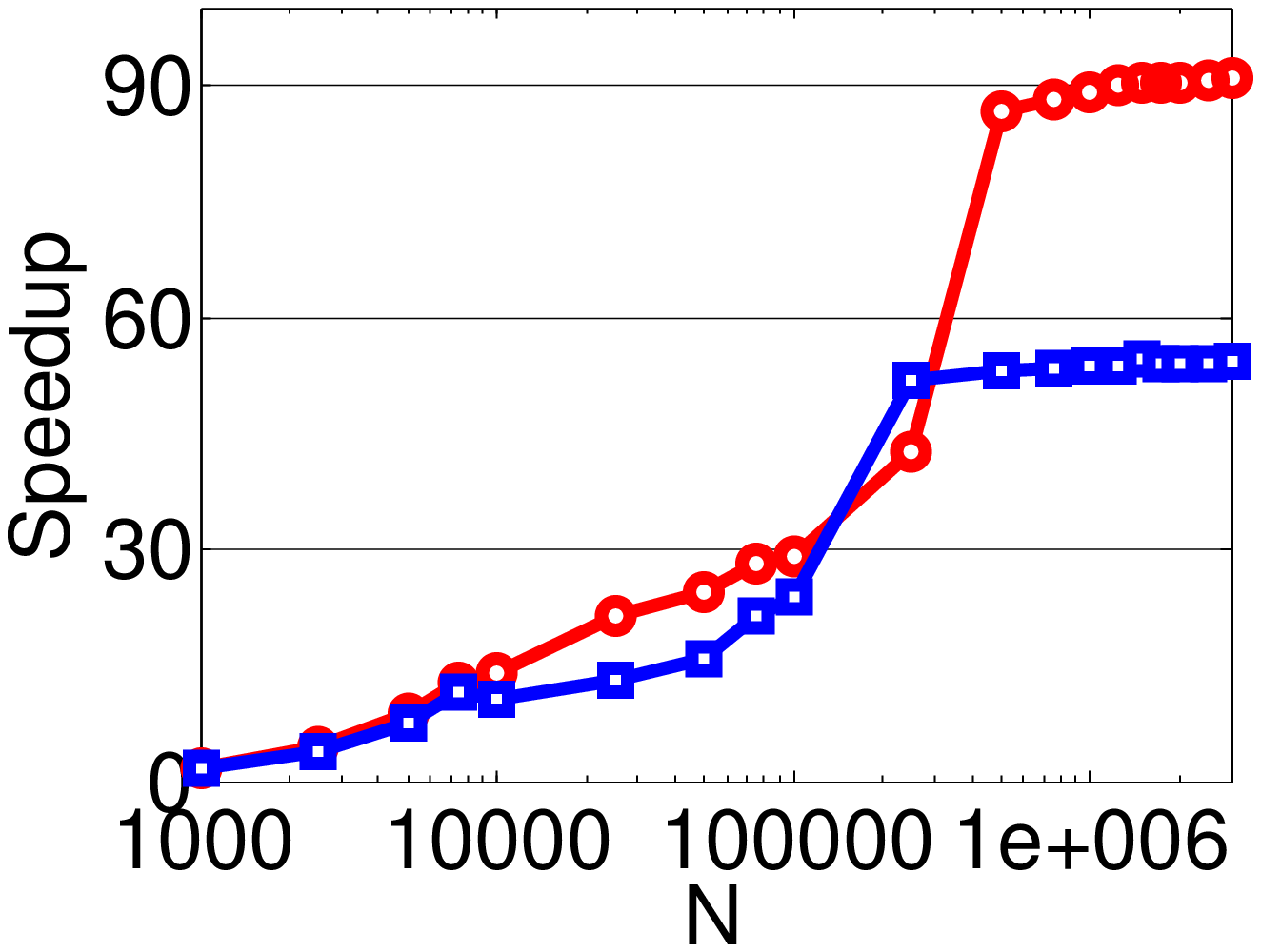} \\
\includegraphics[width=2.8in]{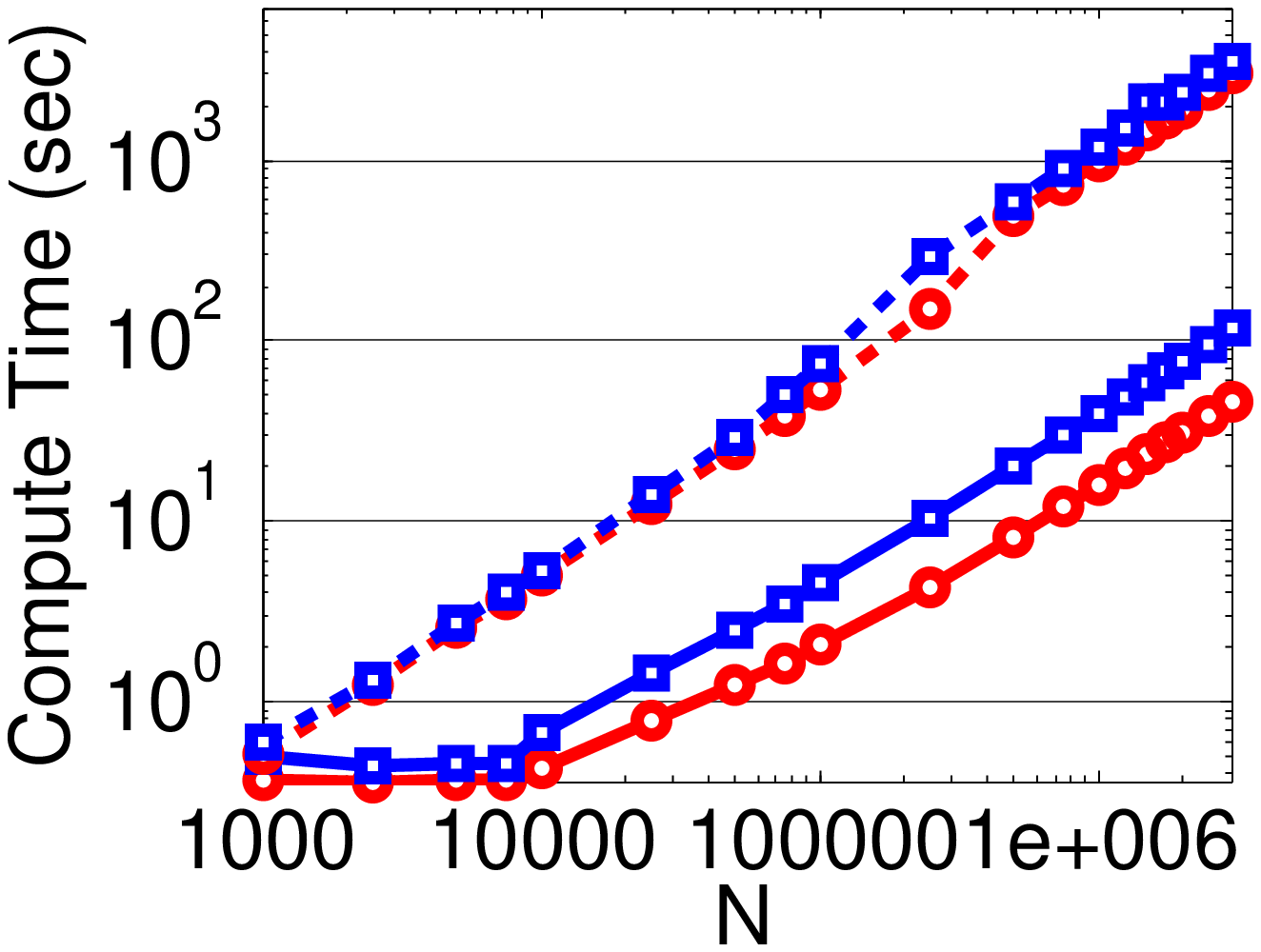}
\includegraphics[width=2.8in]{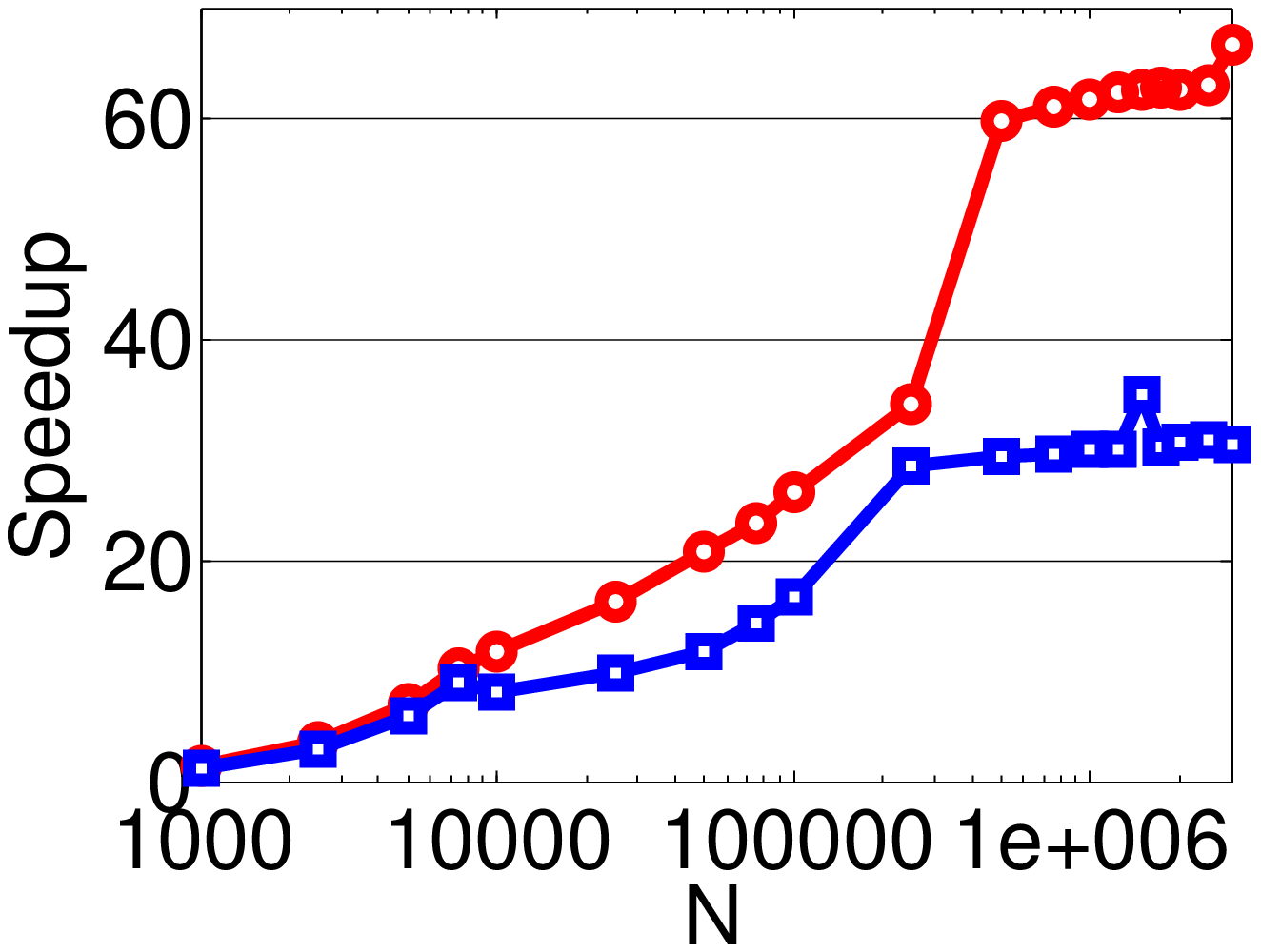}
\caption[Computation times and speedups of the one-dimensional NLSEmagic CUDA MEX integrators.]{Computation times (left) and speedups (right) of the one-dimensional NLSEmagic CUDA MEX integrators (solid) versus the serial MEX integrators (dashed) for simulating the dark soliton solution described in Sec.~\ref{s:examples} with an end-time of $t=50$. The results are shown for both the RK4+CD scheme (top) and the RK4+2SHOC scheme (bottom) using both single (red circles) and double (blue squares) precision. \label{f:speedup1D}}
\end{figure}

For the one-dimensional results, we use the dark soliton solution in Sec.~\ref{s:examples} with an end-time of $t=50$.  The solution is plotted five times during the simulation yielding a chunk-size of $5,000$, which is well over the efficiency requirements as discussed in Sec.~\ref{s:chunk-size}.  The soliton is simulated with a grid size which varies from $N=1,000$ to $N=3,000,000$.  Validation of the codes is done by comparing the solution to the known exact solution.  The simulation compute-times and speedups compared to the serial MEX integrators are shown in Fig.~\ref{f:speedup1D} and Table~\ref{t:speedup1D}.
\begin{table}[p!] 
\centering 
\caption[One-Dimensional GPU Speedup Results]{(Color online) Subset of the one-dimensional timing results from Fig.~\ref{f:speedup1D}.  The results are shown for the RK4+CD scheme in single (s) precision (best results) and for the RK4+2SHOC in double (d) precision (weakest results).}
\begin{tabular}{|l||r|r||r|r|||c|c|} \cline{2-7}
\multicolumn{1}{c|}{ \;} &  \multicolumn{2}{|c||}{GPU Time (sec)} & \multicolumn{2}{|c|||}{CPU Time (sec)} & \multicolumn{2}{|c|}{SPEEDUP} \\ \hline
$N$      & CD (s)  & 2SHOC (d) &    CD (s) & 2SHOC (d) & CD (s)  & 2SHOC (d) \\ \hline \hline
1000     &  0.21 &   0.49  &    0.35  &    0.61 &  1.69 & 1.23  \\ \hline
10000    &  0.24 &   0.67  &    3.39  &    5.38 & 14.07 & 7.99  \\ \hline
100000   &  1.21 &   4.52  &   35.15  &   75.31 & 29.12 & 16.65 \\ \hline
1000000  &  9.41 &  39.72  &  838.21  & 1192.27 & 89.08 & 30.01 \\ \hline
3000000  & 27.65 & 117.32  & 2515.56  & 3581.40 & 90.98 & 30.53 \\ \hline
\end{tabular}
\label{t:speedup1D}
\end{table} 
The best results are those of using the CD+RK4 scheme in single precision where we observed speedups around $90$ for the larger grid sizes.  Since, as was shown in Fig.~\ref{f:mexspeedup}, the serial integrators are about $12$ times faster than the MATLAB script codes for those resolutions, the NLSEmagic CUDA MEX codes for the one-dimensional CD+RK4 scheme in single precision for large resolutions are about \emph{1000 times faster} than the equivalent MATLAB script code.  In terms of actual time for the simulation tested, this would equate to taking roughly $30$ seconds using the GPU code, $40$ minutes using the serial MEX code, and $8$ \emph{hours} $20$ minutes using the MATLAB script code.  Assuming a perfect OpenMP implementation of a quad-core CPU of equivalent specifications, the GPU integrators still achieve a maximum speedup of over $20$.

Since the block size for the integrators was chosen to be 512, it is understandable that the compute time for the CUDA MEX codes stays quite low until resolutions of $10,000$ or so.  This is because the GPU being used has 16 MPs, and therefore resolutions up to $8000$ can be computed in one sweep of the GPU MPs, while higher resolutions require multiple blocks to be computed on the same MP.

It is noticeable that the double-precision performance is almost half that of the single precision.  This is partly due to what was mentioned in Sec.~\ref{s:gpuphys} that the GeForce cards have their double precision FLOP count artificially reduced by three-quarters, making the FLOP count one-eighth that of the single precision performance.  In addition, memory transfer is considered to be a large factor in code performance, and since double-precision variables take twice the memory space as single precision, a reduction in performance is understandable.

It is also apparent that the speedup when using the 2SHOC scheme is lower than the CD scheme.  This is also understandable because in the 2SHOC scheme, kernels which only compute the $D$ array (the standard second-order Laplacian) are needed, and they have a smaller amount of floating-point operations than the kernels which compute the full NLSE and RK4 step.  Since the amount of computations in the $D$ kernel is so small but the amount of required memory transfers is comparable to those of $F(\Psi)$, the speedup is smaller.  As will be shown in Secs.~\ref{s:speedup2d} and \ref{s:speedup3d}, in higher dimensions, this issue is somewhat minimized due to the added number of points in the second-order Laplacian stencil (the speedup reduction is lowered from $27\%$ in one dimension to $16\%$ in three dimensions in single precision, and from $45\%$ to $0\%$ in double precision).

\subsection{Two-Dimensional Speedup Results}
\label{s:speedup2d}
For the two-dimensional tests, we use the unoptimized steady-state dark vortex approximate solution of Sec.~\ref{s:examples}.  Since there is no analytical solution to the dark vortex, and moreover, since we are using only an approximation of the true solution, we cannot record the error of the runs.  Therefore, in order to validate the simulations, we displayed near-center cell data points for each frame and compared them with those from the output of the non-CUDA MEX codes, and checked that they were equivalent.

Like the one-dimensional tests, the solution is plotted five times during the simulation yielding a chunk-size of $5,000$, which is well over the efficiency requirements discussed in Sec.~\ref{s:chunk-size}. The vortex solution is simulated with a total grid size which varies from about ${\bf N}=N*M\approx 1,000$ to ${\bf N} \approx 3,000,000$.  The $N$ and $M$ dimensions are determined by taking the floor of the square-root of ${\bf N}$ and are sometimes slightly altered due to the block requirements of using the MSD boundary condition (see Sec.~\ref{s:cudamexspec}).  

\begin{figure}[p!]
\centering
\includegraphics[width=2.8in]{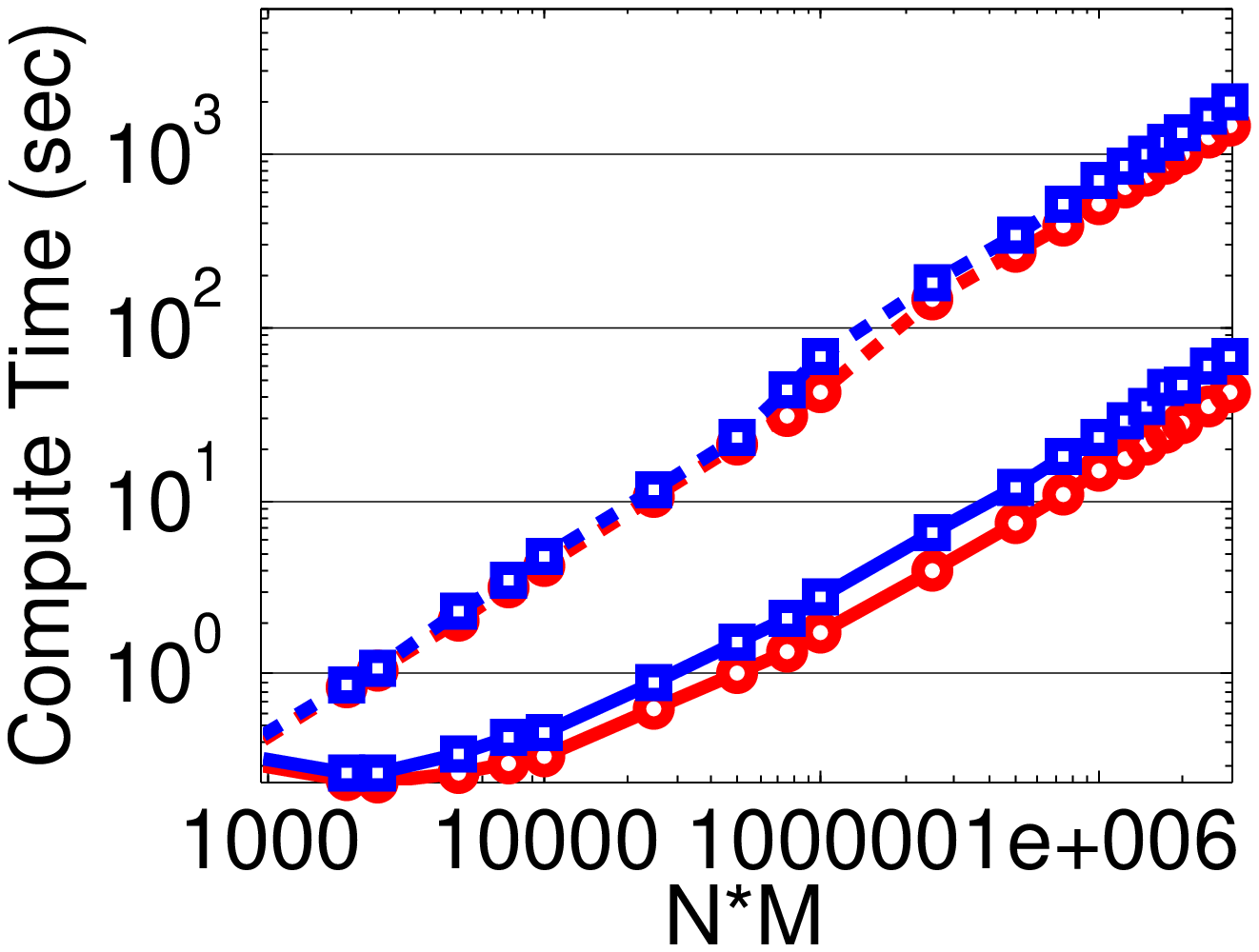}
\includegraphics[width=2.8in]{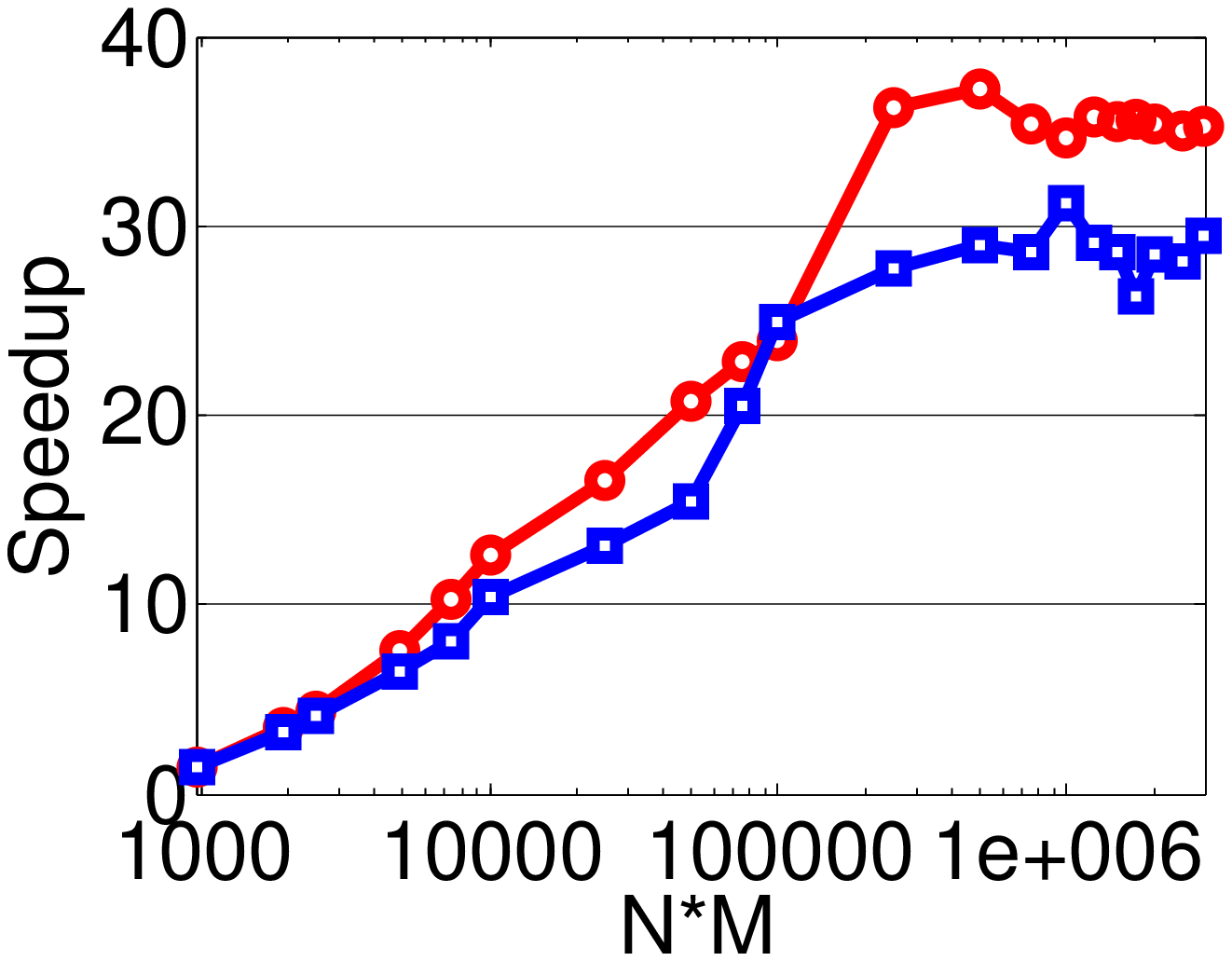} \\
\includegraphics[width=2.8in]{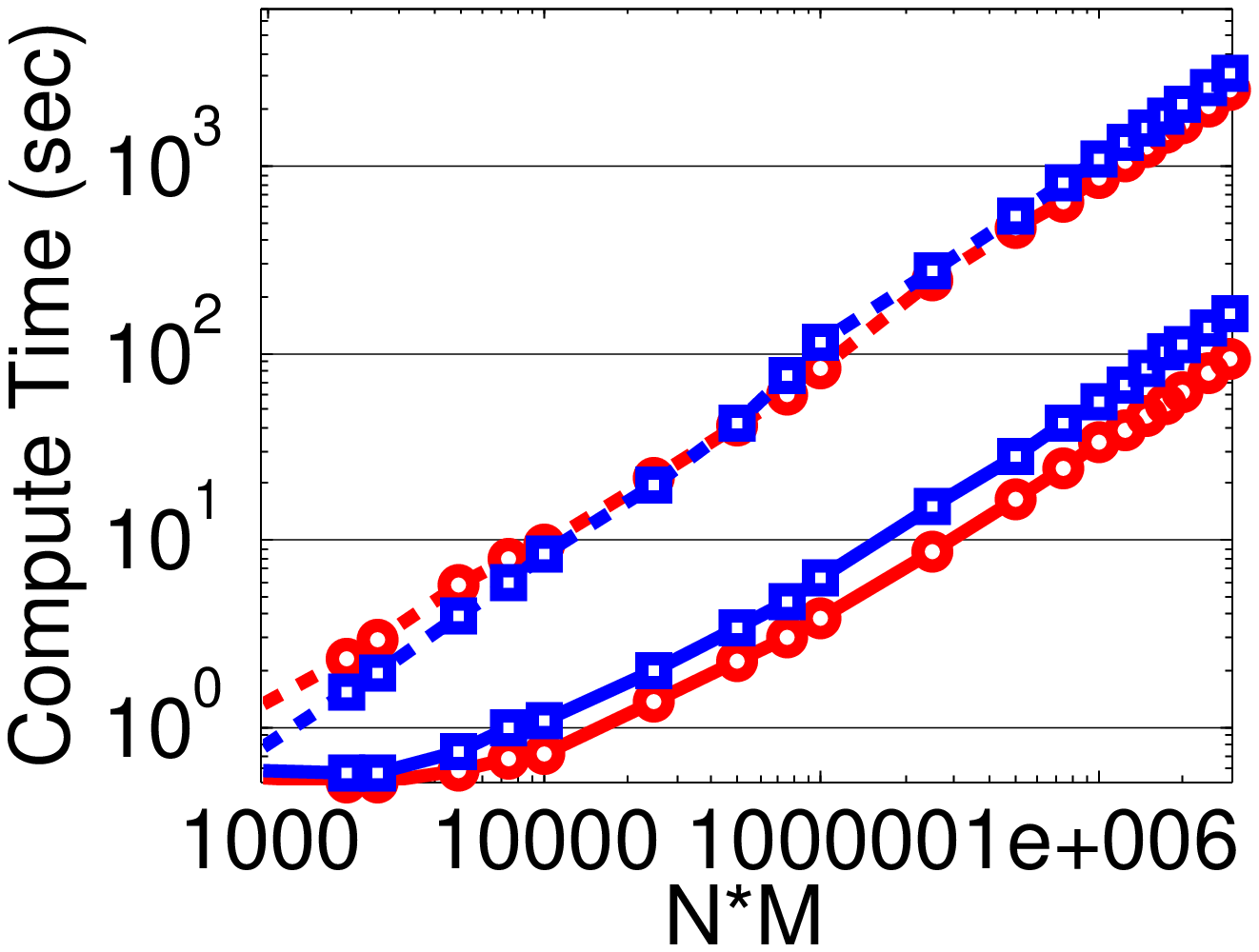}
\includegraphics[width=2.8in]{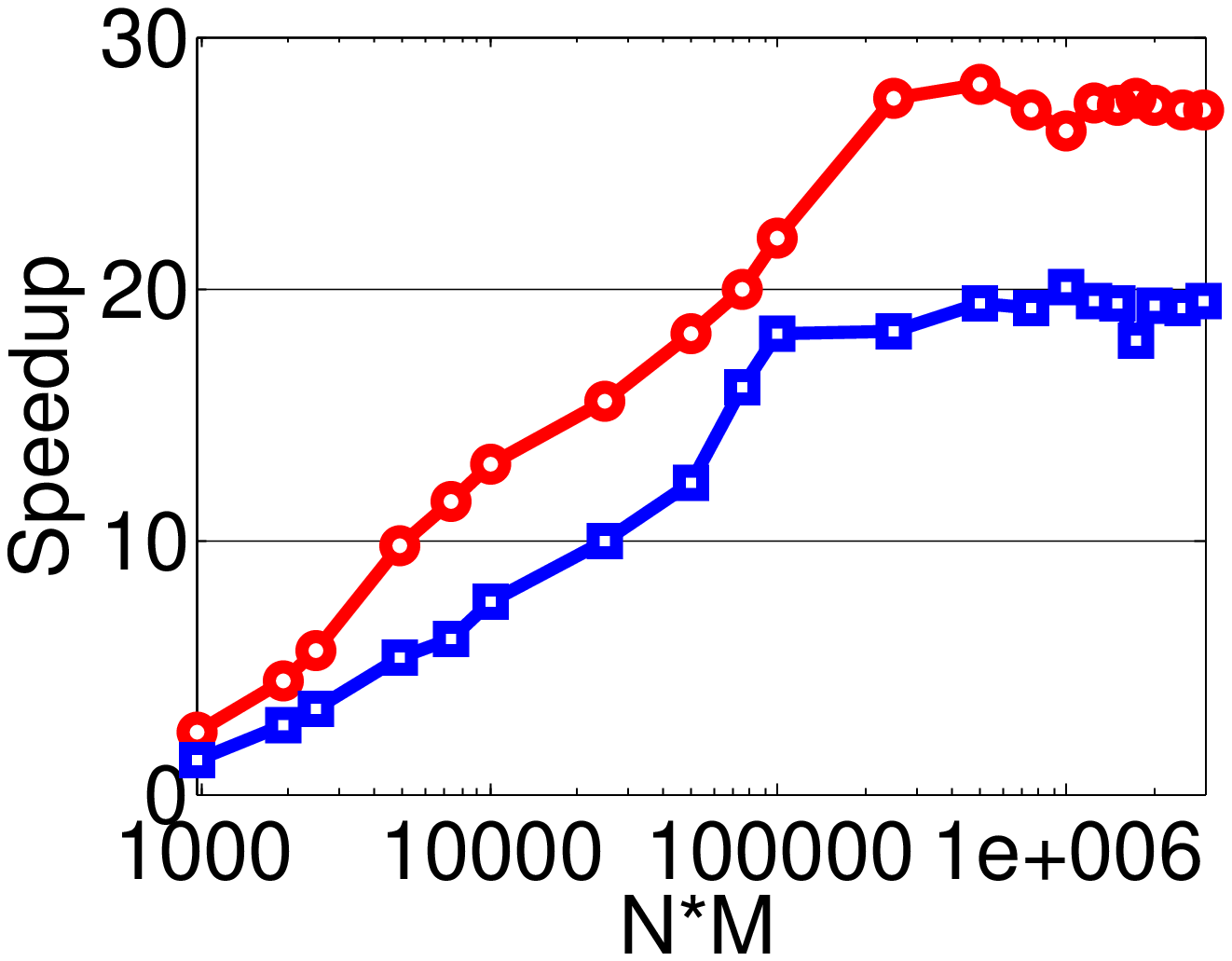}
\caption[Computation times and speedups of the two-dimensional NLSEmagic CUDA MEX integrators.]{(Color online) Computation times (left) and speedups (right) of the two-dimensional NLSEmagic CUDA MEX integrators (solid) versus the serial MEX integrators (dashed) for simulating the approximate dark vortex solution described in Sec.~\ref{s:examples} with an end-time of $t=50$. The results are shown for both the RK4+CD scheme (top) and the RK4+2SHOC scheme (bottom) using both single (red circles) and double (blue squares) precision. \label{f:speedup2D}}
\end{figure}

The simulation compute-times and speedups compared to the serial MEX integrators are shown in Fig.~\ref{f:speedup2D} and Table~\ref{t:speedup2D}.
\begin{table}[p!] 
\centering 
\caption[Two-Dimensional GPU Speedup Results]{Subset of the two-dimensional timing results from Fig.~\ref{f:speedup2D}.  The results are shown for the RK4+CD scheme in single (s) precision (best results) and for the RK4+2SHOC in double (d) precision (weakest results).}
\begin{tabular}{|l||r|r||r|r|||c|c|} \cline{2-7}
\multicolumn{1}{c|}{ \;} &  \multicolumn{2}{|c||}{GPU Time (sec)} & \multicolumn{2}{|c|||}{CPU Time (sec)} & \multicolumn{2}{|c|}{SPEEDUP} \\ \hline
$N*M$   & CD (s)& 2SHOC (d) &    CD (s) & 2SHOC (d) & CD (s)  & 2SHOC (d) \\ \hline \hline
961     &   0.29 &   0.59    &    0.42  &    0.78 &  1.43 &  1.32 \\ \hline
10000   &   0.34 &   1.09    &    4.25  &    8.35 & 12.61 &  7.65 \\ \hline
99856   &   1.76 &   6.25    &   42.20  &  113.96 & 23.99 & 18.23 \\ \hline
1000000 &  14.94 &  54.65    &  516.65  & 1094.72 & 34.59 & 20.03 \\ \hline
2999824 &  42.15 & 162.79    & 1485.88  & 3173.51 & 35.26 & 19.49 \\ \hline
\end{tabular}
\label{t:speedup2D}
\end{table} 
There is noticeably less speedup in the two-dimensional codes when compared to the speedup of the one-dimensional codes.  A possible explanation for the cause of the reduced speedup is that in two dimensions there are many more boundary cells versus interior cells of the CUDA blocks (in one dimension there are $O(1)$ while in a two-dimensional square block, there are $O(N)$).  Since, as described in detail in Sec.~\ref{s:cudacode2d}, block cell boundaries require additional global memory accesses (especially the corners), a decrease in performance was excepted.  Also, the number of total boundary grid points is much higher in two dimensions, in which case there are more threads computing grid boundary conditions than in one dimension.  This can cause less speedup in the codes since the MSD boundary condition requires an extra block-synchronization not needed in the internal scheme.  Slightly better speedup results could be obtained by choosing a problem that has Dirichlet boundary conditions, but it is important to show results that apply to a more general range of applications.

Despite the reduction in speedup of the two-dimensional codes, the speedup observed is still quite high, especially considering the cost of the GPU card.  In fact, in the RK4+CD test of ${\bf N}=1732^2=2999824$, the GPU code took about $40$ seconds, while the serial MEX code took over $24$ minutes.  Based on the MEX speedups of Sec.~\ref{s:mexspeedup}, the equivalent MATLAB script code would be expected to take almost \emph{$5$ hours} to complete the same simulation.  In this case, assuming a perfect OpenMP implementation of a quad-core CPU, the GPU integrators achieve a maximum speedup of around $8$.  Although lower than the one-dimensional case, this is still a significant improvement.  

It is important to note that resolutions for large ${\bf N}$ values (where the best speedups are observed) are more common in two dimensions.  Thus, the higher speedup results are expected to be more common in actual applications than those in one dimension. 


\subsection{Three-Dimensional Speedup Results}
\label{s:speedup3d}
For the three-dimensional timings, we use the unoptimized approximate dark vortex ring solution described in Sec.~\ref{s:examples}.  As was the case in the two-dimensional tests, we cannot record the error of the simulations.  Therefore, to validate the simulations, we once again displayed near-center cell data points for each frame and compared them with those from the output of the non-CUDA MEX codes.

Like the one- and two-dimensional tests, the solution is plotted five times during the simulation yielding a chunk-size of $5000$, which is once again well over the efficiency requirements discussed in Sec.~\ref{s:chunk-size}. The vortex ring is simulated with a total grid size which varies from ${\bf N}=N*M*L \approx 10,000$ to ${\bf N} \approx 3,000,000$.  The $N$, $M$, and $L$ dimensions are determined by taking the floor of the third-root of ${\bf N}$ and then slightly adjusted as needed to comply with the block requirements of using the MSD boundary condition mentioned in Sec.~\ref{s:cudamexspec}.  Although for the one- and two-dimensional tests we started with simulations of size ${\bf N}\approx1000$, in the three-dimensional case this was not possible due to the size of the vortex ring and the chosen spatial step-size (the vortex ring overlaps the grid boundaries when ${\bf N}\approx 1000$). 

\begin{figure}[p!]
\centering
\includegraphics[width=2.8in]{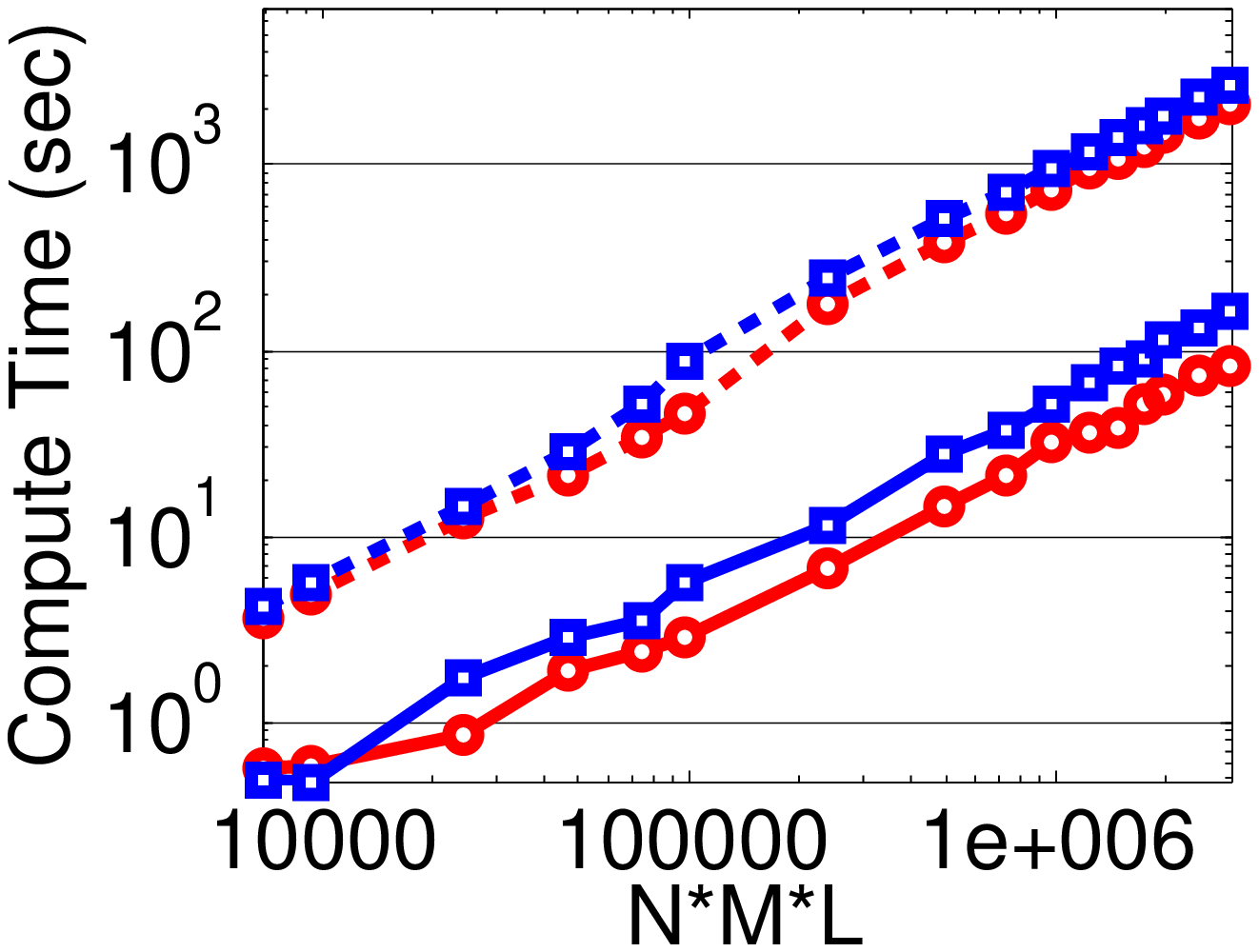}
\includegraphics[width=2.8in]{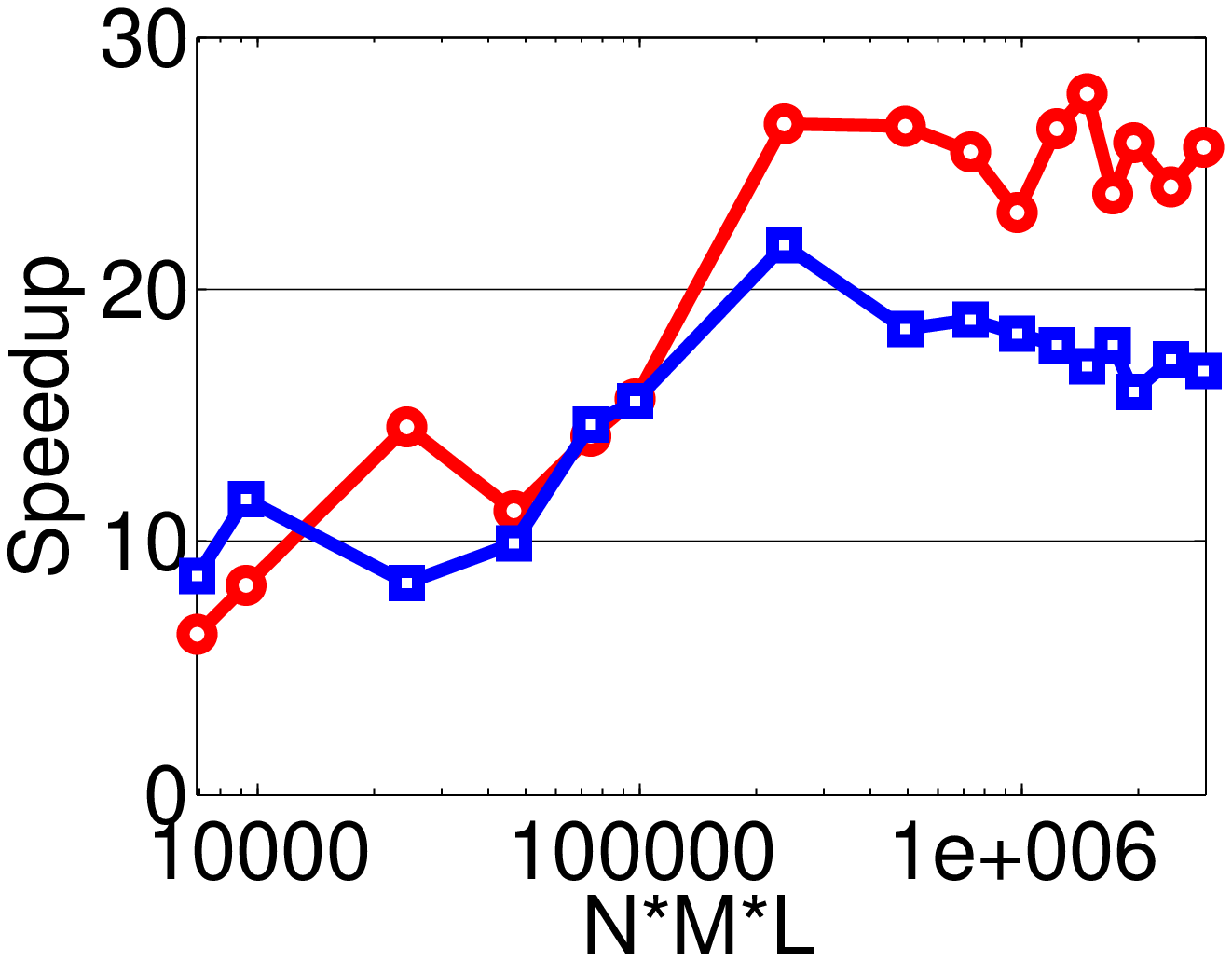} \\
\includegraphics[width=2.8in]{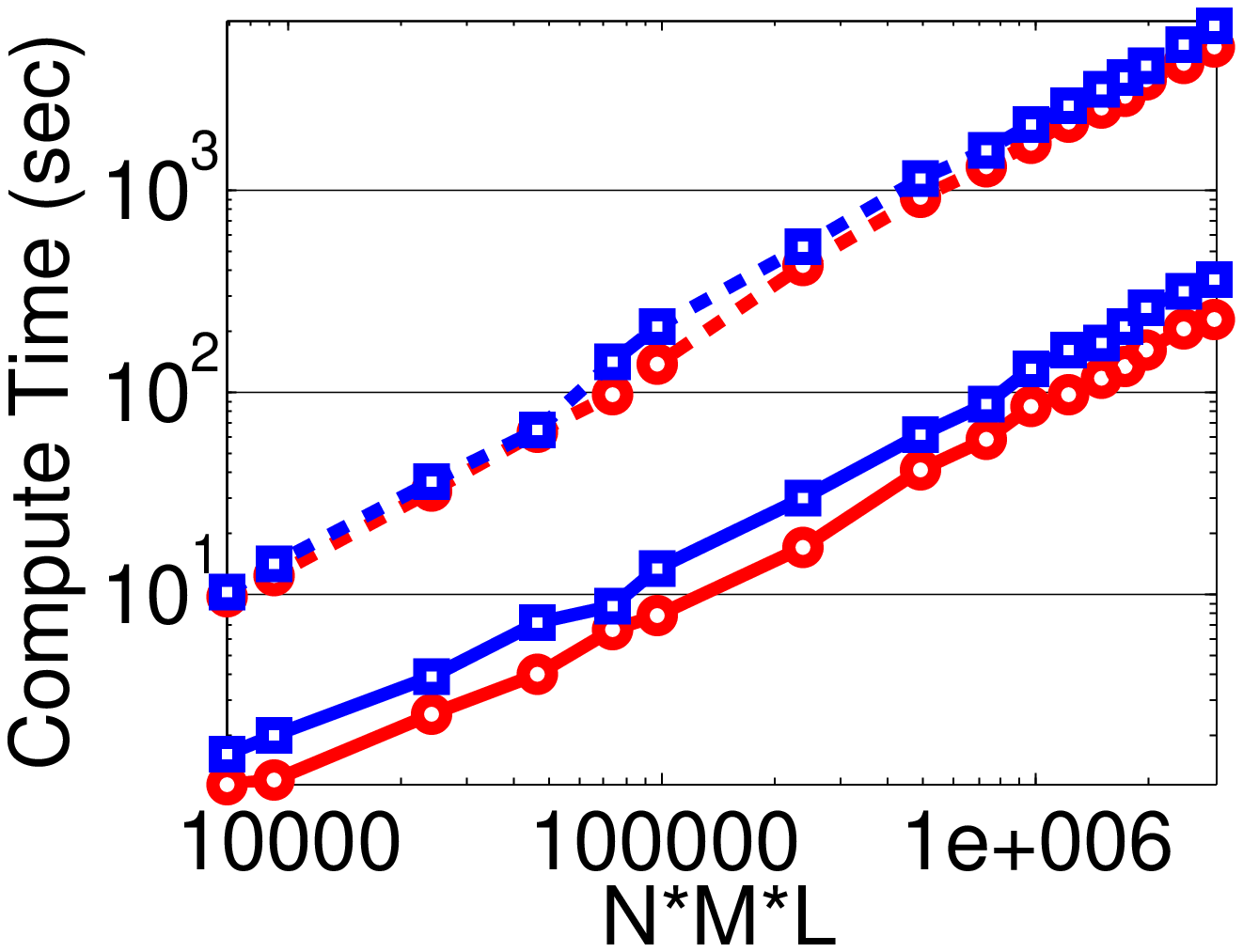}
\includegraphics[width=2.8in]{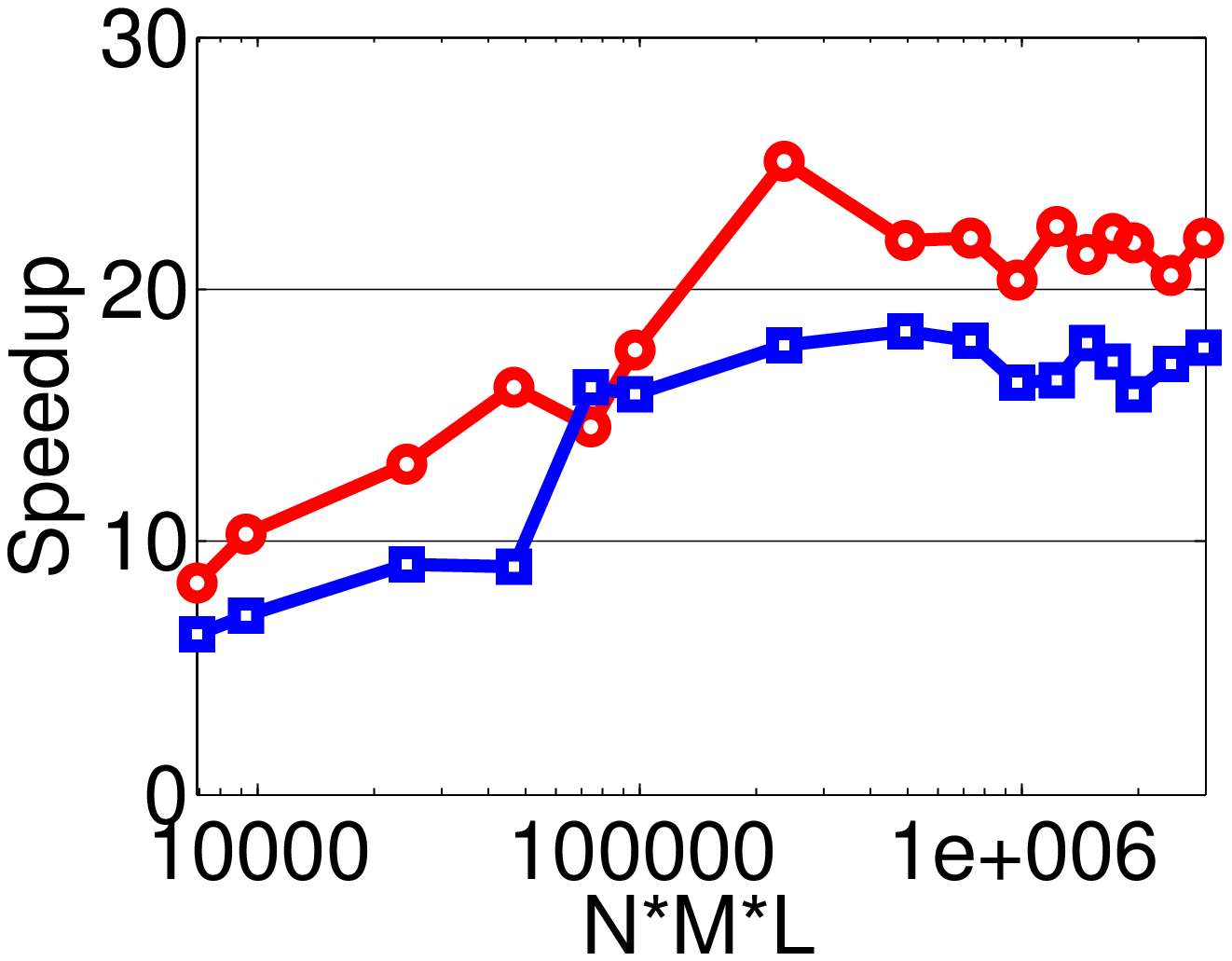}
\caption[Computation times and speedups of the three-dimensional NLSEmagic CUDA MEX integrators.]{(Color online) Computation times (left) and speedups (right) of the three-dimensional NLSEmagic CUDA MEX integrators (solid) versus the serial MEX integrators (dashed) for simulating the approximate dark vortex ring solution described in Sec.~\ref{s:examples} with an end-time of $t=50$. The results are shown for both the RK4+CD scheme (top) and the RK4+2SHOC scheme (bottom) using both single (red circles) and double (blue squares) precision. \label{f:speedup3D}}
\end{figure}

The simulation compute-times and speedups compared to the serial MEX integrators are shown in Fig.~\ref{f:speedup3D} and Table~\ref{t:speedup3D}.
\begin{table}[p!] 
\centering 
\caption[Three-Dimensional GPU Speedup Results]{Subset of the three-dimensional timing results from Fig.~\ref{f:speedup3D}.  The results are shown for the RK4+CD scheme in single (s) precision (best results) and for the RK4+2SHOC in double (d) precision (weakest results).}
\begin{tabular}{|l||r|r||r|r|||c|c|} \cline{2-7}
\multicolumn{1}{c|}{ \;} &  \multicolumn{2}{|c||}{GPU Time (sec)} & \multicolumn{2}{|c|||}{CPU Time (sec)} & \multicolumn{2}{|c|}{SPEEDUP} \\ \hline
$N*M*L$ & CD (s) & 2SHOC (d) &  CD (s) & 2SHOC (d) & CD (s) & 2SHOC (d) \\ \hline \hline
9261    &   0.59 &   2.00    &    4.85 &    14.17  & 8.29   &  7.10 \\ \hline
97336   &   2.90 &  13.34    &   45.43 &   211.23  & 15.65  & 15.84 \\ \hline
970299  &  31.85 & 131.74    &  733.46 &  2146.78  & 23.03  & 16.30 \\ \hline
2985984 &  82.14 & 365.47    & 2106.77 &  6460.61  & 25.65  & 17.68 \\ \hline
\end{tabular}
\label{t:speedup3D}
\end{table} 
We see that, once again, the higher dimensionality has reduced the speedup performance of the GPU codes.  A possible explanation is that the three-dimensional codes have even more boundary cells in the CUDA block ($O(N^2)$) than the two-dimensional case, and those cells require even more global memory accesses (as shown in Sec.~\ref{s:cudacode3d}, the 2SHOC scheme required up to $7$ accesses for the corner cells).  The additional number of grid boundary cells can also be a factor when using the MSD boundary condition as was explained in Sec.~\ref{s:speedup2d}.  However, in should be noted that the decrease in speedup is less (in percent) than the decrease from the one- to the two-dimensional codes.

Despite the reduction in performance compared to the two-dimensional results, the three-dimensional codes still exhibit very good speedups, especially considering the cost and portability of the GPU card.  For example, in the RK4+CD test of ${\bf N}=149^3=2985984$, the GPU code took about $1$ minute $20$ seconds, while the serial MEX code took over $34$ minutes.  Based on the MEX speedups of Sec.~\ref{s:mexspeedup}, the equivalent MATLAB script code would take over \emph{$5$ hours} to complete the same simulation.  In this case, the GPU integrators achieve a maximum speedup of around $6$ compared to a theoretical perfect quad-core OpenMP implementation.  Although smaller than in the lower-dimensional cases, this speedup can be quite significant in large three-dimensional simulations.  Also, in such problems, even moderately resolved solutions require very large total grid sizes.  Therefore, the larger speedups will be the most common in practical applications.  


All the speedup tests presented in this chapter were on grids which were nearly equal-sized in each dimension (i.e. $N \times N$ or $N \times N \times N$).  Depending on the sizes, the block structure could be more or less optimized in terms of filling the edge blocks with more or less capacity.  This explains why the speedups in two dimensions and (all the more so in three dimensions) are not smooth over changes in ${\bf N}$, but rather fluctuate.  Therefore, certain specific grid sizes have the potential to be more efficient.  As an example, we have simulated a vortex ring solution in a grid with dimensions $87 \times 87 \times 203 \approx 1,500,000$ for an end-time of $t=100$ and time-step size of $k=0.03$ (yielding a chunk-size of only $56$ which is on the low end of the efficiency requirements from Sec.~\ref{s:chunk-size}) with the RK4+2SHOC scheme and the MSD boundary condition in double precision, and saw a speedup of over $26$.  This is much higher than the comparable speedup shown in Fig.~\ref{f:speedup3D} for the same total grid size and scheme options.  Thus, the actual speedup in any given problem may be higher (or admittedly, lower) than those shown in Fig.~\ref{f:speedup3D}.

We wish to reiterate that the speedup tests performed in this section are to show how an inexpensive GPU upgrade of a researcher's workstation can result in a great speedup over a serial implementation of the NLSE codes.  In addition, we have noted that the GPU integrators should still yield a speedup of over $5$ for large three-dimensional problems compared to a perfect OpenMP implementation of the integrators on a quad-core CPU.  The speedup results of the NLSEmagic integrators will undoubtedly change with the release of new CPU and GPU architectures, but currently, in terms of both cost and performance, the GPU implementations seem to be the most efficient.

\section{The NLSEmagic code package}
\label{s:pack}
The codes described in this paper are packaged together and collectively named NLSEmagic: Nonlinear Schr{\"o}dinger Equation Multidimensional MATLAB-based GPU-accelerated Integrators using Compact high-order schemes.  The basic NLSEmagic code package contains all the serial and CUDA MEX integrators described in this paper as well as MATLAB script driver codes which are used as examples of how to use the integrators.  These scripts are called {\tt NLSEmagic1D.m}, {\tt NLSEmagic2D.m}, and {\tt NLSEmagic3D.m}, and include the files {\tt makeNLSEmagic1D.m}, {\tt makeNLSEmagic2D.m}, and {\tt makeNLSEmagic3D.m} which contain the commands to compile the integrators. In order to make setting up the codes as easy as possible, we also make available pre-compiled binaries of the serial and CUDA MEX integrators for 32- and 64-bit Windows.  All configuration files are included as well as an installation guide and a setup guide for compiling and running CUDA codes through the MATLAB MEX interface.

An important reason for distributing codes is to allow others to reproduce research results.  The issue of reproducible research when using numerical simulations is a topic of great concern in the scientific community \cite{COMP_PARADIGM,COMP_Reproduce,COMP_Reproduce2}.  Because many codes in use are written to run on specific hardware, and are typically not available to others, it is very difficult to validate numerical results without taking a  large amount of time to reproduce an equivalent code to test the simulations.  This is highly unfeasible, especially when the codes require parallelization and large computer clusters to run.  

As an example of reproducibility, we have included what we call `full research scripts' to the NLSEmagic package.  These driver scripts contain all the code necessary to reproduce the results computed in this paper (as well as those in the author's dissertation \cite{ME_DISS}).  This includes the initial conditions (vortices, vortex rings, etc.), as well as code to visualize, track, analyze, and save images and movies of the simulations.  They therefore also contain the script integrators used in the MEX speedup calculations of Sec.~\ref{s:mexspeedup}, which can be used to develop new numerical methods.  The full research driver scripts are called {\tt NLSE1D.m}, {\tt NLSE2D.m}, and {\tt NLSE3D.m}, and rely on other script codes which are also included in the download package.

The full-research and basic driver scripts of NLSEmagic make use of some freely available third-party MATLAB codes.  One such code is called {\tt vol3D} \cite{MAT_VOL3D} which is used to produce the volumetric renderings of the three-dimensional NLSE simulations.  Another code included is {\tt nsoli} \cite{OPT_NSOLI_BOOK} which is a Newton-Krylov nonlinear equations solver used to find numerically exact vortices and vortex rings.  Finally, a code called {\tt ezyfit} \cite{MAT_EZYFIT} is used to perform simple least-squares curve fitting to formulate models of the numerical results.
 
The NLSEmagic code package is distributed on a dedicated website free of charge \footnote{{\tt http://www.nlsemagic.com}}.  A Facebook users group is also maintained for posting updates and discussions \footnote{{\tt http://www.facebook.com/nlsemagic}}.  As of this writing, the website has received over $950$ unique hits from $50$ countries which indicates its interest to many researchers.  The code packages and website contain all the documentation necessary to setup, compile, install, and run the various codes.

\section{Conclusion}
\label{s:conclusion}
In this paper we have described the implementation of a new code package for simulating the multi-dimensional nonlinear Schr{\"o}dinger equation utilizing GPU acceleration called NLSEmagic.  The codes use an explicit finite-difference scheme using fourth-order Runge-Kutta for the time-stepping and both a second-order central differencing and a two-step compact fourth-order differencing for the Laplacian.  The codes are interfaced with MATLAB through MEX-compiled codes written in C and CUDA for ease of use and efficiency.  We have shown that the CUDA codes exhibit very good speedup results even when using an inexpensive off-the-shelf GPU card.  The NLSEmagic code package is distributed as a free download at {\tt www.nlsemagic.com}.  

{\tiny All trademarks and trade names including AMD, ATI, EVGA, Mathworks, MATLAB, MEX, Intel, NVIDIA, CUDA, GeForce, Quadro, Tesla, Fermi, Kepler, Microsoft, Windows, Visual C++, Facebook, and PGI are the property of their respective holders.}

\section*{Acknowledgments}
This research was supported by NSF-DMS-0806762 and the Computational Science Research Center (CSRC) at San Diego State University.  We gratefully acknowledge Professor Ricardo Carretero for his facilitation of this project and would like to thank Mohammad Abouali for his advise on optimizing the CUDA codes.

\appendix
\label{a:gpu}
\section{Specifications of the CPU and GPU used in this paper}
Here we show the specifications of the GPU and CPU that were used for the results in this paper.  The CPU has the following relevant specifications:
\begin{verbatim}
Operating System:           MS Windows 7 Enterprise 64-bit
CPU Name:                   Intel Core i3 540
CPU Clock Speed:            3.07 Ghz
Cores:                      2
Threads:                    4		
L1 Data cache:              2 x 32 KB
L1 Instruction cache:       2 x 32 KB
L2 cache:                   2 x 256 KB
L3 cache:                   4 MB		
DDR3 RAM:                   4.0 GB
Memory Clock rate:          669 Mhz		

Performance Information (Computed with QwikMark)
------------------------------------------------
CPU Core Performance:       61 Gflop/s
Memory Bandwidth:           6 GB/s
\end{verbatim}
The GPU has the following relevant specifications:
\begin{verbatim}
GPU Name:                                      GeForce GTX 580 (EVGA)
CUDA Driver Version / Runtime Version:         4.1 / 4.1
CUDA Capability Major/Minor version number:    2.0
Total amount of global memory (GDDR5 RAM):     1536 MB
(16) Multiprocessors x (32) CUDA Cores/MP:     512 CUDA Cores
GPU Clock Speed:                               1.59 Ghz
Memory Clock rate:                             2025.00 Mhz
Memory Bandwidth (GB/sec)                      192.4
L2 Cache Size:                                 786432 bytes
Total amount of shared memory per block:       49152 bytes
Total number of registers available per block: 32768
Maximum number of threads per block:           1024
Concurrent copy and execution:                 Yes with 1 copy engine
Run time limit on kernels:                     Yes
Concurrent kernel execution:                   Yes
Device has ECC support enabled:                No

Performance Information (Computed with CUDA-Z)
----------------------------------------------
Memory Copy
  Host Pinned to Device:          596.208 MB/s
  Host Pageable to Device:        532.289 MB/s
  Device to Host Pinned:          594.033 MB/s
  Device to Host Pageable:        531.315 MB/s
GPU Core Performance
  Single-precision Float:      1622440 Mflop/s
  Double-precision Float:       204026 Mflop/s
  32-bit Integer:               814731 Mflop/s
  24-bit Integer:               813770 Mflop/s
\end{verbatim}

\def\myitemsep{5pt}
\bibliographystyle{cpc}
\bibliography{CUDA}  

\begin{thebibliography}{10}

\bibitem{BEC_RCBOOK}
Kevrekidis, P., Frantzeskakis, D., and Carretero-Gonz{\'{a}}lez, R.,
\newblock {\em Emergent Nonlinear Phenomena in {B}ose-{E}instein Condensates:
  Theory and Experiment}, volume~45,
\newblock Springer Series on Atomic, Optical, and Plasma Physics, 2008.

\bibitem{ME_MI}
Caplan, R.~M., Hoq, Q.~E., Carretero-Gonz{\'{a}}lez, R., and Kevrekidis, P.~G.,
\newblock Opt. Comm. {\bf 282} (2009) 1399.

\bibitem{NLSE_nlpdebook}
Debnath, L.,
\newblock {\em Nonlinear Partial Differential Equations for Scientists and
  Engineers},
\newblock Birkhauser Boston, second edition, 2005.

\bibitem{CUDA_FD_2004}
Krakiwsky, S., Turner, L., and Okoniewski, M.,
\newblock Graphics processor unit {GPU} acceleration of finite-difference
  time-domain ({FDTD}) algorithm,
\newblock in {\em Proceedings of the 2004 International Symposium on Circuits
  and Systems}, volume~5, pages 265--268, 2004.

\bibitem{CUDA_FD_maxwell}
Adams, S., Payne, J., and Boppana, R.,
\newblock HPCMP User Group Conf.  (2007) 334.

\bibitem{CUDA_FD_2008}
Balevic, A. et~al.,
\newblock Accelerating simulations of light scattering based on
  finite-difference time-domain method with general purpose {GPU}s,
\newblock in {\em Proceedings of Computational Science and Engineering '08},
  pages 327--334, 2008.

\bibitem{CUDA_FD_2009}
Micikevicius, P.,
\newblock 3{D} finite difference computation on {GPU}s using {CUDA},
\newblock in {\em Proceedings of 2nd Workshop on General Purpose Processing on
  Graphics Processing Units}, pages 79--84, 2009.

\bibitem{CUDA_FD_3Dwave}
Mich{\'e}a, D. and Komatitsch, D.,
\newblock Geo. J. Inter. {\bf 182} (2010) 389.

\bibitem{CUDA_FD_GL}
Hawick, K.~A. and Playne, D.~P.,
\newblock Massey University Tech. Report  (2010) CSTN.

\bibitem{CODE_GP_FORTRAN}
Muruganandam, P. and Adhikari, S.~K.,
\newblock Comput. Phys. Commun. {\bf 180} (2009) 1888.

\bibitem{CODE_GP_C}
Vudragovic, D., Vidanovic, I., Balaz, A., Muruganandam, P., and Adhikari,
  S.~K.,
\newblock Computer Physics Communications {\bf 183} (2012) 2021.

\bibitem{CUDA_matlabgpu}
Mathworks,
\newblock http://www.mathworks.com/discovery/matlab-gpu.html  (2011).

\bibitem{CUDA_matlabgpuOLD}
NVIDIA,
\newblock http://developer.download.nvidia.com  (2007).

\bibitem{ME_2SHOC}
Caplan, R.~M. and Carretero-Gonz{\'a}lez, R.,
\newblock Submitted to Appl. Math and Comp.  (2012) arXiv:1109.1027.

\bibitem{cudadoc_pg}
NVIDIA,
\newblock http://developer.nvidia.com/nvidia-gpu-computing-documentation
  (2011).

\bibitem{CUDA_OpenCL}
Stone, J.~E., Gohara, D., and Shi, G.,
\newblock Comp. in Sci. and Engi. {\bf 12} (2010) 66.

\bibitem{CUDA_OpenCLvsCUDA1}
Komatsu, K. et~al.,
\newblock The Fifth International Workshop on Automatic Performance Tuning
  (iWAPT2010)  (2010).

\bibitem{CUDA_OpenCLvsCUDA2}
Fang, J., Varbanescu, A.~L., and Sips, H.,
\newblock A comprehensive performance comparison of {CUDA} and {O}pen{CL},
\newblock in {\em Proceedings of Parallel Processing '11}, pages 216--225,
  2011.

\bibitem{CUDA_OpenCLvsCUDAold}
Karimi, K., Dickson, N.~G., and Hamze, F.,
\newblock ArXiv e-prints  (2010) arXiv:1005.2581.

\bibitem{CUDA_CU2CL}
Martinez, G., Feng, W., and Gardner, M.,
\newblock {CU2CL}: A {CUDA}-to-{O}pen{CL} translator for multi- and many-core
  architectures,
\newblock Technical report, Virginia Tech, 2011.

\bibitem{CUDA_x86}
Group, T.~P.,
\newblock PGInsider {\bf 2} (2010) a1.

\bibitem{CUDA_MCUDA}
Stratton, J., Stone, S., and Hwu, W.,
\newblock {\em {MCUDA}: An Efficient Implementation of {CUDA} Kernels for
  {M}ulti-core {CPU}s}, volume 5335 of {\em Lecture Notes in Computer Science},
\newblock Springer Berlin / Heidelberg, 2008.

\bibitem{CUDA_PTXonAMDandIntel}
Vetter, J.~S. et~al.,
\newblock Comp. in Sci. and Engi. {\bf 13} (2011) 90.

\bibitem{NUM_Mimetic}
Castillo, J., Hymanb, J., Shashkov, M., and Steinberg, S.,
\newblock Appl. Num. Math. {\bf 37} (2001) 171–187.

\bibitem{RK4}
Butcher, J.,
\newblock Appl. Num. Math. {\bf 20} (1996) 247.

\bibitem{FD_STBNLSEDISS}
Dai, W.,
\newblock SIAM Jour. Num. Analy. {\bf 29} (1992) 174.

\bibitem{ME_RK4STB}
Caplan, R.~M. and Carretero-Gonz{\'a}lez, R.,
\newblock Submitted to App. Num. Math.  (2012) arXiv:1107.4810.

\bibitem{ME_MSD}
Caplan, R.~M. and Carretero-Gonz{\'a}lez, R.,
\newblock Submitted to Appl. Math and Comp.  (2012) arXiv:1110.0569.

\bibitem{SOL_Bright_Gray_Dark_Opt}
Kivshar, Y.~S. and Luther-Davies, B.,
\newblock Phys. Rep. {\bf 298} (1998) 81.

\bibitem{BEC_DYN_NONLIN}
R.~Carretero-Gonz{\'{a}}lez, D. F.-z.-k.-k. and Kevrekidis, P.,
\newblock Nonlinearity {\bf 21} (2008) R139.

\bibitem{ME_DISS}
Caplan, R.~M.,
\newblock {\em Study of Vortex Ring Dynamics in the Nonlinear {S}chr{\"o}dinger
  Equation utilizing {GPU}-Accelerated High-Order Compact Numerical
  Integrators},
\newblock PhD thesis, Claremont Graduate University and San Diego State
  University, 2012.

\bibitem{VR_NLSE_VEL_71}
Roberts, P.~H. and Grant, J.,
\newblock J. Phys. A: Gen. Phys. {\bf 4} (1971) 55.

\bibitem{matlabprimer8th}
Davis, T.,
\newblock {\em MATLAB Primer, Eighth Edition},
\newblock CRC Press, 2010.

\bibitem{CUDA_MemAccess}
Jang, B., Schaa, D., Mistry, P., and Kaeli, D.,
\newblock IEEE Trans. on Par. and Dist. Sys. {\bf 22} (2011) 105.

\bibitem{COMP_PARADIGM}
Post, D.~E. and Votta, L.~G.,
\newblock Phys. Today {\bf 58} (2005) 35.

\bibitem{COMP_Reproduce}
Donoho, D.~L., Maleki, A., Rahman, I.~U., Shahram, M., and Stodden, V.,
\newblock Comp. Sci. Engi. {\bf 11} (2009) 8.

\bibitem{COMP_Reproduce2}
Diethelm, K.,
\newblock Comp. in Sci. Engi. {\bf 14} (2012) 64.

\bibitem{MAT_VOL3D}
Woodford, O. and Conti, J.,
\newblock http://www.mathworks.com/matlabcentral/fileexchange/22940-vol3d-v2
  (2011).

\bibitem{OPT_NSOLI_BOOK}
Kelley, C.~T.,
\newblock {\em Solving Nonlinear Equations with {N}ewton's Method},
\newblock Fundamentals of Algorithms, SIAM, 2003.

\bibitem{MAT_EZYFIT}
Moisy, F.,
\newblock http://www.fast.u-psud.fr/ezyfit/  (2010).

\end{thebibliography}
\end{document}